\def\half {\mbox{$\textstyle {1 \over 2}$}}

\documentclass[12pt,a4paper]{article}
\usepackage{amsmath,amssymb,epsfig,array,lscape}

\renewcommand{\title}[1]{{\Large\bf\mbox{}\\\medskip#1\bigskip\medskip\\ }}
\renewcommand{\author}[1]{{\large #1\smallskip\\ }}
\newcommand{\address}[1]{{\em #1\medskip\\ }}
\def\Z{{\mathbb Z}}
\def\N{{\mathbb N}}
\def\zbar{\overline{z}}
\def\wbar{\overline{w}}
\def\hbar{\overline{h}}
\def\Tr{\mbox{Tr}}

\def\ket#1{| #1\rangle}
\def\V{{\cal V}}

\def\P{{\cal P}}
\def\M{{\cal M}}

\def\smallstar{\mbox{\scriptsize $\bigstar$}}
\def\smat#1{\mbox{\scriptsize{\mbox{$\begin{pmatrix}#1\end{pmatrix}$}}}}

\newcommand{\tpf}[3]{{\phi \! \begin{pmatrix} \scriptstyle #1 \\[-2mm]  \scriptstyle #2 , \, #3 \end{pmatrix} }}

\begin{document}
\hfill\today
\begin{center}
\title{Critical RSOS and Minimal Models II: \\
Building Representations of the\\ 
Virasoro Algebra and Fields}
\author{ Giovanni Feverati\footnote{
Email: feverati@ms.unimelb.edu.au} and Paul A. Pearce\footnote{
Email: P.Pearce@ms.unimelb.edu.au}}
\address{Department of Mathematics and Statistics\\
University of Melbourne\\Parkville, Victoria 3010, Australia}
\end{center}
\begin{abstract}
We consider $s\ell(2)$ minimal conformal field theories and the dual parafermion models. Guided by results for the critical $A_L$ Restricted Solid-on-Solid (RSOS) models and its Virasoro modules expressed in terms of paths, we propose a general level-by-level algorithm to build matrix representations of the Virasoro generators and chiral vertex operators (CVOs). We  implement our scheme for the critical Ising, tricritical Ising,  3-state Potts and Yang-Lee theories on a cylinder and confirm that it is consistent with the known two-point functions for the CVOs and energy-momentum tensor. Our algorithm employs a distinguished basis which we call  the $L_1$-basis. We relate the states of this canonical basis level-by-level to orthonormalized Virasoro states.
\end{abstract}

\section{Introduction}
\setcounter{equation}{0}

In this series of papers we study the relation between the critical Restricted Solid-on-Solid (RSOS) lattice models~\cite{ABF} and the minimal conformal field theories (CFT)~\cite{BPZ,FMS}. In Part I of this series (FPI)~\cite{FPI} we argued that, for these theories on a cylinder, the Hilbert space of physical states consists of fermionic states built from one-dimensional configurational RSOS paths on the lattice. As this series of papers will show, this allows us to translate much of the insight gained from studying finite-size critical RSOS lattice models by Yang-Baxter techniques~\cite{BaxBook} across to the study of the associated CFTs. 

For some time now, workers in the field have been trying to build matrix representations of the Virasoro algebra based on paths~\cite{FNO,RosgenV} and to make sense of a finitized Virasoro algebra~\cite{ItoyamaThacker,KooS}. In this paper, we propose a general algorithm to build matrix representations of the Virasoro generators and chiral vertex operators (CVOs) and implement it for the critical Ising, tricritical Ising,  3-state Potts and Yang-Lee theories on a cylinder. We emphasize that our algorithm constructs these matrices level-by-level and, while our algorithm appears to be consistent level-by-level, we do not have expressions for these matrices valid at arbitrary level.

The layout of the paper is as follows. In Section~2, we recall some relevant facts about CFT, the Virasoro algebra, Virasoro states, primary fields and their two-point functions. In Section~3, we present our general algorithm to build matrix representations of the Virasoro generators and primary fields. We implement this algorithm for the critical Ising, tricritical Ising,  3-state Potts and Yang-Lee theories on a cylinder in Sections~4--7. 
We conclude in Section~8 with a discussion of open questions for further research.

\section{Conformal Field Theory}
\setcounter{equation}{0}

\subsection{Minimal models and $\Z_k$ parafermions}
The $s\ell(2)$ minimal models~\cite{BPZ} $\mathcal{M}(p',p)$ with $p,p'$ coprime have central charges
\begin{equation}\label{centralcharge}
c= 1- \frac{6 \, (p-p')^2}{p \, p'}
\end{equation}
The conformal weights are
\begin{equation}
h=h_{r,s}={(rp-sp')^2-(p-p')^2\over 4pp'},\qquad r=1,2,\ldots,p'-1;\qquad s=1,2,\ldots,p-1
\end{equation}
and the Virasoro characters are
\begin{equation}
\chi_h(q)={q^{-c/24+h}\over (q)_\infty}
\sum_{k=-\infty}^\infty \big(q^{k(kpp'+rp-sp')}-q^{(kp+s)(kp'+r)}\big)
\label{bosonchar}
\end{equation}
where
\begin{equation}
(q)_n=\prod_{k=1}^n (1-q^k)
\label{qfactorial}
\end{equation}
The minimal models are unitary if $p-p'=\pm 1$ and non-unitary otherwise.
We consider only the diagonal $A$-type series with $p'<p$ and use the critical Ising ${\cal M}(3,4)$, tricritical Ising ${\cal M}(4,5)$ and Yang-Lee theories ${\cal M}(2,5)$ as prototypical examples.

The  $s\ell(2)$ $\Z_k$ parafermion models~\cite{ZamFat} are dual to the minimal models and have central charges
\begin{equation}
c={2(k-1)\over k+2},\qquad k=2,3,\ldots
\end{equation}
We consider only the diagonal $A$-type series and we use the $\Z_3$ or hard hexagon model~\cite{BaxHH,BaxBook} as the prototypical example. The hard hexagon model is in the universality class of the 3-state Potts model so we refer to this as the 3-state Potts CFT. Generally, the characters of the $\Z_k$ models are string functions but, for the $\Z_3$ model, these are easily related to the Virasoro characters of the $\M(5,6)$ model.

The minimal and $\Z_k$ parafermion models are rational and admit a finite number of primary fields $\phi(z)=\phi^{(h)}(z)$. We view these theories as arising from the continuum scaling limit of the $A_L$ RSOS models~\cite{ABF} with $L=p-1$ and $L=k+1$ respectively.

\subsection{Virasoro algebra and Virasoro states}

The Virasoro algebra
\begin{equation}
\mbox{Vir}=\langle L_n,n\in\Z\rangle 
\end{equation}
 is an infinite dimensional complex Lie algebra associated with conformal symmetry. The generators $L_n$ satisfy the commutation relations
\begin{equation}
[L_n,L_m]=(n-m)L_{n+m}+{c\over 12}\,n(n^2-1)\delta_{n,-m}
\end{equation}
where the central element $c$ is the central charge.  
These relations remain unchanged under an orthogonal change of basis 
$L_n\mapsto L_n'=U^TL_nU$. 
The Virasoro generators are the modes of the energy-momentum tensor
\begin{equation} \label{generators}
T(z)=\sum_{n\in\Z} L_n\, z^{-n-2}
\end{equation}
On a cylinder with prescribed boundary conditions, which is the case of primary concern here, there is just one copy of the Virasoro algebra. For bulk theories on the torus, however, there is a second copy $\overline{\mbox{Vir}}$ of Virasoro which is the antiholomorphic counterpart.

For rational CFTs, the Hilbert space ${\cal H}$ of states on which Vir acts is naturally decomposed into a finite direct sum  of  irreducible highest weight representations (Virasoro modules)
\begin{equation} \label{hilbert}
{\cal H}= \oplus_{h} \V_h
\end{equation}
where the sum is over the conformal weights $h$ of the primary fields $\phi(z)=\phi^{(h)}(z)$. 
The vacuum $\ket 0$ and primary (highest weight) states $\ket h$ are characterised by
\begin{equation}
L_0\ket h=h\ket h;\qquad L_n\ket 0=0,\ \ n\ge -1;\qquad L_n\ket h=0,\ \ n>0
\end{equation}
Moreover, there is a one-to-one correspondence between primary fields and primary states induced by 
\begin{equation} \label{prim.states}
\lim_{z\to 0} \phi^{(h)}(z)\ket 0 = \ket h
\end{equation}
The vacuum state $\ket 0$ with $h=0$ corresponds to the identity operator.

The generically reducible highest weight representation of Vir (Verma module) is the linear span of Virasoro states in the canonical form
\begin{equation} \label{linear_span}
L_{-n_j}L_{-n_{j-1}} \ldots L_{-n_1}\ket h,\qquad n_j\ge n_{j-1}\ge \cdots \ge n_1\ge 1
\end{equation}
If its maximal proper submodule is quotiented out, we are led to the irreducible Virasoro module  
$\V_h=\V_{c,h}$ and the states  (\ref{linear_span}) are no longer linearly independent due to the existence of null vectors.
The generic Virasoro module in the $h=0$ vacuum sector is shown in Figure~1. Typically, for given $c$ and $h$, some states at a given level enter in a vanishing non-trivial linear combination that is the null vector. 
Surprisingly, it seems that a complete set of linearly independent Virasoro states is not known even for the Ising model, although it is known~\cite {FNO} for the  Yang-Lee theory $\M(2,5)$ and the whole family $\M(2,2n+3)$, $n\ge 1$. Note that under an orthogonal change of basis 
($L_n\mapsto L_n'=U^TL_nU$, $\ket h\mapsto {\ket h}'=U^T\ket h$), Virasoro states transform into Virasoro states.

\begin{figure}[htb]
\hspace{.5in}
\begin{picture}(320,240)
\put(0,220){$\phantom{2}1$}\put(40,220){$\ket 0$}
\put(0,200){$\phantom{2}0$}\put(40,200){--}
\put(0,180){$\phantom{2}q^2$}\put(40,180){$L_{-2}\ket 0$}
\put(0,160){$\phantom{2}q^3$}
\put(40,160){$L_{-3}\ket 0$}
\put(0,140){$2q^4$}
\put(40,140){$L_{-4}\ket 0$}
\put(120,140){$L_{-2}^2\ket 0$}
\put(0,120){$2q^5$}
\put(40,120){$L_{-5}\ket 0$}
\put(120,120){$L_{-3}L_{-2}\ket 0$}
\put(0,100){$4q^6$}
\put(40,100){$L_{-6}\ket 0$}
\put(120,100){$L_{-4}L_{-2}\ket 0$}
\put(200,100){$L_{-3}^2\ket 0$}
\put(280,100){$L_{-2}^3\ket 0$}
\put(0,80){$4q^7$}
\put(40,80){$L_{-7}\ket 0$}
\put(120,80){$L_{-5}L_{-2}\ket 0$}
\put(200,80){$L_{-4}L_{-3}\ket 0$}
\put(280,80){$L_{-3}L_{-2}^2\ket 0$}
\put(0,60){$7q^8$}
\put(40,60){$L_{-8}\ket 0$}
\put(120,60){$L_{-6}L_{-2}\ket 0$}
\put(200,60){$L_{-5}L_{-3}\ket 0$}
\put(280,60){$L_{-4}^2\ket 0$}
\put(60,40){$L_{-4}L_{-2}^2\ket 0$}
\put(160,40){$L_{-3}^2L_{-2}\ket 0$}
\put(260,40){$L_{-2}^4\ket 0$}
\put(0,20){$8q^9$}
\put(50,20){$\ldots$}
\put(140,20){$\ldots$}
\put(220,20){$\ldots$}
\put(300,20){$\ldots$}
\put(0,0){$12q^{10}$}
\put(50,0){$\ldots$}
\put(140,0){$\ldots$}
\put(220,0){$\ldots$}
\put(300,0){$\ldots$}
\end{picture}
\caption{Virasoro module $\V_0$ of Virasoro states in the vacuum $h=0$ sector. The generic Virasoro character is $\chi_0(q)=\prod_{n=2}^\infty (1-q^n)^{-1}
=1+q^2+q^3+2q^4+2q^5+4q^6+4q^7+7q^8+8q^9+12q^{10}+\cdots$. In this sector, there is a null vector $L_{-1}\ket 0=0$ at level  $\ell=1$. For the minimal theories $\M(p',p)$, further null vectors appear. For example, for the Ising model $\M(3,4)$, there is one null vector at level $6$ and $7$ and two at level $8$. For $\M(4,5)$, the first null vector enters at  level $12$. In practice it is often convenient to truncate the states at some fixed level.}
\end{figure}

With reference to the vectors (\ref{linear_span}), the module $\V_h$ is graded according to the level
\begin{equation}
\V_h=
\mathop{\oplus}_{\ell=0}^\infty \V_{h,\ell},\qquad \ell=\sum_{i=1}^j n_i
\end{equation}
The Virasoro character $\chi_h(q)$, which is the generating function for the spectrum of the Virasoro module $\V_h$, is
\begin{equation} \label{main_character}
\chi_h(q)= \Tr_{\V_h} q^{L_0-c/24}=q^{-c/24+h}\sum_{\ell\in \N} d_\ell\, q^\ell,\qquad d_\ell\ge 0
\end{equation}
where $q$ is the modular parameter and the degeneracy $d_\ell=d^h_\ell=\mbox{dim}\,\V_{h,\ell}$ is the dimension of the space of states at level $\ell$.

\subsection{Fields}
Consider a bulk primary field $\phi^{(h)}(z,\zbar)$. 
Radial quantization imposes the following definition of adjoint
on the real surface $\bar{z}=z^{*}$
\begin{equation}
\phi^{(h)}(z,\zbar)^\dagger=\zbar^{\,-2h}z^{-2\hbar}\phi^{(h)}(1/\zbar,1/z)\label{Hermitian}
\end{equation}
for bulk (quasi-)primary fields ($z$ and $\bar{z}$ are independent complex variables, $z^*$ is the complex conjugate of $z$).
The decoupling between holomorphic and antiholomorphic degrees of freedom allows to 
drop the dependence on the antiholomorphic coordinate $\zbar$ and consistently use chiral fields. 
The full picture is obtained after restoring the antiholomorphic dependence.
The chiral form of the definition of adjoint
\begin{equation}
\phi^{(h)}(z)^\dagger= {(z^{*})}^{-2h} \phi^{(h)}(1/z^{*})\label{chiralHermitian}
\end{equation}
is used to define the dual space (bras)
\begin{equation} \label{bras}
\langle h | = \lim_{z \rightarrow 0}  \langle 0 | \phi^{(h)} (z) ^{\dagger} = \lim_{u \rightarrow \infty} u^{2h}  \langle 0 | \phi^{(h)} (u)  .
\end{equation}
Unlike the generators $L_n$, a chiral primary field $\phi^{(h)} (z)$ intertwines between different 
Virasoro modules, its action being dictated by the fusion rules
\begin{equation}
\phi^{(h_i)} \times \phi^{(h_j)} = \sum _k {\mathcal{N}_{ij}}^{k} \phi^{(h_k)}
\end{equation}
If the fusion coefficient is ${\mathcal{N}}_{ij}{}^{k} \neq 0$ then the corresponding restriction of the field 
$\phi^{(h_i)}$, called chiral vertex operator (CVO), is non-zero
\begin{equation}
\tpf{h_i}{h_k}{h_j} (z): \quad  \mathcal{V}_{h_j} \rightarrow \mathcal{V}_{h_k}  
\end{equation} 
As a consequence of global conformal invariance we have the expression for the two-point function
\begin{equation} \label{2point}
\langle \phi^{(h)} (z) \phi^{(h)} (w) \rangle \equiv \langle 0 | \phi^{(h)} (z) \phi^{(h)} (w) |0\rangle=
\frac{1}{(z-w)^{2 h}}
\end{equation}
corresponding to the trivial fusion coefficients ${\mathcal{N}_{h \, 0}}^{h}= {\mathcal{N}_{h \, h}}^{0}=1$ and to the non-zero CVOs:
\begin{eqnarray}
\tpf{h}{h}{0} (z) : & & \mathcal{V}_{0} \rightarrow \mathcal{V}_{h} \\
\tpf{h}{0}{h} (z) : & & \mathcal{V}_{h} \rightarrow \mathcal{V}_{0} 
\end{eqnarray}
The nonvanishing of the generic fusion coefficient  ${\mathcal{N}_{h \, h_1}}^{h_2} \neq 0$ corresponds
to a non-zero  three-point function
\begin{eqnarray} \label{3point}
 \langle \phi^{(h_2)}(u) \phi^{(h)} (z) \phi^{(h_1)} (w) \rangle &\equiv &
\langle 0| \phi^{(h_2)}(u) \phi^{(h)} (z) \phi^{(h_1)} (w) |0\rangle \\ 
&=&
\displaystyle\frac{\sqrt{C_{h_2,h,h_1}}}{(u-z)^{h_2+h-h_1} (u-w)^{h_2+h_1-h} (z-w)^{h+h_1-h_2}}    \nonumber
\end{eqnarray}
The two-point relation (\ref{2point}) is contained in  (\ref{3point}) if  we set $h_1$ or $h_2$ to zero with $C_{0,h,h_1}=C_{h_2,h,0}=1$. 

Using (\ref{prim.states}) and (\ref{bras})  in the three-point function, we have
\begin{equation} \label{limit}
\langle h_2 | \phi^{(h)} (z) | h_1 \rangle= \lim_{\rule{0mm}{1.7mm} u \rightarrow \infty \atop w \rightarrow 0} u^{2 h_2}
\langle 0| \phi^{(h_2)}(u) \phi^{(h)} (z) \phi^{(h_1)} (w) |0\rangle=  
\frac{\sqrt{C_{h_2,h,h_1}}}{z^{h+h_1-h_2}}  
\end{equation}   
forcing the following Laurent expansion 
\begin{equation} \label{laurent}
\tpf{h}{h_2}{h_1} (z)=
z^{h_2-h_1-h} \sum_{k\in\Z} \tpf{h}{h_2}{h_1} _{k+h_1-h_2}\, z^{-k}
\end{equation}
with  the operator modes 
\begin{equation}
\tpf{h}{h_2}{h_1} _{k+h_1-h_2}=\oint {dz\over 2\pi i}\, z^{k-1} \, \tpf{h}{h_2}{h_1} (z) \, z^{h_1+h-h_2} ,
\qquad k \in \Z
\end{equation}
It is  not possible to write a unique Laurent expansion for the various restrictions 
of a field.  
The analytic behaviour of $ \phi^{(h)} (z) | h_1 \rangle $ as $z \rightarrow 0$
and of  $\langle h_2 | \phi^{(h)} (z) $ as $z \rightarrow \infty$
requires the constraints
\begin{eqnarray} \label{constraint}
\tpf{h}{h_2}{h_1} _{k+h_1-h_2}|h_1\rangle = 0, & & \quad k > 0 \\
\langle h_2 | \tpf{h}{h_2}{h_1} _{k+h_1-h_2} = 0, & & \quad k < 0 \nonumber
\end{eqnarray}
In particular, for the primary states the modes vanish if $k>0$ giving
\begin{equation}
|h\rangle=\phi ^{(h)}(0)|0\rangle= \tpf{h}{h}{0} _{-h}|0\rangle,\qquad \langle h|h'\rangle=\delta_{h,h'}
\end{equation}

Under infinitesimal coordinate transformations $z\mapsto z+\varepsilon_m z^{m+1}$, the primary fields must satisfy~\cite{BPZ}
\begin{equation} \label{primary}
\delta_m \phi^{(h)}(z)=[L_m,\phi^{(h)}(z)]=z^{m+1}{\partial\over\partial z}\phi^{(h)}(z)+h(m+1)z^m\phi^{(h)}(z)
\end{equation}
and consequently the modes satisfy
\begin{equation} 
[L_m, \, \tpf{h}{h_2}{h_1} _{k+h_1-h_2} \, ]=[-m(1-h)-(k+h_1-h_2)] \, \tpf{h}{h_2}{h_1} _{m+k+h_1-h_2}
\label{primary_modes}
\end{equation}

The chiral vertex operators (CVOs) are used in the quasi-particle approach~\cite{spinons} to construct quasi-particle states by repeated application of appropriate modes of primary fields 
\begin{equation}
\tpf{j_N}{i_N}{i_{N-1}}_{-n_N-h_N+h_{N-1}} \ldots \tpf{j_2}{i_2}{i_1}_{-n_2-h_2+h_1}  
\tpf{j_1}{i_1}{0}_{-n_1-h_1} |0\rangle 
\end{equation} 
In general this set of states is overcomplete. It is supposed that a linearly independent set can be obtained by applying appropriate restrictions such as those arising from fusion and braiding.
In this paper we concentrate on understanding the CVOs which are the building blocks of the quasi-particle approach.

\subsection{Two and three-point functions}

We can put the mode expansions inside the two-point function (\ref{2point}) and use the constraint (\ref{constraint})
to reduce the number of terms 
\begin{equation} \label{laur-tay}
\langle 0 | \phi^{(h)} (z) \phi^{(h)} (w) |0\rangle= z^{-2 h} \sum_{k \ge 0} 
\langle 0 | \tpf{h}{0}{h}_{k+h} \tpf{h}{h}{0}_{-k-h} |0\rangle \left( \frac{w}{z} \right)^k  
\end{equation}
This expression is intimately related to the Laurent-Taylor expansion 
\begin{equation} \label{laur-tay2}
(1-u)^{-2h}=\sum_{\ell\in\Z} a^h_\ell\, u^\ell,\qquad \mbox{$a^h_\ell=0$ for $\ell<0$}
\end{equation}
where
\begin{equation}  \label{laur-tay3}
a^h_\ell = \langle 0 | \tpf{h}{0}{h}_{\ell+h} \tpf{h}{h}{0}_{-\ell-h} |0\rangle 
\end{equation}
which we later verify for small $|\ell|$ from explicit matrices for the chiral modes. 
Of course the full result is obtained by restoring the antiholomorphic dependence
\begin{eqnarray}
\langle \phi(z,\zbar)\phi(w,\wbar)\rangle
=(z-w)^{-2h}(\zbar-\wbar)^{-2h}
\end{eqnarray}
%
The structure constants and operator product expansions involve the calculation of three-point functions, that can be 
accessed in a completely similar way. 
The mode expansion  in (\ref{limit}) together with (\ref{constraint}) yields
\begin{equation}
\lim_{z\rightarrow 0} z^{h+h_1-h_2} \langle h_2 | \phi^{(h)} (z) | h_1 \rangle=
\langle h_2 | \tpf{h}{h_2}{h_1} _{h_1-h_2} | h_1 \rangle \equiv \sqrt{C_{h_2,h,h_1}}
\end{equation}
where we see that the structure constant explicitly enters in the mode coupled with $z^0$ (fundamental mode).

The energy momentum tensor is not a primary field, but still its two-point function 
\begin{equation}
\langle 0 | T(z) T(w) |0\rangle=  \frac{c/2}{(z-w)^4}  
\end{equation}
is fixed by conformal invariance. 
We can proceed as we did for the fields.
The mode expansion is now
\begin{equation}
\langle 0 | T(z) T(w) |0\rangle= z^{-4} \sum_{n \ge 0} \langle 0 | L_{n+2} L_{-n-2} |0\rangle \left(\frac{w}{z}\right)^n
\end{equation}
The relevant Laurent-Taylor expansion is
\begin{equation}
 (1-u)^{-4} = \sum_{\ell=0}^\infty a_\ell\, u^\ell = 1+4 u+10 u^2+20 u^3+35 u^4+56 u^5 + \cdots
\label{Ttaylor}
\end{equation}
where
\begin{equation}
\frac{c}{2} \,  a_n =  \langle 0 | L_{n+2} L_{-n-2} |0\rangle
\end{equation}
Again we later verify this for small $n$ from explicit matrices for the Virasoro generators.

\section{Algorithm for Building Matrix Representations}
\setcounter{equation}{0}

In this section we present a general algorithm for building, level by level, matrix representations of the Virasoro generators and primary fields. We work in a distinguished basis which we call the $L_1$-basis. The algorithm itself relates the vectors of the $L_1$-basis to the basis of Virasoro states (\ref{linear_span}). Applying the algorithm for generic $c$ also yields general expressions for the null vectors, level by level, for arbitrary minimal models $\M(p',p)$. Alternatively, these could be obtained level-by-level from the null vectors of the Gram matrix for Virasoro states.

We first present our general algorithm for the Virasoro generators in the $L_1$-basis. In the subsequent subsection, we implement this algorithm for generic $c$ in the $h=0$ vacuum sector pointing out its salient features and discuss the relation between the $L_1$-basis and the basis of Virasoro states. 

We implement our algorithm systematically for the critical Ising, tricritical Ising, $3$-state Potts and Yang-Lee theories in Sections 4--7. In each case the algorithm was implemented using Mathematica~\cite{Wolfram}.

\subsection{Matrix algorithms for Virasoro generators in $L_1$-basis}
\label{algorithm}

Because states belonging to different Virasoro modules are orthogonal, the matrices representing the Virasoro generators $L_n$ 
can be obtained separately within each module $\V_h$
\begin{equation}
\label{sum_Verma}
L_n= \displaystyle \oplus_h L_{n}^h
\end{equation}
Usually we work within a fixed module and suppress the index $h$.
To obtain matrix representations of Vir, we will take 
$L_{-n}=L^T_n$. For unitary theories $L_n$ will be real so that $L_{-n}=L^T_n=L^\dagger_n$ whereas for non-unitary theories $L_n$ will be complex with $L_{-n}=L^T_n\ne L^\dagger_n$.
Suppose we can find a diagonal matrix $L_0$ and matrices $L_1=L_{-1}^T$, $L_2=L_{-2}^T$ satisfying
\begin{eqnarray} 
&[L_1,L_0]=L_1,\qquad [L_2,L_0]=2L_2&\nonumber\\[4pt]
&[L_1,L_{-1}]=2L_0,\qquad [L_2,L_{-1}]=3L_1,\qquad [L_2,L_{-2}]=4L_0+c/2&
\label{Lcommutators}
\end{eqnarray}
Then we can extend this to a matrix representation of the full Virasoro algebra by recursively defining
\begin{equation}
L_n=L_{-n}^T={1\over n-2}\,[L_{n-1},L_1],\qquad n\ge 3
\end{equation}

Note that, if the generators $L_n$ satisfy the Virasoro algebra, then the generators
\begin{equation}
L_n'=U^TL_nU,\qquad U^T=U^{-1}
\label{orthogonal}
\end{equation}
also satisfy the Virasoro algebra where $U=\oplus U^h$ is orthogonal and preserves the sectors  (\ref{sum_Verma}). This corresponds to the freedom of an orthogonal change of basis. 
To specify a unique matrix representation of the Virasoro algebra one must first specify a distinguished {\em canonical} basis. 
The generators $L_{-n}$ act naturally on the basis (\ref{linear_span}) of Virasoro states. In particular, the grading of these states imposes a block structure on the matrix representatives of the Virasoro generators
\begin{equation}
L^h_{-n}=\oplus_\ell L^h_{-n,\ell},\qquad L^h_{-n,\ell}:\quad \V_{h,\ell}\rightarrow \V_{h,\ell+n}
\end{equation}
The orthogonal matrix (\ref{orthogonal}) must respect this grading so that 
\begin{equation} \label{graded_orthog}
U^h=
\mathop{\oplus}_{\ell=0}^\infty U^h_\ell 
\end{equation}

Within $\V_{h,\ell}$, the Virasoro states  (\ref{linear_span}) with $\ell=\sum_{i=1}^j n_i$ are degenerate. 
However, they are not orthogonal and typically they contain null vectors, so a basis of Virasoro states (\ref{linear_span}) is not a good basis to choose in seeking matrix 
representations of the Virasoro generators. One possibility, is to use the Gram-Schmidt process to orthonormalize the Virasoro states with the null vectors removed. A better alternative, at least for unitary minimal models, is to use the basis of fermion states of FPI~\cite{FPI}.
Since the fermion basis is a complete orthonormal basis of physical states we do not have to worry about removing null vectors. 
To obtain matrix representations of the Virasoro generators, we therefore start working in a fixed  orthonormal basis of states. 

With such a choice of basis, the structure of the matrices $L_1=L_{-1}^T$ and $L_2=L_{-2}^T$ is 
\begin{equation}
L_1=\mbox{\scriptsize $
\bordermatrix{
&d_0&d_{1}&d_{2}&\cdots&d_{\ell-1}&&&&d_\ell&&&&&&\cr
d_0&\mbox{\bf 0}&\bigstar&\mbox{\bf 0}&&&&&&&&&&&&&\cr 
d_{1}&&\mbox{\bf 0}&\bigstar&&&&&&&&&&&&&\cr 
d_{2}&&&\mbox{\bf 0}&&&&&&&&&&&&&\cr 
\;\vdots&&&&\ddots&&&&&&&&&&&&\cr 
&&&&&&*&0&0&0&0&0&0&&&&\cr 
&&&&&&0&*&0&0&0&0&0&&&&\cr 
d_{\ell-1}&&&&&&0&0&*&0&0&0&0&&&&\cr 
&&&&&&0&0&0&*&0&0&0&&&&\cr 
&&&&&&&&&&&&&&&&\cr 
&&&&&&&&&&&&&&&& }$}\end{equation}
\begin{equation}
L_2=\mbox{\scriptsize $
\bordermatrix{
&d_0&d_{1}&d_{2}&\cdots&d_{\ell-1}&&&&d_\ell&&&&&&\cr
d_0&\mbox{\bf 0}&\mbox{\bf 0}&\bigstar&&&&&&&&&&&&&\cr 
d_{1}&&\mbox{\bf 0}&\mbox{\bf 0}&&&&&&&&&&&&&\cr 
d_{2}&&&\mbox{\bf 0}&&&&&&&&&&&&&\cr 
\;\vdots&&&&\ddots&&&&&&&&&&&&\cr 
&&&&&&*&*&*&*&*&0&0&&&&\cr 
d_{\ell-2}&&&&&&*&*&*&*&*&*&0&&&&\cr 
&&&&&&*&*&*&*&*&*&*&&&&\cr 
&&&&&&&&&&&&&&&&\cr 
&&&&&&&&&&&&&&&\cr 
&&&&&&&&&&&&&&&& }$}
\end{equation}
where $d_{\ell} = d_{\ell}^h=\mbox{dim}\,\V_{h,\ell}$ as in (\ref{main_character}), $\smallstar$ denotes non-zero off-diagonal blocks, the non-zero $\ell$-blocks in position $\ell$ are of sizes $d_{\ell-1}\times d_\ell$ and $d_{\ell-2}\times d_\ell$ respectively and
\begin{equation}
\chi_h(q)=q^{-c/24+h}\sum_{\ell=0}^\infty d_\ell\, q^\ell
\end{equation}
We allow for arbitrary entries in each non-zero block, substitute into the basic Virasoro commutators (\ref{Lcommutators}) and proceed to solve the equations level-by-level. At a given level, we find that the equations are underdetermined and only involve dot products between  rows in the $\ell$-blocks. This freedom corresponds to an allowed change of orthogonal basis implemented by the orthogonal matrix $U^h_\ell$ and is removed by demanding that the entries above the leading diagonal vanish in the $(d_{l-1}+d_{\ell-2})\times d_\ell$ matrix $B_\ell$ formed by placing the $\ell$ block of $L_1$ above the $\ell$-block of $L_2$. In this basis, we find that the entries below the leading diagonal of the $\ell$-block of $L_1$ automatically vanish. There remains a residual freedom associated with the choice of sign of each of the basis elements. We remove this by further demanding that the first non-zero entry at the top of each column of $B_\ell$ is positive. If this is pure imaginary, as can occur in non-unitary cases, we demand that the imaginary part of this non-zero entry of $B_\ell$ is positive. This algorithm thus fixes a distinguished {\em canonical} basis which we call the {\it $L_1$-basis} because of the diagonal form of the blocks of $L_1$. The $L_1$-basis should be viewed as the fermionic basis of paths rotated by the orthogonal transformation $U$. Notice that only the level degeneracies, which can be read off from the Virasoro character, and the orthonormality of the starting basis is essential in this algorithm --- the precise description of the starting basis is unimportant.
\goodbreak

\subsection{Generic minimal Virasoro matrices and $L_1$ basis}

Let us consider the minimal models $\M(p',p)$ with $p=L+1$ large. It is then natural to consider fermionic paths on $A_\infty$ and to treat the central charge $c$ as a parameter. We will see that this case is generic in the sense that specific minimal models can be obtained by specializing $c$. 

In the vacuum $h=0$ sector the generic Virasoro character is
\begin{eqnarray}
&&\chi_0(q)\;=\;{q^{-c/24}(1-q)\over \prod_{n=1}^\infty (1-q^n)}\\
&=&q^{-c/24}\left(1 + q^2 + q^3 + 2\,q^4 + 2\,q^5 + 4\,q^6 + 4\,q^7 + 7\,q^8 + 8\,q^9 + 12\,q^{10} + \mbox{O}(q^{11})\right)\nonumber
\end{eqnarray}
with degeneracies
\begin{equation}
\{d_\ell\}=\{1,0,1,1,2,2,4,4,7,8,12,\ldots\}
\end{equation}
Treating $c$ as a parameter, and applying our algorithm, we build the following representation of the Virasoro generators in the $L_1$-basis:
\begin{eqnarray}
L_0=\mbox{Diagonal}(0,2,3,4,4,5,5,6,6,6,6,7,7,7,7,8,8,8,8,8,8,8,\ldots)
\end{eqnarray}
\begin{eqnarray}
&&\qquad L_1\;=\;\mbox{\scriptsize{\mbox{$
\left(\begin{array}{cccccccccccc}
.&0&.&.&.&.&.&.&.&.&.&\\
.&.&2&.&.&.&.&.&.&.&.&\\
.&.&.&\sqrt{10}&0&.&.&.&.&.&.&\\
.&.&.&.&.&3\sqrt{2}&0&.&.&.&.&\\ 
.&.&.&.&.&0&2\sqrt{2}&.&.&.&.&\\
&.&.&.&.&.&.&2\sqrt{7}&0&0&0&\\
.&.&.&.&.&.&.&0&3\sqrt{2}&0&0&\\
.&.&.&.&.&.&.&.&.&.&.&\\
.&.&.&.&.&.&.&.&.&.&.&\\
.&.&.&.&.&.&.&.&.&.&.&\\
&&&&&&&&&&&\ddots
\end{array}\right)
$}}}\\[8pt]
&=&\smat{0}\oplus\smat{2}\oplus\smat{\sqrt{10}&0}\oplus\smat{3\sqrt{2}&0\cr 0&2\sqrt{2}}
\oplus\smat{2\sqrt{7}&0&0&0\cr 0&3\sqrt{2}&0&0}
\oplus\smat{2\sqrt{10}&0&0&0\cr 0&\sqrt{30}&0&0\cr 0&0&2\sqrt{3}&0\cr 0&0&0&2\sqrt{3}}
\nonumber\\[4pt]
&&\oplus\smat{3\sqrt{6}&0&0&0&0&0&0\cr 0&2\sqrt{11}&0&0&0&0&0\cr 
0&0&\sqrt{26}&0&0&0&0\cr 0&0&0&\sqrt{26}&0&0&0\cr }
\oplus\smat{
\sqrt{70}&0&0&0&0&0&0&0\cr 
0&2\sqrt{15}&0&0&0&0&0&0\cr
0&0&\sqrt{42}&0&0&0&0&0\cr
0&0&0&\sqrt{42}&0&0&0&0\cr
0&0&0&0&4&0&0&0\cr
0&0&0&0&0&4&0&0\cr
0&0&0&0&0&0&4&0}\oplus\ldots
\nonumber
\end{eqnarray}
\begin{eqnarray}
&&\mbox{}\hspace{-.25in}\mbox{} L_2\;=\;\mbox{\scriptsize{\mbox{$
\left(\begin{array}{cccccccccccc}
.&\sqrt{c\over 2}&.&.&.&.&.&.&.&.&.&\\
.&.&.&3\sqrt{2\over 5}&\sqrt{c+{22\over 5}}&.&.&.&.&.&.&\\
.&.&.&.&.&{7\over\sqrt{5}}&\sqrt{{1\over 2}(c+{22\over 5})}&.&.&.&.&\\
.&.&.&.&.&.&.&8\sqrt{2\over 7}&{1\over 3}\sqrt{{5\over 2}(c+{22\over 5})}&
{1\over 3}\sqrt{2(c+{29\over 70})}&0&\\ 
.&.&.&.&.&.&.&0&2&{4\over \sqrt{5}}\sqrt{{c+{22\over 5}\over c+{29\over 70}}}&
\sqrt{3(c-{1\over 2})(c+{68\over 7})\over 2(c+{29\over 70})}&\\
.&.&.&.&.&.&.&.&.&.&.&\\
.&.&.&.&.&.&.&.&.&.&.&\\
.&.&.&.&.&.&.&.&.&.&.&\\
&&&&&&&&&&&\ddots
\end{array}\right)
$}}}\nonumber
\end{eqnarray}
\begin{eqnarray}
\mbox{}
&=&\smat{\sqrt{c\over 2}}\oplus\smat{3\sqrt{2\over 5}&\sqrt{c+{22\over 5}}}
\oplus\smat{{7\over\sqrt{5}}&\sqrt{{1\over 2}(c+{22\over 5})}}
\oplus\smat{
8\sqrt{2\over 7}&{1\over 3}\sqrt{{5\over 2}(c+{22\over 5})}&
{1\over 3}\sqrt{2(c+{29\over 70})}&0\cr
0&2&{4\over \sqrt{5}}\sqrt{{c+{22\over 5}\over c+{29\over 70}}}&
\sqrt{3(c-{1\over 2})(c+{68\over 7})\over 2(c+{29\over 70})}}\\[4pt]
&&\mbox{}\hspace{-.6in}\mbox{}
\oplus\smat{9\sqrt{5\over 14}&{1\over\sqrt{6}}\sqrt{c+{22\over 5}}&{1\over\sqrt{3}}\sqrt{c+{29\over 70}}&0\cr
0&{13\over\sqrt{15}}&{8\over\sqrt{30}}\sqrt{c+{22\over 5}\over c+{29\over 70}}&
\sqrt{(c-{1\over 2})(c+{68\over 7})\over (c+{29\over 70})}}\nonumber\\[4pt]
&&\mbox{}\hspace{-.6in}\mbox{}
\oplus\!\smat{
5\sqrt{5\over 3}&\!\!\!\sqrt{{7\over 66}(c+{22\over 5})}&
\sqrt{{14\over 39}(c+{29\over 70})}&0&\sqrt{{5\over 143}(c+{11\over 105})}&0&0\cr
0&14\sqrt{6\over 55}&\!\!\!\!\!8\sqrt{3(c+{22\over 5})\over 130(c+{29\over 70})}&
\!\!\!3\sqrt{(c-{1\over 2})(c+{68\over 7})\over 13(c+{29\over 70})}&
{10\over 3\sqrt{1001}}\sqrt{c+{22\over 5}\over c+{11\over 105}}&
{1\over 3}\sqrt{13(1050c^2+3305c-251)\over 105(c+{11\over 105})}&0\cr
0&0&3\sqrt{6\over 13}&0&
\!\!{82\over 3}\sqrt{11(c+{29\over 70})\over 455(c+{11\over 105})}&
{1\over 30}\sqrt{13(c+{22\over 5})(1050c^2+3305c-251)
\over 21(c+{29\over 70})(c+{11\over 105})}&0\cr
0&0&0&3\sqrt{6\over 13}&0&
\!\!\!\!\!\!36\sqrt{70(c-{1\over 2})(c+{68\over 7})(c+{11\over 105})\over 
13(c+{29\over 70})(1050c^2+3305c-251)}&
\!\!\!\!\!\!10\sqrt{21(c+{3\over 5})(c+{46\over 3})(c+{29\over 70})
\over (1050c^2+3305c-251)} 
}\nonumber\\[4pt]
&&\mbox{}\hspace{-.6in}\mbox{}\oplus\ldots\nonumber
\end{eqnarray}

We find that these matrices are generic in the sense that they can be reduced to produce the correct matrices for any special minimal value of $c$. More explicitly, if we specialize $c$ to an allowed central charge for a minimal theory, say $c=1/2$, then starting on the left we seek the first zero column of $L_2$ (excluding the first and third columns which are identically zero for all $c$). For $c=1/2$ this occurs in column~11 corresponding to the known single null (unphysical) vector at level~6. We then remove the rows and columns of $L_0$, $L_1$ and $L_2$ corresponding to this position and repeat the process. For $c=1/2$, we find the next zero column of $L_2$ at the original position 15, corresponding to the single null vector at level~7, so we remove this row and column. At the next stage we remove rows and columns at the orginal positions 19 and 22, corresponding to the two null vectors at level~8, and so on. This process exactly produces the Virasoro generators of the critical Ising model $\M(3,4)$ as given in Section~4. The same process works for the Yang-Lee theory with $c=-22/5$ as given in Section~7 and other special values of $c$ corresponding to minimal models. Indeed, the matrix $L_2$ exhibits a remarkable structure of factors of the form $\sqrt{c-c_{p',p}}$ where $c_{p',p}$ is the central charge of the minimal model $\M(p',p)$.


The elements of the $L_1$-basis can be obtained in terms of the Virasoro states from the generic minimal Virasoro matrices $L_{-n}=L_n^T$, $n=1,2$ by considering the action of these matrices on the fundamental basis formed by the column vectors 
$e_i=\{\delta_{i,j}\}_{j=1,2,\ldots}$. We identify $e_1$ with 
the vacuum $\ket 0$. Using the action of the Virasoro algebra on the states, the next states are identified as
\begin{eqnarray}
L_{-2} \,e_1 \equiv L_{-2}\, \ket 0 = \sqrt{c\over2}\, e_2 &\Rightarrow & 
e_2=\sqrt{2\over c}\,L_{-2} \,\ket 0 \\
L_{-1}\, e_2 \equiv \sqrt{2\over c}\, L_{-3}\, \ket 0 = 2 \,e_3 & \Rightarrow & 
e_3=\frac{1}{\sqrt{2 c}}\,L_{-3}\,\ket 0 
\end{eqnarray}
This procedure continues for higher levels leading to the expressions in Figure~9. These expressions hold for both unitary and non-unitary models. 
The orthonormality of the $L_1$ basis is guaranteed by the orthonormality of the fundamental basis $e_i$.
\begin{figure}[htb]
\begin{tabular}{r@{\hspace{8mm}}l@{\hspace{8mm}}l}
$1$ & $|0\rangle$ & \\[7pt]
$q^2$ & $\sqrt{\frac{2}{c}} L_{-2}|0\rangle$ & \\[7pt]
$q^3$ & $\frac{1}{\sqrt{2\, c}}  L_{-3}|0\rangle $ & \\[7pt]
$2q^4$ & $\frac{1}{\sqrt{5\, c}}  L_{-4}|0\rangle$ &
   $ -\frac{\sqrt{2} ( 3\, L_{-4} - 5\, L_{-2}^{2} ) |0\rangle } 
    {5 \sqrt{c\, (c+ \frac{22}{5} )}}   $ \\[7pt]
$2q^5$ & $\frac{1}{\sqrt{10\, c}} L_{-5}|0\rangle $ &
  $-\frac{ ( 2\, L_{-5} - 5\, L_{-3} L_{-2}) |0\rangle }
     {5\, \sqrt{c \, (c+ \frac{22}{5} )}} $ \\[7pt]
$4q^6$ & $ \sqrt{\frac{2}{35 \,c}}  L_{-6}|0\rangle $ &
$ -\frac{ (8 L_{-6} - 10 L_{-4} L_{-2} -5 L_{-3}^{2} ) |0\rangle }
     {15 \sqrt{2 \,c\, (c+ \frac{22}{5})}}  $ \\[7pt]
& $ \frac{ ( 20\, L_{-6} + 56\, L_{-4} L_{-2} - 35\, L_{-3}^2 ) |0\rangle}
    {42 \, \sqrt{10\, c\, (c+\frac{29}{70})} } $ & 
$ - \frac{\left( (78 +60 \,c)  L_{-6} + (201 +126 \,c) L_{-4} L_{-2} 
 - 93\, L_{-3}^2 - (29 + 70 \,c) L_{-2}^3 \right) |0\rangle  }
{35\, \sqrt{3\,c\,(c-\frac{1}{2})(c+\frac{22}{5})(c+\frac{68}{7})(c+\frac{29}{70})}}$
\\[7pt] $4q^7$ & \ldots & \ldots 
\end{tabular}
\caption{$L_1$-basis of orthonormal states for the generic Virasoro module $\mathcal{V}_{0}$ in terms of Virasoro states.  The central charge $c$ is treated as a parameter. 
Unphysical (null) states appear if  $c$ is set to the value of the central charge of a minimal model. 
For the Ising model with $c=1/2$ an unphysical state appears at level $\ell=6$. The apparent singularity of the state is an artifact of the normalization. 
Actually, the numerator is a null state. Further null states appear for the non-unitary cases with 
$c=-22/5$ and $c=-68/7$ corresponding to $\mathcal{M}(2,5)$ and $\mathcal{M}(2,7)$.}
\end{figure}

\subsection{Matrix algorithm for fields}

An algorithm can similarly be applied level-by-level to obtain matrix representations of the primary fields 
$\phi^{(h)}(z).$ 
In the vector space $ \V_{h_1} \oplus \V_{h_2} $ we consider the action
of the two CVOs
\begin{eqnarray}
\tpf{h}{h_2}{h_1}(z): &  &  \V_{h_1} \rightarrow \V_{h_2}  \\ 
\tpf{h}{h_1}{h_2}(z): &  &  \V_{h_2} \rightarrow \V_{h_1}  
\end{eqnarray}
The Virasoro generators and fields have the following block structure when acting on 
$\V_{h_1} \oplus \V_{h_2} $
\begin{equation} \label{directsum}
L_n=\begin{pmatrix}
L^{h_1}_n & \mbox{\bf 0}\cr \mbox{\bf 0} & L^{h_2}_n
\end{pmatrix}, \qquad 
\phi^{(h)}(z)=
\begin{pmatrix}
\mbox{\bf 0}& \tpf{h}{h_1}{h_2} (z) \cr \tpf{h}{h_2}{h_1} (z) & \mbox{\bf 0}
\end{pmatrix}
\end{equation}
The two CVOs can be worked out separately because they do not mix under the commutation relation 
due to the block structure of the generators $L_n.$ 
So we consider $\tpf{h}{h_2}{h_1} (z)$ and expand the field in modes 
\begin{equation}
 \Phi(z)= \begin{pmatrix}
\mbox{\bf 0} & \mbox{\bf 0} \cr \tpf{h}{h_2}{h_1} (z) & \mbox{\bf 0}
\end{pmatrix},   \qquad
 \Phi_k = 
\begin{pmatrix}
\mbox{\bf 0} & \mbox{\bf 0} \cr \tpf{h}{h_2}{h_1}_{k+h_1-h_2} & \mbox{\bf 0}
\end{pmatrix} 
\end{equation}
Working in the $L_1$-basis for $\V_{h_1}$ and $\V_{h_2}$, the block structure of the mode $\tpf{h}{h_2}{h_1}_{h_1-h_2}$ is
\begin{equation}
\tpf{h}{h_2}{h_1}_{h_1-h_2} =\mbox{\scriptsize $
\bordermatrix{
&d^{h_1}_0&d^{h_1}_1&d^{h_1}_2&\cdots&&&d^{h_1}_\ell&&&&&\cr
\rule{0mm}{5mm} d^{h_2}_0&\bigstar&\mbox{\bf 0}&\mbox{\bf 0}&&&&&&&&&&\cr 
\rule{0mm}{5mm} d^{h_2}_{1}&\mbox{\bf 0}&\bigstar&\mbox{\bf 0}&&&&&&&&&&\cr 
\rule{0mm}{5mm} d^{h_2}_{2}&\mbox{\bf 0}&\mbox{\bf 0}&\smallstar&&&&&&&&&&\cr 
\;\vdots&&&&\ddots&&&&&&&&&\cr 
&&&&&&*&*&*&*&&&&\cr 
&&&&&&*&*&*&*&&&&\cr 
d^{h_2}_{\ell}&&&&&&*&*&*&*&&&&\cr 
&&&&&&*&*&*&*&&&&\cr 
&&&&&&*&*&*&*&&&&\cr 
&&&&&&*&*&*&*&&&&\cr
&&&&&&&&&&&\ddots&& }$}
\end{equation}
where
\begin{equation}
\chi_{h_1}(q)=q^{-c/24+h_1}\sum_{\ell=0}^\infty d^{h_1}_\ell\, q^\ell,\qquad
\chi_{h_2}(q)=q^{-c/24+h_2}\sum_{\ell=0}^\infty d^{h_2}_{\ell}\, q^\ell,\qquad 
d^{h_1}_\ell, d^{h_2}_\ell\ge 0
\end{equation}
Given the mode $\Phi_{0}$ we use (\ref{primary_modes}) to generate the remaining modes by the action of the known matrix representation of $L_n$
\begin{equation}
\label{recursion_phi}
\Phi_{n}=\frac{1}{n(h-1)+h_2-h_1}\, [L_n,\Phi_{0}], \qquad n \neq 0
\end{equation}
With the higher modes defined in this way, we solve the linear equations 
\begin{equation}
[L_1,\Phi_{1}]=(h-2+h_2-h_1) \Phi_{2}, \quad [L_1,\Phi_{-1}]=(h+h_2-h_1) \Phi_{0}, \quad
[L_2,\Phi_{-1}]=(2h-1+h_2-h_1) \Phi_{1} \qquad 
\end{equation}
level-by-level for the entries of $\Phi_{0}$.
We find that these three equations are consistent and sufficient to determine all the entries of the modes so that no higher level equations are required. 

\section{Critical Ising model}
\label{CIM}

The critical Ising model $\M(3,4)$ with $c=1/2$ is realized as the critical RSOS model on $A_3$. It is essentially equivalent to the free fermion model and is well understood. We present it in this section to make contact with previous work and to illustrate the new features of our representation matrices in a familiar setting. In particular, we discuss the matrix representations of the Virasoro generators and fields in both the usual fermion basis and our $L_1$-basis.

The free fermion (Ising) model is usually discussed in the context of a superconformal theory with Neveu-Schwarz ($h=0,1/2$) and Ramond ($h=1/16$) sectors. The mode expansion of the free fermion field is given by
\begin{equation}
\label{majorana_field}
\psi(z)=\sum_k b_k\, z^{-k-1/2},\qquad \mbox{$k\in$}
\begin{cases}
\mbox{$\Z-1/2$,\quad NS}\\[2pt]
\mbox{$\Z$,\phantom{$\mbox{}-1/2$}\quad R}
\end{cases}
\end{equation}
with $b_{-k}=b_k^\dagger$. This is the primary field $\phi^{(1/2)}(z)=\psi (z)$ in the Neveu-Schwarz sector.
The Virasoro generators have simple bilinear expressions in terms of normal ordered fermion modes 
\begin{eqnarray}
L_0&=&
\begin{cases}
\displaystyle\sum_{k >0 } k\, b_{-k} b_k,&\mbox{NS:}\quad k\in\Z-\half\\
\displaystyle\sum_{k >0 } k\, b_{-k} b_k+\frac{1}{16},\qquad\mbox{} &\mbox{\phantom{N}R:}\quad k\in\Z
\end{cases}\\
L_n&=& \half \sum _k (k+\half) : b_{n-k}b_k :
\qquad : b_nb_m : =
\begin{cases}
\phantom{-}b_n b_m, &n<m\\
-b_m b_n,&n>m
\end{cases}
\end{eqnarray}
with
\begin{equation} \label{fermion_vir}
[L_n,b_{k}]=(-\frac{n}{2}+k) b_{n+k}
\end{equation}
Simple expressions for the generators $L_n$ as bilinear forms only appear to hold for the Ising model. 


The critical Ising model has three independent sectors corresponding to $(r,s)=(1,1)$, $(1,3)$ and $(1,2)$ and boundary conditions $(\mbox{left},\mbox{right})=(+,+)$, $(-,+)$ and $(\mbox{Free},+)$ respectively. The basis of physical fermion states in these $h=0,1/2,1/16$ sectors is
\begin{eqnarray}
\begin{array}{rrll}
\V_0: \quad\qquad& \displaystyle b_{-{j_{2n}\over 2}} \cdots b_{-{j_2\over 2}} 
b_{-{j_1\over 2}}|0\rangle,  &\quad j_k \in 2\N-1,\qquad\qquad &
E =  \frac{1}{2} \sum_k j_k \in \mathbb{N}-1 \\[10pt]
\V_{1/2}: \quad\qquad& \displaystyle b_{-{j_{2n+1}\over 2}} \cdots b_{-{j_2\over 2}}
b_{-{j_1\over 2}}|0\rangle,&\quad  j_k \in 2\N-1, &
E = \frac{1}{2} \sum_k j_k  \in \mathbb{N}-\frac{1}{2}  \\[10pt]
\V_{1/16}: \quad\qquad&\displaystyle  b_{-{j_{n}}} \cdots b_{-{j_2}}
b_{-{j_1}}\ket{\mbox{\small $1\over 16$}}, &\quad j_k \in \mathbb{N}, &
E = \sum_k j_k+\frac{1}{16} \in \mathbb{N}-\frac{15}{16} 
\end{array}
\end{eqnarray} 
The $h=0$ and $h=1/2$ sectors combine into one superconformal sector with
$\ket {1/2}=b_{-{1\over 2}}\ket 0$, $\ket 0=b_{{1\over 2}}\ket  {1/2}$ and total fermion number operator
\begin{equation}
F=\sum_{j=1}^\infty b_{-{j\over 2}}b_{j\over 2}
\end{equation}
The projectors onto the three Virasoro modules are
\begin{eqnarray}
\P_h=
\begin{cases}
\displaystyle\half \big(1+(-1)^F\big)\prod_{j\in\N} (1-b_{-j}b_j),&\quad h=0\\[16pt]
\displaystyle\half \big(1-(-1)^F\big)\prod_{j\in\N} (1-b_{-j}b_j),&\quad h=1/2\\[16pt]
\displaystyle\prod_{j\in\N-1/2} (1-b_{-j}b_j),&\quad h=1/16
\end{cases}
\end{eqnarray}

\subsection{Ising model: $\mathcal{V}_{0}$}
The $q$-series for the Ising character in the vacuum $h=0$ sector  is
\begin{equation}
\chi_{0}(q)=1 + q^2 + q^3 + 2\,q^4 + 2\,q^5 + 3\,q^6 + 3\,q^7 + 5\,q^8 + 5\,q^9 + 7\,q^{10} +  8\,q^{11} + 11\,q^{12} + O(q^{13})
\end{equation}
The form for the energy is independent of which basis 
we use
\begin{equation}  
L_0 = \mbox{Diagonal} ( 0, 2, 3, 4, 4, 5, 5, 6, 6, 6, 7, 7, 7, 8, 8, 8, 8, 8,\ldots )
\end{equation}
The states in the fermionic basis are ordered as
\begin{equation}
\begin{array}{c}
\mathcal{B}_{0} = \{ |0\rangle,~b_{-\frac{3}{2}} b_{-\frac{1}{2}} |0\rangle, 
~b_{-\frac{5}{2}} b_{-\frac{1}{2}} |0\rangle, ~b_{-\frac{7}{2}} b_{-\frac{1}{2}} |0\rangle,
~b_{-\frac{5}{2}} b_{-\frac{3}{2}} |0\rangle, ~b_{-\frac{9}{2}} b_{-\frac{1}{2}} |0\rangle,  \\[4pt] 
~b_{-\frac{7}{2}} b_{-\frac{3}{2}} |0\rangle, ~b_{-\frac{11}{2}} b_{-\frac{1}{2}} |0\rangle,
~b_{-\frac{9}{2}} b_{-\frac{3}{2}} |0\rangle, ~b_{-\frac{7}{2}} b_{-\frac{5}{2}} |0\rangle, ~\ldots \} 
\end{array}
\end{equation}
In this fermionic basis 
\begin{equation}
L_1=\smat{ 
. & . & . & . & . & . & . & . & . & . & . & . & . & . & . & . & . & . &  \\ 
. & . & 2 & . & . & . & . & . & . & . & . & . & . & . & . & . & . & . &  \\ 
. & . & . & 3 & 1 & . & . & . & . & . & . & . & . & . & . & . & . & . &  \\ 
. & . & . & . & . & 4 & 1 & . & . & . & . & . & . & . & . & . & . & . &  \\ 
. & . & . & . & . & 0 & 3 & . & . & . & . & . & . & . & . & . & . & . &  \\ 
. & . & . & . & . & . & . & 5 & 1 & 0 & . & . & . & . & . & . & . & . &  \\ 
. & . & . & . & . & . & . & 0 & 4 & 2 & . & . & . & . & . & . & . & . &  \\
. & . & . & . & . & . & . & . & . & . & 6 & 1 & 0 & . & . & . & . & . &  \\ 
. & . & . & . & . & . & . & . & . & . & 0 & 5 & 2 & . & . & . & . & . &  \\ 
. & . & . & . & . & . & . & . & . & . & 0 & 0 & 4 & . & . & . & . & . &  \\ 
. & . & . & . & . & . & . & . & . & . & . & . & . & 7 & 1 & 0 & 0 & 0 &  \\ 
. & . & . & . & . & . & . & . & . & . & . & . & . & 0 & 6 & 2 & 0 & 0 &  \\ 
. & . & . & . & . & . & . & . & . & . & . & . & . & 0 & 0 & 5 & 3 & 0 &  \\ 
 &  &  &  &  &  &  &  &  &  &  &  &  &  &  &  &  &  & \ddots }
\end{equation}
\begin{equation}
L_2=\frac{1}{2}\;\smat{ 
. & 1 & . & . & . & . & . & . & . & . & . & . & . & . & . & . & . & . &  \\ 
. & . & . & 5 & -3 & . & . & . & . & . & . & . & . & . & . & . & . & . &  \\ 
. & . & . & . & . & 7 & 0 & . & . & . & . & . & . & . & . & . & . & . &  \\ 
. & . & . & . & . & . & . & 9 & 0 & 3 & . & . & . & . & . & . & . & . &  \\ 
. & . & . & . & . & . & . & 0 & 7 & -5 & . & . & . & . & . & . & . & . &  \\ 
. & . & . & . & . & . & . & . & . & . & 11 & 0 & 3 & . & . & . & . & . &  \\ 
. & . & . & . & . & . & . & . & . & . & 0 & 9 & 0 & . & . & . & . & . &  \\ 
. & . & . & . & . & . & . & . & . & . & . & . & . & 13 & 0 & 3 & 0 & 0 &  \\ 
. & . & . & . & . & . & . & . & . & . & . & . & . & 0 & 11 & 0 & 5 & 0 &  \\ 
. & . & . & . & . & . & . & . & . & . & . & . & . & 0 & 0 & 9 & -7 & 1 &  \\ 
  &  &   &   &   &   &   &   &   &   &   &   &   &   &   &   &  &  & \ddots }
\end{equation}
We write the same matrices in the $L_1$ basis. The order of the states is the same as 
in Figure~2 after removal of the null vectors.
\begin{equation}
L_1 = \smat{ 
. & . & . & . & . & . & . & . & . & . & . & . & . & . & . & . & . & . & \\
. & . & 2 & . & . & . & . & . & . & . & . & . & . & . & . & . & . & . & \\
. & . & . & {\sqrt{10}} & 0 & . & . & . & . & . & . & . & . & . & . & . & . & . & \\ 
. & . & . & . & . & 3\,{\sqrt{2}} & 0 & . & . & . & . & . & . & . & . & . & . & . & \\ 
. & . & . & . & . & 0 & 2\,{\sqrt{2}} & . & . & . & . & . & . & . & . & . & . & . & \\ 
. & . & . & . & . & . & . & 2\,{\sqrt{7}} & 0 & 0 & . & . & . & . & . & . & . & . & \\ 
. & . & . & . & . & . & . & 0 & 3\,{\sqrt{2}} & 0 & . & . & . & . & . & . & . & . & \\ 
. & . & . & . & . & . & . & . & . & . & 2\,{\sqrt{10}} & 0 & 0 & . & . & . & . & . & \\ 
. & . & . & . & . & . & . & . & . & . & 0 & {\sqrt{30}} & 0 & . & . & . & . & . & \\ 
. & . & . & . & . & . & . & . & . & . & 0 & 0 & 2\,{\sqrt{3}} & . & . & . & . & . & \\ 
. & . & . & . & . & . & . & . & . & . & . & . & . & 3\,{\sqrt{6}} & 0 & 0 & 0 & 0 & \\ 
. & . & . & . & . & . & . & . & . & . & . & . & . & 0 & 2\,{\sqrt{11}} & 0 & 0 & 0 & \\ 
. & . & . & . & . & . & . & . & . & . & . & . & . & 0 & 0 & {\sqrt{26}} & 0 & 0 & \\
& & & & & & & & & & & & & & & & & & \ddots }\qquad
\end{equation}
\begin{eqnarray}
&  L_2= \smat{ 
. & \frac{1}{2} & . & . & . & . & . & . & . & . & . & . & . &  \\[5pt] 
. & . & . & 3\,{\sqrt{\frac{2}{5}}} & \frac{7}{{\sqrt{10}}} & . & . & . & . & . & . & . & . \\[5pt] 
. & . & . & . & . & \frac{7}{{\sqrt{5}}} & \frac{7}{2\,{\sqrt{5}}} & . & . & . & . & . & . \\[5pt] 
. & . & . & . & . & . & . & 8\,{\sqrt{\frac{2}{7}}} & \frac{7}{6} & \frac{8}{3\,{\sqrt{35}}} & . & . & . \\[5pt] 
. & . & . & . & . & . & . & 0 & 2 & \frac{7}{2} {\sqrt{\frac{7}{5}}} & . & . & .  \\[5pt] 
. & . & . & . & . & . & . & . & . & . & 9\,{\sqrt{\frac{5}{14}}} & \frac{7}{2\,{\sqrt{15}}} & 4\,  {\sqrt{\frac{2}{105}}}  \\[5pt] 
. & . & . & . & . & . & . & . & . & . & 0 & \frac{13}{{\sqrt{15}}} & 7\,
   {\sqrt{\frac{7}{30}}} } & \nonumber \\[5pt]
 & \oplus ~ \smat{ 
5\,{\sqrt{\frac{5}{3}}} & \frac{7}{2}{\sqrt{\frac{7}{165}}} & \frac{8}{{\sqrt{195}}} & {\sqrt{\frac{127}{6006}}} & 0 \\[5pt] 
0 & 14\,{\sqrt{\frac{6}{55}}} & 7\,{\sqrt{\frac{21}{130}}} & \frac{70}{{\sqrt{54483}}} & \frac{16}{3\,{\sqrt{127}}} \\[5pt] 
0 & 0 & 3\,{\sqrt{\frac{6}{13}}} & 656\,{\sqrt{\frac{11}{173355}}} & \frac{91}{3} {\sqrt{\frac{7}{635}}} }  \oplus \ldots & 
\end{eqnarray}

\subsection{Ising model: $\mathcal{V}_{1/2}$}
The $q$ series for the Ising character in the $h=1/2$ sector is
\begin{equation}
q^{-{1\over 2}}\chi_{1\over2}(q)=1 + q + q^2 + q^3 + 2\,q^4 + 2\,q^5 + 3\,q^6 + 4\,q^7 + 5\,q^8 + 6\,q^9 + 
  8\,q^{10} + 9\,q^{11} + 12\,q^{12} + O(q^{13})
\end{equation}
The energy is the same in both bases
\begin{equation}
L_0 = \frac{1}{2} + \mbox{Diagonal} (0, 1, 2, 3, 4, 4, 5, 5, 6, 6, 6, 7, 7, 7, 7,\ldots)
\end{equation} 
The fermionic basis is 
\begin{equation}
\begin{array}{c}
\mathcal{B}_{1/2}= \{ |1/2\rangle, ~b_{-\frac{3}{2}} b_{1 \over 2}|1/2\rangle,  
~b_{-\frac{5}{2}} b_{1 \over 2}|1/2\rangle,~b_{-{7 \over 2}} b_{1 \over 2}|1/2\rangle,
~b_{-{9 \over 2}} b_{1 \over 2}|1/2\rangle,~b_{-{5 \over 2}} b_{-{3 \over 2}}|1/2\rangle, \\[5pt]
~b_{-{11 \over 2}} b_{1 \over 2}|1/2\rangle,~b_{-{7 \over 2}} b_{-{3 \over 2}}|1/2\rangle, 
~b_{-{13 \over 2}} b_{1 \over 2}|1/2\rangle,~b_{-{9 \over 2}} b_{-{3 \over 2}}|1/2\rangle, ~b_{-{7 \over 2}} b_{-{5 \over 2}}|1/2\rangle, 
\\[5pt]
~b_{-{15 \over 2}} b_{1 \over 2}|1/2\rangle,~b_{-{11 \over 2}} b_{-{3 \over 2}}|1/2\rangle, ~b_{-{9 \over 2}} b_{-{5 \over 2}}|1/2\rangle,
~b_{-{7 \over 2}} b_{-{5 \over 2}} b_{-\frac{3}{2}} b_{1 \over 2}|1/2\rangle, 
~\ldots \}
\end{array}
\end{equation}
Notice the appearence of the positive mode $b_{1/2}$ in the basis itself. 
In this basis we have
\begin{equation}
L_1=\smat{
. & 1 & . & . & . & . & . & . & . & . & . & . & . & . & . &  \\ 
. & . & 2 & . & . & . & . & . & . & . & . & . & . & . & . &  \\ 
. & . & . & 3 & . & . & . & . & . & . & . & . & . & . & . &  \\ 
. & . & . & . & 4 & 0 & . & . & . & . & . & . & . & . & . &  \\ 
. & . & . & . & . & . & 5 & 0 & . & . & . & . & . & . & . &  \\ 
. & . & . & . & . & . & 0 & 3 & . & . & . & . & . & . & . &  \\ 
. & . & . & . & . & . & . & . & 6 & 0 & 0 & . & . & . & . &  \\
. & . & . & . & . & . & . & . & 0 & 4 & 2 & . & . & . & . &  \\ 
. & . & . & . & . & . & . & . & . & . & . & 7 & 0 & 0 & 0 &  \\ 
. & . & . & . & . & . & . & . & . & . & . & 0 & 5 & 2 & 0 &  \\ 
. & . & . & . & . & . & . & . & . & . & . & 0 & 0 & 4 & 1 &  \\ 
 &  &  &  &  &  &  &  &  &  &  &  &  &  &  & \ddots }
\end{equation}
\begin{equation}
L_2=\frac{1}{2} \;\smat{
. & . & 3 & . & . & . & . & . & . & . & . & . & . & . & . &  \\ 
. & . & . & 5 & . & . & . & . & . & . & . & . & . & . & . &  \\ 
. & . & . & . & 7 & 1 & . & . & . & . & . & . & . & . & . &  \\ 
. & . & . & . & . & . & 9 & 1 & . & . & . & . & . & . & . &  \\ 
. & . & . & . & . & . & . & . & 11 & 1 & 0 & . & . & . & . &  \\
. & . & . & . & . & . & . & . & 0 & 7 & -5 & . & . & . & . &  \\ 
. & . & . & . & . & . & . & . & . & . & . & 13 & 1 & 0 & 0 &  \\ 
. & . & . & . & . & . & . & . & . & . & . & 0 & 9 & 0 & -3 &  \\ 
 &  &  &  &  &  &  &  &  &  &  &  &  &  &  & \ddots }
\end{equation}
By comparison, using the $ L_1$ basis, gives
\begin{equation}
L_1=\smat{
   . & 1 & . & . & . & . & . & . & . & . & . & . & . & . & . & \\ 
   . & . & 2 & . & . & . & . & . & . & . & . & . & . & . & . & \\ 
   . & . & . & 3 & . & . & . & . & . & . & . & . & . & . & . & \\ 
   . & . & . & . & 4 & 0 & . & . & . & . & . & . & . & . & . & \\ 
   . & . & . & . & . & . & 5 & 0 & . & . & . & . & . & . & . & \\ 
   . & . & . & . & . & . & 0 & 3 & . & . & . & . & . & . & . & \\ 
   . & . & . & . & . & . & . & . & 6 & 0 & 0 & . & . & . & . & \\ 
   . & . & . & . & . & . & . & . & 0 & 2\,{\sqrt{5}} & 0 & . & . & . & . & \\ 
   . & . & . & . & . & . & . & . & . & . & . & 7 & 0 & 0 & 0 & \\ 
   . & . & . & . & . & . & . & . & . & . & . & 0 & {\sqrt{33}} & 0 & 0 & \\ 
   . & . & . & . & . & . & . & . & . & . & . & 0 & 0 & {\sqrt{13}} & 0 & \\  
     &   &   &   &   &   &   &   &   &   &   &   &   &   &   & \ddots }
\end{equation}
\begin{equation}
L_2= \smat{
   . & . & \frac{3}{2} & . & . & . & . & . & . & . & . & . & . & . & . & \\[5pt] 
   . & . & . & \frac{5}{2} & . & . & . & . & . & . & . & . & . & . & . & \\[5pt] 
   . & . & . & . & \frac{7}{2} & \frac{1}{2} & . & . & . & . & . & . & . & . & . & \\[5pt] 
   . & . & . & . & . & . & \frac{9}{2} & \frac{1}{2} & . & . & . & . & . & . & . & \\[5pt] 
   . & . & . & . & . & . & . & . & \frac{11}{2} & \frac{1}{{\sqrt{5}}} & \frac{1}
   {2\,{\sqrt{5}}} & . & . & . & . & \\[5pt] 
   . & . & . & . & . & . & . & . & 0 & \frac{9}{2\,{\sqrt{5}}} & \frac{17}
   {2\,{\sqrt{5}}} & . & . & . & . & \\[5pt] 
   . & . & . & . & . & . & . & . & . & . & . & \frac{13}{2} & {\sqrt{\frac{5}
      {33}}} & \frac{1}{2} {\sqrt{\frac{5}{13}}} & \frac{1}{{\sqrt{429}}} & \\[5pt] 
   . & . & . & . & . & . & . & . & . & . & . & 0 & \frac{29}{2} {\sqrt{\frac{3}{55}}} & 
    \frac{51}{2\,{\sqrt{65}}} & -7\, {\sqrt{\frac{3}{143}}} & \\[5pt]  
   &   &   &   &   &   &   &   &   &   &   &   &   &   &   & \ddots }
\end{equation}

\subsection{Ising model: $\mathcal{V}_{1/16}$}
The $q$-series for the Ising character in the $h=1/16$ sector is
\begin{equation}
q^{-{1\over 16}}\chi_{1 \over 16}(q)= 
1 + q + q^2 + 2\,q^3 + 2\,q^4 + 3\,q^5 + 4\,q^6 + 5\,q^7 + 6\,q^8 + 
8\,q^9 + 10\,q^{10} + 12\,q^{11} + 15\,q^{12} + O(q^{13})\ \ 
\end{equation}
with energy 
\begin{equation}
L_0 = \frac{1}{16} + \mbox{Diagonal} 
(0, 1, 2, 3, 3, 4, 4, 5, 5, 5, 6, 6, 6, 6, 7, 7, 7, 7, 7,\ldots)
\end{equation} 
In the $L_1$ basis we find
\setlength{\arraycolsep}{1mm}
\begin{equation}
L_1=\frac{1}{2 \sqrt{2} } \,\smat{
 . & 1 & . & . & . & . & . & . & . & . & . & . & . & . & . & . & . & . & . & \\ 
 . & . & 3\,{\sqrt{2}} & . & . & . & . & . & . & . & . & . & . & . & . & . & . & . & . & \\ 
 . & . & . & {\sqrt{51}} & 0 & . & . & . & . & . & . & . & . & . & . & . & . & . & . & \\ 
 . & . & . & . & . & 10 & 0 & . & . & . & . & . & . & . & . & . & . & . & . & \\ 
 . & . & . & . & . & 0 & 7 & . & . & . & . & . & . & . & . & . & . & . & . & \\ 
 . & . & . & . & . & . & . & {\sqrt{165}} & 0 & 0 & . & . & . & . & . & . & . & . & . & \\ 
 . & . & . & . & . & . & . & 0 & {\sqrt{114}} & 0 & . & . & . & . & . & . & . & . & . & \\ 
 . & . & . & . & . & . & . & . & . & . & {\sqrt{246}} & 0 & 0 & 0 & . & . & . & . & . & \\ 
 . & . & . & . & . & . & . & . & . & . & 0 & {\sqrt{195}} & 0 & 0 & . & . & . & . & . & \\ 
 . & . & . & . & . & . & . & . & . & . & 0 & 0 & 9 & 0 & . & . & . & . & . & \\ 
 . & . & . & . & . & . & . & . & . & . & . & . & . & . & 7\, {\sqrt{7}} & 0 & 0 & 0 & 0 & \\ 
 . & . & . & . & . & . & . & . & . & . & . & . & . & . & 0 & 2\,{\sqrt{73}} & 0 & 0 & 0 & \\ 
 . & . & . & . & . & . & . & . & . & . & . & . & . & . & 0 & 0 & {\sqrt{178}} & 0 & 0 & \\ 
 . & . & . & . & . & . & . & . & . & . & . & . & . & . & 0 & 0 & 0 & {\sqrt{97}} & 0 & \\  
  & & & & & & & & & & & & & & & & & & & \ddots }
\end{equation}
\begin{eqnarray}
& L_2 = \smat{
   . & . & \frac{1}
   {{\sqrt{2}}} & . & . & . & . & . & . & . & . & . & . & . \\[5pt] 
   . & . & . & \frac{19}{{\sqrt{102}}} & \frac{7}
   {{\sqrt{51}}} & . & . & . & . & . & . & . & . & . \\[5pt] . & . & 
   . & . & . & 7\,{\sqrt{\frac{3}{17}}} & {\sqrt{\frac{6}
      {17}}} & . & . & . & . & . & . & . \\[5pt] . & . & . & . & . & 
   . & . & 17\,{\sqrt{\frac{3}{55}}} & {\sqrt{\frac{3}
      {19}}} & 56\,
   {\sqrt{\frac{2}{53295}}} & . & . & . & . \\[5pt] . & . & . & . & 
   . & . & . & 0 & 7\,{\sqrt{\frac{3}{38}}} & 13\,
   {\sqrt{\frac{55}{969}}} & . & . & . & . \\[5pt] 
   . & . & . & . & 
   . & . & . & . & . & . & 67\,{\sqrt{\frac{5}{902}}} & 2\,
   {\sqrt{\frac{5}{247}}} & \frac{112}{9}\sqrt{\frac{10}{10659}}
   & \frac{128}{9}  {\sqrt{\frac{5}{27183}}} \\[5pt] 
   . & . & . & . & . & . & . & . & . & . & 0 & \frac{163}
   {{\sqrt{2470}}} & \frac{91}{9} {\sqrt{\frac{55}{969}}} & 
    \frac{-31}{9} \sqrt{\frac{82}{3315}}}   & \nonumber \\[8pt]
& \oplus \;\smat{
   83\,{\sqrt{\frac{3}{574}}} & 5\,{\sqrt{\frac{33}{18031}}} & 
    \frac{560}{9\,{\sqrt{28747}}} & \frac{640}{9} {\sqrt{\frac{11}{878917}}}
   & 5504\,{\sqrt{\frac{2}{4124717905}}} \\[5pt] 0 & 179\,
   {\sqrt{\frac{3}{4745}}} & \frac{91}{9}  {\sqrt{\frac{55}{1513}}}
    & \frac{-62}{9}  {\sqrt{\frac{779}{107185}}} & 1736\,
   {\sqrt{\frac{14}{203557507}}} \\[5pt] 0 & 0 & \frac{27}
   {{\sqrt{178}}} & 0 & 7\,{\sqrt{\frac{49567}{93005}}} } \oplus \ldots 
\end{eqnarray}
The corresponding energy momentum tensor (up to order $q^6$) is shown in Figure~3.

\subsection{Fields}
The chiral operator $\phi^{(1/2)}(z)$ corresponds to the fermion field in the NS
sector. Its action in $\mathcal{V}_{0} \oplus \mathcal{V}_{1/16} \oplus \mathcal{V}_{1/2}$ is described by the block structure
\begin{equation}
\phi^{(1/2)}(z)= \left( \begin{array}{c|c|c} 0 & 0& \smallstar \\ \cline{1-3} 
0& \smallstar & 0 \\ \cline{1-3} 
\smallstar & 0 & 0 \end{array} \right)
\end{equation}
A careful analysis of the two and three-point functions yields the following relations between the blocks (the chiral remnant of the 
self-adjointness property of the full conformal field $\varepsilon(z,\bar{z}) \equiv \phi^{(1/2)}(z,\bar{z})$)
\begin{equation}
\tpf{1/2}{0}{1/2} (z) = z^{-1} \left[ \tpf{1/2}{1/2}{0} (\frac{1}{z}) \right] ^{T} , \qquad  
\tpf{1/2}{1/16}{1/16} (z) = z^{-1} \left[ \tpf{1/2}{1/16}{1/16} (\frac{1}{z}) \right] ^{T}
\end{equation}
In the $L_1$  basis, the lower-left block and the central block are shown in Figures~4 and 5.

The chiral operator $\phi^{(1/16)}(z)$ corresponds to the magnetic field of Ising in the R
sector. Its action in $\mathcal{V}_{0} \oplus \mathcal{V}_{1/16} \oplus  \mathcal{V}_{1/2}$ is described by the block structure
\begin{equation}
\phi^{(1/16)}(z)= \left( \begin{array}{c|c|c} 0 & \smallstar & 0 \\ \cline{1-3} 
\smallstar & 0 & \smallstar \\ \cline{1-3} 
0 & \smallstar & 0 \end{array} \right)
\end{equation}

\clearpage
\thispagestyle{empty}
\setlength{\arraycolsep}{1mm}
\begin{figure}[p]
\mbox{}\hspace{1.1in}\mbox{}
\rotatebox{90}{
$  \mbox{}\hspace{-4mm} T(z) =\smat{ 
   \frac{1}{16\,z^2} & \frac{1}{2\,{\sqrt{2}}\,z^3} & \frac{1}{{\sqrt{2}}\,z^4} & \frac{8}
   {{\sqrt{51}}\,z^5} & \frac{-7}{2\,{\sqrt{102}}\,z^5} & \frac{8\,{\sqrt{\frac{2}{51}}}}
   {z^6} & \frac{-5}{{\sqrt{51}}\,z^6} & \frac{64}{{\sqrt{935}}\,z^7} & \frac{-20}
   {{\sqrt{323}}\,z^7} & \frac{-7}{2\,{\sqrt{2090}}\,z^7} & \frac{896}
   {{\sqrt{115005}}\,z^8} & \frac{-56\,{\sqrt{\frac{10}{12597}}}}{z^8} & \frac{-49}
   {9\,{\sqrt{1045}}\,z^8} & \frac{5\,{\sqrt{\frac{5}{1066}}}}{9\,z^8} & \\[5pt] \frac{1}
   {2\,{\sqrt{2}}\,z} & \frac{17}{16\,z^2} & \frac{3}{2\,z^3} & \frac{19}
   {{\sqrt{102}}\,z^4} & \frac{7}{{\sqrt{51}}\,z^4} & \frac{16}{{\sqrt{51}}\,z^5} & \frac{31}
   {2\,{\sqrt{102}}\,z^5} & \frac{56\,{\sqrt{\frac{2}{935}}}}{z^6} & \frac{25\,
     {\sqrt{\frac{2}{323}}}}{z^6} & \frac{21}{{\sqrt{1045}}\,z^6} & \frac{64\,
     {\sqrt{\frac{22}{10455}}}}{z^7} & \frac{116\,{\sqrt{\frac{5}{12597}}}}{z^7} & \frac{91\,
     {\sqrt{\frac{11}{190}}}}{18\,z^7} & \frac{-40\,{\sqrt{\frac{5}{533}}}}{9\,z^7} & \\[5pt] \frac{1}
   {{\sqrt{2}}} & \frac{3}{2\,z} & \frac{33}{16\,z^2} & \frac{{\sqrt{\frac{51}{2}}}}
   {2\,z^3} & 0 & \frac{7\,{\sqrt{\frac{3}{17}}}}{z^4} & \frac{{\sqrt{\frac{6}{17}}}}{z^4} & 
    \frac{72\,{\sqrt{\frac{2}{935}}}}{z^5} & \frac{63}{2\,{\sqrt{646}}\,z^5} & \frac{-28}
   {{\sqrt{1045}}\,z^5} & \frac{296\,{\sqrt{\frac{6}{38335}}}}{z^6} & \frac{31\,
     {\sqrt{\frac{15}{4199}}}}{z^6} & \frac{-217\,{\sqrt{\frac{2}{1045}}}}{9\,z^6} & \frac{80\,
     {\sqrt{\frac{5}{533}}}}{9\,z^6} & \\[5pt] \frac{8\,z}{{\sqrt{51}}} & \frac{19}{{\sqrt{102}}} & 
    \frac{{\sqrt{\frac{51}{2}}}}{2\,z} & \frac{49}{16\,z^2} & 0 & \frac{5}
   {{\sqrt{2}}\,z^3} & 0 & \frac{17\,{\sqrt{\frac{3}{55}}}}{z^4} & \frac{{\sqrt{\frac{3}{19}}}}
   {z^4} & \frac{56\,{\sqrt{\frac{2}{53295}}}}{z^4} & \frac{208}{{\sqrt{2255}}\,z^5} & \frac{{
       \sqrt{\frac{95}{26}}}}{2\,z^5} & \frac{-28\,{\sqrt{\frac{19}{2805}}}}{9\,z^5} & \frac{-320\,
     {\sqrt{\frac{10}{27183}}}}{9\,z^5} & \\[5pt] \frac{-7\,z}{2\,{\sqrt{102}}} & \frac{7}
   {{\sqrt{51}}} & 0 & 0 & \frac{49}{16\,z^2} & 0 & \frac{7}{2\,{\sqrt{2}}\,z^3} & 0 & \frac{7\,
     {\sqrt{\frac{3}{38}}}}{z^4} & \frac{13\,{\sqrt{\frac{55}{969}}}}{z^4} & 0 & \frac{56}
   {{\sqrt{1235}}\,z^5} & \frac{104\,{\sqrt{\frac{110}{969}}}}{9\,z^5} & \frac{217\,
     {\sqrt{\frac{41}{3315}}}}{18\,z^5} & \\[5pt] 8\,{\sqrt{\frac{2}{51}}}\,z^2 & \frac{16\,z}
   {{\sqrt{51}}} & 7\,{\sqrt{\frac{3}{17}}} & \frac{5}{{\sqrt{2}}\,z} & 0 & \frac{65}
   {16\,z^2} & 0 & \frac{{\sqrt{\frac{165}{2}}}}{2\,z^3} & 0 & 0 & \frac{67\,
     {\sqrt{\frac{5}{902}}}}{z^4} & \frac{2\,{\sqrt{\frac{5}{247}}}}{z^4} & \frac{112\,
     {\sqrt{\frac{10}{10659}}}}{9\,z^4} & \frac{128\,{\sqrt{\frac{5}{27183}}}}{9\,z^4} & \\[5pt] 
    \frac{-5\,z^2}{{\sqrt{51}}} & \frac{31\,z}{2\,{\sqrt{102}}} & {\sqrt{\frac{6}{17}}} & 0 & 
    \frac{7}{2\,{\sqrt{2}}\,z} & 0 & \frac{65}{16\,z^2} & 0 & \frac{{\sqrt{57}}}
   {2\,z^3} & 0 & 0 & \frac{163}{{\sqrt{2470}}\,z^4} & \frac{91\,{\sqrt{\frac{55}{969}}}}
   {9\,z^4} & \frac{-31\,{\sqrt{\frac{82}{3315}}}}{9\,z^4} & \\[5pt] \frac{64\,z^3}{{\sqrt{935}}} & 56\,
   {\sqrt{\frac{2}{935}}}\,z^2 & 72\,{\sqrt{\frac{2}{935}}}\,z & 17\,
   {\sqrt{\frac{3}{55}}} & 0 & \frac{{\sqrt{\frac{165}{2}}}}{2\,z} & 0 & \frac{81}
   {16\,z^2} & 0 & 0 & \frac{{\sqrt{123}}}{2\,z^3} & 0 & 0 & 0 & \\[5pt] \frac{-20\,z^3}
   {{\sqrt{323}}} & 25\,{\sqrt{\frac{2}{323}}}\,z^2 & \frac{63\,z}{2\,{\sqrt{646}}} & {\sqrt{
       \frac{3}{19}}} & 7\,{\sqrt{\frac{3}{38}}} & 0 & \frac{{\sqrt{57}}}{2\,z} & 0 & \frac{81}
   {16\,z^2} & 0 & 0 & \frac{{\sqrt{\frac{195}{2}}}}{2\,z^3} & 0 & 0 & \\[5pt] \frac{-7\,z^3}
   {2\,{\sqrt{2090}}} & \frac{21\,z^2}{{\sqrt{1045}}} & \frac{-28\,z}{{\sqrt{1045}}} & 56\,
   {\sqrt{\frac{2}{53295}}} & 13\,{\sqrt{\frac{55}{969}}} & 0 & 0 & 0 & 0 & \frac{81}
   {16\,z^2} & 0 & 0 & \frac{9}{2\,{\sqrt{2}}\,z^3} & 0 & \\[5pt] \frac{896\,z^4}
   {{\sqrt{115005}}} & 64\,{\sqrt{\frac{22}{10455}}}\,z^3 & 296\,{\sqrt{\frac{6}{38335}}}\,
   z^2 & \frac{208\,z}{{\sqrt{2255}}} & 0 & 67\,{\sqrt{\frac{5}{902}}} & 0 & \frac{{\sqrt{123}}}
   {2\,z} & 0 & 0 & \frac{97}{16\,z^2} & 0 & 0 & 0 & \\[5pt] -56 z^4 \sqrt{\frac{10}{12597}} & 116\,
   {\sqrt{\frac{5}{12597}}}\,z^3 & 31\,{\sqrt{\frac{15}{4199}}}\,z^2 & \frac{{\sqrt{\frac{95}
         {26}}}\,z}{2} & \frac{56\,z}{{\sqrt{1235}}} & 2\,{\sqrt{\frac{5}{247}}} & \frac{163}
   {{\sqrt{2470}}} & 0 & \frac{{\sqrt{\frac{195}{2}}}}{2\,z} & 0 & 0 & \frac{97}
   {16\,z^2} & 0 & 0 & \\[5pt] \frac{-49\,z^4}{9\,{\sqrt{1045}}} & \frac{91\,{\sqrt{\frac{11}{190}}}\,
     z^3}{18} & \frac{-217\,{\sqrt{\frac{2}{1045}}}\,z^2}{9} & \frac{-28\,
     {\sqrt{\frac{19}{2805}}}\,z}{9} & \frac{104\,{\sqrt{\frac{110}{969}}}\,z}{9} & \frac{112\,
     {\sqrt{\frac{10}{10659}}}}{9} & \frac{91\,{\sqrt{\frac{55}{969}}}}{9} & 0 & 0 & \frac{9}
   {2\,{\sqrt{2}}\,z} & 0 & 0 & \frac{97}{16\,z^2} & 0 & \\[5pt] \frac{5\,{\sqrt{\frac{5}{1066}}}\,z^4}
   {9} & \frac{-40\,{\sqrt{\frac{5}{533}}}\,z^3}{9} & \frac{80\,{\sqrt{\frac{5}{533}}}\,z^2}
   {9} & \frac{-320\,{\sqrt{\frac{10}{27183}}}\,z}{9} & \frac{217\,{\sqrt{\frac{41}{3315}}}\,z}
   {18} & \frac{128\,{\sqrt{\frac{5}{27183}}}}{9} & \frac{-31\,{\sqrt{\frac{82}{3315}}}}
   {9} & 0 & 0 & 0 & 0 & 0 & 0 & \frac{97}{16\,z^2} & \\[5pt]
   & & & & & & & & & & & & & &\ddots  }
$
}
\mbox{}\hspace{.4in}\mbox{}
\rotatebox{90}{Figure~3: Energy-momentum tensor (in the $L_1$ basis) of the critical Ising model in the $h=1/16$ sector.}
\thispagestyle{empty}
\end{figure}
\clearpage

\setlength{\arraycolsep}{2mm} 
\begin{figure}[p]  
\mbox{}\hspace{1.0in}\mbox{}
\rotatebox{90}{
$ \tpf{1/2}{1/2}{0} (z) = \smat{
   1 & z^{-2} & z^{-3} & \frac{3}{{\sqrt{10}}\,z^4} & \frac{1}{{\sqrt{10}}\,z^4} & \frac{2}
   {{\sqrt{5}}\,z^5} & \frac{1}{{\sqrt{5}}\,z^5} & \frac{{\sqrt{\frac{5}{7}}}}{z^6} & \frac{{
       \sqrt{\frac{5}{2}}}}{3\,z^6} & \frac{1}{3\,{\sqrt{14}}\,z^6} & \frac{3}
   {{\sqrt{14}}\,z^7} & \frac{1}{{\sqrt{3}}\,z^7} & \frac{1}{{\sqrt{42}}\,z^7} & \\[5pt] 
   z & -
     \frac{1}{z}  & 0 & \frac{1}{{\sqrt{10}}\,z^3} & \frac{-3}{{\sqrt{10}}\,z^3} & \frac{1}
   {{\sqrt{5}}\,z^4} & \frac{-2}{{\sqrt{5}}\,z^4} & \frac{3}{{\sqrt{35}}\,z^5} & \frac{-7}
   {3\,{\sqrt{10}}\,z^5} & \frac{-5}{3\,{\sqrt{14}}\,z^5} & \frac{{\sqrt{\frac{2}{7}}}}
   {z^6} & - \frac{1}{{\sqrt{3}}\,z^6}  & \frac{-2\,{\sqrt{\frac{2}{21}}}}
   {z^6} & \\[5pt] 
   z^2 & 0 & - \frac{1}{z}  & - \frac{1}{{\sqrt{10}}\,z^2} 
      & \frac{3}{{\sqrt{10}}\,z^2} & 0 & 0 & \frac{1}{{\sqrt{35}}\,z^4} & \frac{-2\,
     {\sqrt{\frac{2}{5}}}}{3\,z^4} & \frac{5\,{\sqrt{\frac{2}{7}}}}{3\,z^4} & \frac{1}
   {{\sqrt{14}}\,z^5} & - \frac{1}{{\sqrt{3}}\,z^5}  & \frac{5}
   {{\sqrt{42}}\,z^5} & \\[5pt] 
   z^3 & 0 & 0 & \frac{-3}{{\sqrt{10}}\,z} & - \frac{1}
     {{\sqrt{10}}\,z}  & - \frac{1}{{\sqrt{5}}\,z^2}  & \frac{2}
   {{\sqrt{5}}\,z^2} & - \frac{1}{{\sqrt{35}}\,z^3}  & \frac{2\,{\sqrt{\frac{2}{5}}}}
   {3\,z^3} & \frac{-5\,{\sqrt{\frac{2}{7}}}}{3\,z^3} & 0 & 0 & 0 & \\[5pt] z^4 & 0 & 0 & 0 & 0 & 
    \frac{-2}{{\sqrt{5}}\,z} & - \frac{1}{{\sqrt{5}}\,z}  & \frac{-3}
   {{\sqrt{35}}\,z^2} & \frac{7}{3\,{\sqrt{10}}\,z^2} & \frac{5}{3\,{\sqrt{14}}\,z^2} & -
     \frac{1}{{\sqrt{14}}\,z^3}  & \frac{1}{{\sqrt{3}}\,z^3} & \frac{-5}
   {{\sqrt{42}}\,z^3} & \\[5pt] 0 & z^2 & -z & \frac{1}{{\sqrt{10}}} & \frac{-3}
   {{\sqrt{10}}} & 0 & 0 & 0 & 0 & 0 & 0 & 0 & 0 & \\[5pt] z^5 & 0 & 0 & 0 & 0 & 0 & 0 & - 
      \frac{{\sqrt{\frac{5}{7}}}}{z}  & \frac{-{\sqrt{\frac{5}{2}}}}{3\,z} & \frac{-1}
   {3\,{\sqrt{14}}\,z} & - \frac{{\sqrt{\frac{2}{7}}}}{z^2}  & \frac{1}
   {{\sqrt{3}}\,z^2} & \frac{2\,{\sqrt{\frac{2}{21}}}}{z^2} & \\[5pt] 0 & z^3 & 0 & \frac{-3\,z}
   {{\sqrt{10}}} & - \frac{z}{{\sqrt{10}}}  & \frac{1}{{\sqrt{5}}} & \frac{-2}
   {{\sqrt{5}}} & 0 & 0 & 0 & 0 & 0 & 0 & \\[5pt] z^6 & 0 & 0 & 0 & 0 & 0 & 0 & 0 & 0 & 0 & \frac{-3}
   {{\sqrt{14}}\,z} & - \frac{1}{{\sqrt{3}}\,z}  & - \frac{1}
     {{\sqrt{42}}\,z}  & \\[5pt]  
   0 & \frac{2\,z^4}{{\sqrt{5}}} & \frac{z^3}{{\sqrt{5}}} & 
    \frac{-3\,z^2}{5\,{\sqrt{2}}} & \frac{-z^2}{5\,{\sqrt{2}}} & \frac{-4\,z}{5} & \frac{-2\,z}
   {5} & \frac{{\sqrt{7}}}{5} & \frac{-3\,{\sqrt{2}}}{5} & 0 & 0 & 0 & 0 & \\[5pt] 0 & \frac{z^4}
   {{\sqrt{5}}} & \frac{-2\,z^3}{{\sqrt{5}}} & \frac{3\,{\sqrt{2}}\,z^2}{5} & \frac{{\sqrt{2}}\,
     z^2}{5} & \frac{-2\,z}{5} & \frac{-z}{5} & \frac{1}{5\,{\sqrt{7}}} & \frac{1}
   {15\,{\sqrt{2}}} & \frac{-5\,{\sqrt{\frac{5}{14}}}}{3} & 0 & 0 & 0 & \\[5pt] z^
   7 & 0 & 0 & 0 & 0 & 0 & 0 & 0 & 0 & 0 & 0 & 0 & 0 & \\[5pt] 0 & 2\,{\sqrt{\frac{5}{33}}}\,z^5 & 
    \frac{8\,z^4}{{\sqrt{165}}} & \frac{z^3}{5\,{\sqrt{66}}} & \frac{- {\sqrt{\frac{3}
           {22}}}\,z^3 }{5} & \frac{-17\,z^2}{5\,{\sqrt{33}}} & \frac{-2\,
     {\sqrt{\frac{3}{11}}}\,z^2}{5} & \frac{-7\,{\sqrt{\frac{7}{33}}}\,z}{5} & \frac{-3\,
     {\sqrt{\frac{6}{11}}}\,z}{5} & 0 & 2\,{\sqrt{\frac{14}{165}}} & \frac{-6}
   {{\sqrt{55}}} & 0 & \\[5pt] 0 & {\sqrt{\frac{5}{13}}}\,z^5 & \frac{-6\,z^4}{{\sqrt{65}}} & \frac{
      - {\sqrt{\frac{2}{13}}}\,z^3 }{5} & \frac{3\,{\sqrt{\frac{2}{13}}}\,z^3}
   {5} & \frac{14\,z^2}{5\,{\sqrt{13}}} & \frac{2\,z^2}{5\,{\sqrt{13}}} & \frac{-27\,z}
   {5\,{\sqrt{91}}} & \frac{-17\,z}{15\,{\sqrt{26}}} & \frac{-5\,{\sqrt{\frac{5}{182}}}\,z}
   {3} & 2\,{\sqrt{\frac{2}{455}}} & \frac{1}{{\sqrt{195}}} & -5\,
   {\sqrt{\frac{10}{273}}} & \\[5pt] 
   0 & \frac{2\,z^5}{{\sqrt{429}}} & \frac{-5\,z^4}
   {{\sqrt{429}}} & 2\,{\sqrt{\frac{10}{429}}}\,z^3 & -2\,{\sqrt{\frac{30}{143}}}\,z^3 & -2\,
   {\sqrt{\frac{5}{429}}}\,z^2 & 3\,{\sqrt{\frac{15}{143}}}\,z^2 & 2\,{\sqrt{\frac{5}{3003}}}\,
   z & - {\sqrt{\frac{30}{143}}}\,z  & {\sqrt{\frac{66}{91}}}\,z & - \frac{1}
     {{\sqrt{6006}}}  & \frac{1}{{\sqrt{143}}} & -{\sqrt{\frac{11}{182}}} & \\[5pt]
   & & & & & & & & & & & & & \ddots }
$
}
\mbox{}\hspace{.4in}\mbox{}
\rotatebox{90}{\parbox{9.0in}{Figure~4: The lower-left block $\tpf{1/2}{1/2}{0} (z)$  of the primary field $\phi^{(1/2)}(z)$ (in the $L_1$ basis) of the critical Ising model. Notice that the squares of the 
coefficients in the first column give the expansion of $(1-u)^{-1}$ in agreement with the two-point function (\ref{laur-tay2}).}}
\thispagestyle{empty}
\end{figure}

\clearpage

\setlength{\arraycolsep}{1.4mm}
\begin{figure}[p]  
\mbox{}\hspace{1.0in}\mbox{}
\rotatebox{90}{
$ \tpf{1/2}{1/16}{1/16} (z) = z^{-1/2}  
\smat{
    \frac{1}{{\sqrt{2}}} & \frac{1}{z} & z^{-2} & \frac{5\,{\sqrt{\frac{2}{51}}}}{z^3} & 
    \frac{1}{{\sqrt{51}}\,z^3} & \frac{7}{{\sqrt{51}}\,z^4} & \frac{{\sqrt{\frac{2}{51}}}}
   {z^4} & \frac{21\,{\sqrt{\frac{2}{935}}}}{z^5} & \frac{3\,{\sqrt{\frac{2}{323}}}}{z^5} & \frac{1}
   {{\sqrt{1045}}\,z^5} & \frac{7\,{\sqrt{\frac{66}{3485}}}}{z^6} & \\[3mm] z & - \frac{1}
     {{\sqrt{2}}}   & 0 & \frac{1}{{\sqrt{51}}\,z^2} & \frac{-5\,{\sqrt{\frac{2}{51}}}}
   {z^2} & \frac{{\sqrt{\frac{2}{51}}}}{z^3} & \frac{-7}{{\sqrt{51}}\,z^3} & \frac{7}
   {{\sqrt{935}}\,z^4} & \frac{-16}{{\sqrt{323}}\,z^4} & \frac{-9\,{\sqrt{\frac{2}{1045}}}}
   {z^4} & \frac{28\,{\sqrt{\frac{3}{38335}}}}{z^5} & \\[3mm] z^2 & 0 & - \frac{1}{{\sqrt{2}}}
       & - \frac{1}{{\sqrt{51}}\,z}   & \frac{5\,{\sqrt{\frac{2}{51}}}}
   {z} & 0 & 0 & \frac{2}{{\sqrt{935}}\,z^3} & \frac{-7}{{\sqrt{323}}\,z^3} & \frac{21\,
     {\sqrt{\frac{2}{1045}}}}{z^3} & \frac{7\,{\sqrt{\frac{5}{23001}}}}{z^4} & \\[3mm] 5\,
   {\sqrt{\frac{2}{51}}}\,z^3 & \frac{z^2}{{\sqrt{51}}} & - \frac{z}{{\sqrt{51}}} 
      & \frac{-49}{51\,{\sqrt{2}}} & - \frac{10}{51}   & \frac{-10}{51\,z} & \frac
     {35\,{\sqrt{2}}}{51\,z} & \frac{-2\,{\sqrt{\frac{10}{33}}}}{17\,z^2} & \frac{35\,
     {\sqrt{\frac{2}{57}}}}{17\,z^2} & \frac{-14\,{\sqrt{\frac{15}{3553}}}}{z^2} & \frac{{\sqrt
        {\frac{2}{2255}}}}{51\,z^3} & \\[3mm] \frac{z^3}{{\sqrt{51}}} & -5\,{\sqrt{\frac{2}{51}}}\,
   z^2 & 5\,{\sqrt{\frac{2}{51}}}\,z & - \frac{10}{51}   & \frac{49}
   {51\,{\sqrt{2}}} & \frac{-{\sqrt{2}}}{51\,z} & \frac{7}{51\,z} & \frac{-2}
   {17\,{\sqrt{165}}\,z^2} & \frac{7}{17\,{\sqrt{57}}\,z^2} & \frac{-7\,{\sqrt{\frac{6}{17765}}}}
   {z^2} & \frac{-2\,{\sqrt{\frac{5}{451}}}}{51\,z^3} & \\[3mm] \frac{7\,z^4}{{\sqrt{51}}} & 
   {\sqrt{\frac{2}{51}}}\,z^3 & 0 & \frac{-10\,z}{51} & \frac{- {\sqrt{2}}\,z  }
   {51} & \frac{-47}{51\,{\sqrt{2}}} & - \frac{14}{51}   & \frac{-49}
   {17\,{\sqrt{165}}\,z} & \frac{112}{17\,{\sqrt{57}}\,z} & \frac{21\,{\sqrt{\frac{6}{17765}}}}
   {z} & \frac{-247}{51\,{\sqrt{2255}}\,z^2} & \\[3mm] {\sqrt{\frac{2}{51}}}\,z^4 & \frac{-7\,z^3}
   {{\sqrt{51}}} & 0 & \frac{35\,{\sqrt{2}}\,z}{51} & \frac{7\,z}{51} & - \frac{14}{51}
       & \frac{47}{51\,{\sqrt{2}}} & \frac{-7\,{\sqrt{\frac{2}{165}}}}{17\,z} & \frac{16\,
     {\sqrt{\frac{2}{57}}}}{17\,z} & \frac{6\,{\sqrt{\frac{3}{17765}}}}{z} & \frac{-28\,
     {\sqrt{\frac{2}{2255}}}}{51\,z^2} & \\[3mm] 21\,{\sqrt{\frac{2}{935}}}\,z^5 & \frac{7\,z^4}
   {{\sqrt{935}}} & \frac{2\,z^3}{{\sqrt{935}}} & \frac{-2\,{\sqrt{\frac{10}{33}}}\,z^2}{17} & 
    \frac{-2\,z^2}{17\,{\sqrt{165}}} & \frac{-49\,z}{17\,{\sqrt{165}}} & \frac{-7\,
     {\sqrt{\frac{2}{165}}}\,z}{17} & \frac{-829}{935\,{\sqrt{2}}} & \frac{-126\,
     {\sqrt{\frac{2}{1045}}}}{17} & \frac{-42}{55\,{\sqrt{323}}} & \frac{-1762\,
     {\sqrt{\frac{2}{123}}}}{935\,z} & \\[3mm] 3\,{\sqrt{\frac{2}{323}}}\,z^5 & \frac{-16\,z^4}
   {{\sqrt{323}}} & \frac{-7\,z^3}{{\sqrt{323}}} & \frac{35\,{\sqrt{\frac{2}{57}}}\,z^2}{17} & 
    \frac{7\,z^2}{17\,{\sqrt{57}}} & \frac{112\,z}{17\,{\sqrt{57}}} & \frac{16\,
     {\sqrt{\frac{2}{57}}}\,z}{17} & \frac{-126\,{\sqrt{\frac{2}{1045}}}}{17} & \frac{287}
   {323\,{\sqrt{2}}} & \frac{-6}{19\,{\sqrt{935}}} & \frac{-259\,{\sqrt{\frac{2}{128535}}}}
   {17\,z} & \\[3mm] \frac{z^5}{{\sqrt{1045}}} & -9\,{\sqrt{\frac{2}{1045}}}\,z^4 & 21\,
   {\sqrt{\frac{2}{1045}}}\,z^3 & -14\,{\sqrt{\frac{15}{3553}}}\,z^2 & -7\,
   {\sqrt{\frac{6}{17765}}}\,z^2 & 21\,{\sqrt{\frac{6}{17765}}}\,z & 6\,{\sqrt{\frac{3}{17765}}}\,
   z & \frac{-42}{55\,{\sqrt{323}}} & \frac{-6}{19\,{\sqrt{935}}} & \frac{1043}
   {1045\,{\sqrt{2}}} & \frac{-14\,{\sqrt{\frac{3}{13243}}}}{55\,z} & \\[3mm] 7\,
   {\sqrt{\frac{66}{3485}}}\,z^6 & 28\,{\sqrt{\frac{3}{38335}}}\,z^5 & 7\,{\sqrt{\frac{5}{23001}}}\,
   z^4 & \frac{{\sqrt{\frac{2}{2255}}}\,z^3}{51} & \frac{-2\,{\sqrt{\frac{5}{451}}}\,z^3}
   {51} & \frac{-247\,z^2}{51\,{\sqrt{2255}}} & \frac{-28\,{\sqrt{\frac{2}{2255}}}\,z^2}{51} & 
    \frac{-1762\,{\sqrt{\frac{2}{123}}}\,z}{935} & \frac{-259\,{\sqrt{\frac{2}{128535}}}\,z}
   {17} & \frac{-14\,{\sqrt{\frac{3}{13243}}}\,z}{55} & \frac{-98443}{115005\,{\sqrt{2}}} &  \\[3mm]
   & & & & & & & & & & & \ddots } \hspace{-8mm}  $  }
\mbox{}\hspace{.4in}\mbox{}
\rotatebox{90}{Figure~5: The central block $\tpf{1/2}{1/16}{1/16} (z) $ of the primary field $\phi^{(1/2)}(z)$ (in the $L_1$ basis) of the critical Ising model.}
\thispagestyle{empty}
\end{figure}

\clearpage

\begin{figure}[p]
\mbox{}\hspace{.9in}\mbox{}
\rotatebox{90}{
$ \tpf{1/16}{1/16}{0} (z) = \smat{
   1 & \frac{1}{8\,z^2} & \frac{1}{8\,z^3} & \frac{3}{8\,{\sqrt{10}}\,z^4} & \frac{3}
   {64\,{\sqrt{10}}\,z^4} & \frac{1}{4\,{\sqrt{5}}\,z^5} & \frac{3}{64\,{\sqrt{5}}\,z^5} & 
    \frac{{\sqrt{\frac{5}{7}}}}{8\,z^6} & \frac{{\sqrt{\frac{5}{2}}}}{64\,z^6} & \frac{5}
   {512\,{\sqrt{14}}\,z^6} & \\[5pt] \frac{z}{2\,{\sqrt{2}}} & \frac{-15}{16\,{\sqrt{2}}\,z} & \frac{-7}
   {16\,{\sqrt{2}}\,z^2} & \frac{-13}{32\,{\sqrt{5}}\,z^3} & \frac{-93}
   {256\,{\sqrt{5}}\,z^3} & \frac{-3}{8\,{\sqrt{10}}\,z^4} & \frac{-69}
   {128\,{\sqrt{10}}\,z^4} & \frac{-11}{16\,{\sqrt{70}}\,z^5} & \frac{-91}
   {256\,{\sqrt{5}}\,z^5} & \frac{-235}{2048\,{\sqrt{7}}\,z^5} & \\[5pt] \frac{3\,z^2}
   {8\,{\sqrt{2}}} & \frac{35}{64\,{\sqrt{2}}} & \frac{-45}{64\,{\sqrt{2}}\,z} & \frac{-87}
   {128\,{\sqrt{5}}\,z^2} & \frac{713}{1024\,{\sqrt{5}}\,z^2} & \frac{-21}
   {32\,{\sqrt{10}}\,z^3} & \frac{217}{512\,{\sqrt{10}}\,z^3} & \frac{-81}
   {64\,{\sqrt{70}}\,z^4} & \frac{79}{1024\,{\sqrt{5}}\,z^4} & \frac{3055}
   {8192\,{\sqrt{7}}\,z^4} & \\[5pt] \frac{{\sqrt{51}}\,z^3}{32} & \frac{35\,z}{256\,{\sqrt{51}}} & 
    \frac{875}{256\,{\sqrt{51}}} & \frac{-9\,{\sqrt{\frac{255}{2}}}}{256\,z} & \frac{-713\,
     {\sqrt{\frac{15}{34}}}}{2048\,z} & \frac{-11\,{\sqrt{\frac{51}{5}}}}{128\,z^2} & \frac{2139\,
     {\sqrt{\frac{3}{85}}}}{2048\,z^2} & \frac{-43\,{\sqrt{\frac{51}{35}}}}{256\,z^3} & \frac{7967}
   {2048\,{\sqrt{510}}\,z^3} & \frac{-94705}{16384\,{\sqrt{714}}\,z^3} & \\[5pt] 0 & \frac{7\,z}
   {{\sqrt{102}}} & \frac{-7}{2\,{\sqrt{102}}} & 0 & \frac{-5\,{\sqrt{\frac{15}{17}}}}
   {8\,z} & 0 & \frac{-5\,{\sqrt{\frac{15}{34}}}}{16\,z^2} & 0 & \frac{-5\,{\sqrt{\frac{5}{51}}}}
   {16\,z^3} & \frac{-49\,{\sqrt{\frac{7}{51}}}}{128\,z^3} & \\[5pt] \frac{5\,{\sqrt{\frac{51}{2}}}\,z^4}
   {128} & \frac{21\,{\sqrt{\frac{3}{34}}}\,z^2}{1024} & \frac{287\,z}{1024\,{\sqrt{102}}} & 
    \frac{17213}{2048\,{\sqrt{255}}} & \frac{4991\,{\sqrt{\frac{3}{85}}}}{16384} & \frac{-15\,
     {\sqrt{\frac{255}{2}}}}{512\,z} & \frac{-3565\,{\sqrt{\frac{15}{34}}}}{8192\,z} & \frac{-59\,
     {\sqrt{\frac{255}{14}}}}{1024\,z^2} & \frac{4991\,{\sqrt{\frac{5}{51}}}}{16384\,z^2} & 
    \frac{435643}{131072\,{\sqrt{357}}\,z^2} & \\[5pt] 0 & \frac{3\,{\sqrt{\frac{3}{17}}}\,z^2}
   {4} & \frac{31\,z}{8\,{\sqrt{51}}} & -{\sqrt{\frac{10}{51}}} & \frac{5\,{\sqrt{\frac{15}{34}}}}
   {16} & 0 & \frac{-35\,{\sqrt{\frac{15}{17}}}}{64\,z} & 0 & \frac{-35\,{\sqrt{\frac{5}{102}}}}
   {32\,z^2} & \frac{161\,{\sqrt{\frac{7}{102}}}}{256\,z^2} & \\[5pt] \frac{3\,{\sqrt{935}}\,z^5}
   {512} & \frac{21\,{\sqrt{\frac{17}{55}}}\,z^3}{4096} & \frac{1197\,z^2}
   {4096\,{\sqrt{935}}} & \frac{13391\,z}{20480\,{\sqrt{374}}} & \frac{4991\,z}
   {163840\,{\sqrt{374}}} & \frac{77777}{10240\,{\sqrt{187}}} & \frac{204631}
   {163840\,{\sqrt{187}}} & \frac{-225\,{\sqrt{\frac{187}{7}}}}{4096\,z} & \frac{-17825\,
     {\sqrt{\frac{11}{34}}}}{32768\,z} & \frac{-435643\,{\sqrt{\frac{5}{2618}}}}
   {262144\,z} & \\[5pt] 0 & \frac{3\,{\sqrt{\frac{17}{19}}}\,z^3}{16} & \frac{189\,z^2}
   {32\,{\sqrt{323}}} & 5\,{\sqrt{\frac{5}{646}}}\,z & \frac{5\,{\sqrt{\frac{5}{646}}}\,z}
   {64} & -4\,{\sqrt{\frac{5}{323}}} & \frac{565\,{\sqrt{\frac{5}{323}}}}{256} & 0 & \frac{-35\,
     {\sqrt{\frac{95}{34}}}}{128\,z} & \frac{-805\,{\sqrt{\frac{7}{646}}}}{1024\,z} & \\[5pt] 0 & 
    \frac{7\,z^3}{{\sqrt{2090}}} & \frac{-21\,z^2}{2\,{\sqrt{2090}}} & \frac{21\,z}
   {10\,{\sqrt{209}}} & \frac{373\,z}{40\,{\sqrt{209}}} & \frac{-7}{10\,{\sqrt{418}}} & \frac{-373}
   {80\,{\sqrt{418}}} & 0 & 0 & \frac{-189\,{\sqrt{\frac{35}{209}}}}{128\,z} & \\[5pt] \frac{
     {\sqrt{115005}}\,z^6}{2048} & \frac{35\,{\sqrt{\frac{255}{451}}}\,z^4}{16384} & \frac{511\,
     {\sqrt{\frac{51}{2255}}}\,z^3}{16384} & \frac{72093\,{\sqrt{\frac{3}{15334}}}\,z^2}
   {81920} & \frac{14973\,{\sqrt{\frac{3}{15334}}}\,z^2}{655360} & \frac{21679\,
     {\sqrt{\frac{11}{2091}}}\,z}{40960} & \frac{94829\,{\sqrt{\frac{3}{7667}}}\,z}{655360} & 
    \frac{2699239\,{\sqrt{\frac{7}{23001}}}}{81920} & \frac{16684913}
   {655360\,{\sqrt{46002}}} & \frac{435643\,{\sqrt{\frac{35}{46002}}}}{1048576} & \\[5pt] 0 & \frac{5\,
     {\sqrt{\frac{255}{494}}}\,z^4}{32} & \frac{{\sqrt{\frac{4845}{26}}}\,z^3}{64} & \frac{93\,
     {\sqrt{\frac{3}{4199}}}\,z^2}{8} & \frac{15\,{\sqrt{\frac{3}{4199}}}\,z^2}{256} & 29\,
   {\sqrt{\frac{2}{12597}}}\,z & \frac{295\,{\sqrt{\frac{3}{8398}}}\,z}{512} & -16\,
   {\sqrt{\frac{14}{12597}}} & \frac{18215}{512\,{\sqrt{12597}}} & \frac{1127\,
     {\sqrt{\frac{35}{12597}}}}{4096} & \\[5pt] 0 & \frac{35\,{\sqrt{\frac{5}{209}}}\,z^4}{36} & 
    \frac{-7\,{\sqrt{\frac{19}{55}}}\,z^3}{72} & \frac{-217\,z^2}{60\,{\sqrt{418}}} & \frac{373\,
     z^2}{80\,{\sqrt{418}}} & \frac{91\,{\sqrt{\frac{11}{19}}}\,z}{360} & \frac{2611\,z}
   {320\,{\sqrt{209}}} & \frac{-14\,{\sqrt{\frac{7}{209}}}}{45} & \frac{-373}
   {60\,{\sqrt{418}}} & \frac{147\,{\sqrt{\frac{35}{418}}}}{256} & \\[5pt] 0 & \frac{5\,
     {\sqrt{\frac{10}{533}}}\,z^4}{9} & \frac{-10\,{\sqrt{\frac{10}{533}}}\,z^3}{9} & \frac{10\,
     z^2}{3\,{\sqrt{533}}} & \frac{-31\,z^2}{4\,{\sqrt{533}}} & \frac{-10\,{\sqrt{\frac{2}{533}}}\,
     z}{9} & \frac{31\,z}{4\,{\sqrt{1066}}} & \frac{5\,{\sqrt{\frac{2}{3731}}}}{9} & \frac{-31}
   {24\,{\sqrt{533}}} & \frac{279\,{\sqrt{\frac{5}{3731}}}}{64}\\[5pt]
   & & & & & & & & & & \ddots }
$
}
\mbox{}\hspace{.4in}\mbox{}
\rotatebox{90}{\parbox{9.3in}{Figure~6: The left-central block $ \tpf{1/16}{1/16}{0} (z) $ of the primary field $\phi^{(1/16)}(z)$ (in the $L_1$ basis) of the critical Ising model. Notice that the squares of the 
coefficients in the first column give the expansion of $(1-u)^{-1/8}$ in agreement with the two-point function (\ref{laur-tay2}).}}
\thispagestyle{empty}
\end{figure}

\clearpage
\setlength{\arraycolsep}{1.62mm}
\begin{figure}[p]
\mbox{}\hspace{.9in}\mbox{}
\rotatebox{90}{
$  \tpf{1/16}{1/2}{1/16} (z) = z^{3/8} 
\smat{
   \frac{1}{{\sqrt{2}}} & \frac{-3}{4\,z} & \frac{-5}{16\,z^2} & \frac{-65}
   {32\,{\sqrt{102}}\,z^3} & - \frac{1}{{\sqrt{51}}\,z^3}   & \frac{-91\,
     {\sqrt{\frac{3}{17}}}}{256\,z^4} & \frac{-{\sqrt{\frac{3}{34}}}}{2\,z^4} & \frac{-2639}
   {512\,{\sqrt{1870}}\,z^5} & \frac{-29}{8\,{\sqrt{646}}\,z^5} & - \frac{1}
     {{\sqrt{1045}}\,z^5}   & \\[3mm] \frac{z}{2\,{\sqrt{2}}} & \frac{5}{8} & \frac{-21}
   {32\,z} & \frac{-185}{64\,{\sqrt{102}}\,z^2} & \frac{5}{2\,{\sqrt{51}}\,z^2} & \frac{-689}
   {512\,{\sqrt{51}}\,z^3} & \frac{7}{4\,{\sqrt{102}}\,z^3} & \frac{-6279}
   {1024\,{\sqrt{1870}}\,z^4} & \frac{33}{16\,{\sqrt{646}}\,z^4} & \frac{9}
   {2\,{\sqrt{1045}}\,z^4} & \\[3mm] \frac{3\,z^2}{8\,{\sqrt{2}}} & \frac{7\,z}{32} & \frac{81}
   {128} & \frac{-529\,{\sqrt{\frac{3}{34}}}}{256\,z} & \frac{-5\,{\sqrt{\frac{3}{17}}}}
   {8\,z} & \frac{-3955}{2048\,{\sqrt{51}}\,z^2} & \frac{35}{16\,{\sqrt{102}}\,z^2} & \frac{-31837}
   {4096\,{\sqrt{1870}}\,z^3} & \frac{301}{64\,{\sqrt{646}}\,z^3} & \frac{-63}
   {8\,{\sqrt{1045}}\,z^3} & \\[3mm] \frac{5\,z^3}{16\,{\sqrt{2}}} & \frac{9\,z^2}{64} & \frac{55\,z}
   {256} & \frac{3283}{512\,{\sqrt{102}}} & \frac{5}{16\,{\sqrt{51}}} & \frac{-5735\,
     {\sqrt{\frac{3}{17}}}}{4096\,z} & \frac{-35\,{\sqrt{\frac{3}{34}}}}{32\,z} & \frac{-18391\,
     {\sqrt{\frac{5}{374}}}}{8192\,z^2} & \frac{385}{128\,{\sqrt{646}}\,z^2} & \frac{21\,
     {\sqrt{\frac{5}{209}}}}{16\,z^2} & \\[3mm] \frac{35\,z^4}{128\,{\sqrt{2}}} & \frac{55\,z^3}
   {512} & \frac{273\,z^2}{2048} & \frac{8893\,z}{4096\,{\sqrt{102}}} & \frac{5\,z}
   {128\,{\sqrt{51}}} & \frac{148525}{32768\,{\sqrt{51}}} & \frac{175}{256\,{\sqrt{102}}} & 
    \frac{-323337\,{\sqrt{\frac{5}{374}}}}{65536\,z} & \frac{-9135}{1024\,{\sqrt{646}}\,z} & 
    \frac{-63\,{\sqrt{\frac{5}{209}}}}{128\,z} & \\[3mm] \frac{z^4}{128\,{\sqrt{2}}} & \frac{-35\,z^3}
   {512} & \frac{315\,z^2}{2048} & \frac{-1995\,{\sqrt{\frac{3}{34}}}\,z}{4096} & \frac{189\,
     {\sqrt{\frac{3}{17}}}\,z}{128} & \frac{4389\,{\sqrt{\frac{3}{17}}}}{32768} & \frac{-297\,
     {\sqrt{\frac{3}{34}}}}{256} & \frac{-1197\,{\sqrt{\frac{11}{170}}}}{65536\,z} & \frac{891}
   {1024\,{\sqrt{646}}\,z} & \frac{-123\,{\sqrt{\frac{11}{95}}}}{128\,z} & \\[3mm] \frac{63\,z^5}
   {256\,{\sqrt{2}}} & \frac{91\,z^4}{1024} & \frac{405\,z^3}{4096} & \frac{3635\,
     {\sqrt{\frac{3}{34}}}\,z^2}{8192} & \frac{{\sqrt{\frac{3}{17}}}\,z^2}{256} & \frac{100849\,z}
   {65536\,{\sqrt{51}}} & \frac{49\,z}{512\,{\sqrt{102}}} & \frac{3590887}
   {131072\,{\sqrt{1870}}} & \frac{4487}{2048\,{\sqrt{646}}} & \frac{63}
   {256\,{\sqrt{1045}}} & \\[3mm] \frac{3\,z^5}{256\,{\sqrt{2}}} & \frac{-81\,z^4}{1024} & \frac{385\,z^3}
   {4096} & \frac{1575\,{\sqrt{\frac{3}{34}}}\,z^2}{8192} & \frac{189\,{\sqrt{\frac{3}{17}}}\,z^2}
   {256} & \frac{-25137\,{\sqrt{\frac{3}{17}}}\,z}{65536} & \frac{783\,{\sqrt{\frac{3}{34}}}\,z}
   {512} & \frac{44289\,{\sqrt{\frac{11}{170}}}}{131072} & \frac{-22869}{2048\,{\sqrt{646}}} & 
    \frac{41\,{\sqrt{\frac{11}{95}}}}{256} &  \\[3mm] 
   &&&&&&&&&&\ddots }
$ } 
\mbox{}\hspace{.4in}\mbox{}
\rotatebox{90}{Figure~7: The bottom-central block $ \tpf{1/16}{1/2}{1/16} (z)$ of the primary field $\phi^{(1/16)}(z)$ (in the $L_1$ basis) of the critical Ising model.}
\thispagestyle{empty}
\end{figure}

\clearpage\newpage
\noindent
Again, the full field $\sigma(z,\bar{z}) \equiv \phi^{(1/16)}(z,\bar{z})$ is self-adjoint with the following 
relations between blocks
\begin{equation}
\tpf{1/16}{0}{1/16} (z) = z^{-1/8} \left[ \tpf{1/16}{1/16}{0} (\frac{1}{z}) \right] ^{T} , \qquad  
\tpf{1/16}{1/16}{1/2} (z) = z^{-1/8} \left[ \tpf{1/16}{1/2}{1/16} (\frac{1}{z}) \right] ^{T}
\end{equation}
The left-central and the bottom-central blocks in the $L_1$  basis are shown in Figures~6 and 7.

\section{Tricritical Ising model}

In this section we build matrix representations of the Virasoro algebra and fields for the tricritical Ising model.

\subsection{Tricritical Ising model: $\V_0$}
The $q$ series for the Virasoro character in the $h=0$ sector is
\begin{equation}
\chi_0(q) \; = \; 1 + q^2 + q^3 + 2\,q^4 + 2\,q^5 + 4\,q^6 + 4\,q^7 + 7\,q^8 + 8\,q^9 + 
12\,q^{10} + 14\,q^{11} + 20\,q^{12} +  \mbox{O}(q^{13})
\end{equation}
The first discrepancy with the generic $A_{\infty}$ case is at level $12$.
We compute the matrices in the $L_1$ basis, truncating the energy
such that terms corresponding to $q^{10}$ are dropped. We thus build the following $30\times 30$ matrices:
\begin{equation} \label{L0_vacuum}
L_0=\mbox{Diagonal}(0,2,3,4,4,5,5,6,6,6,6,7,7,7,7,8,8,8,8,8,8,8,9,9,9,9,9,9,9,9, \ldots)
\end{equation}
\begin{eqnarray} \label{L0_1}
&\qquad L_1\;=\;\mbox{\scriptsize{\mbox{$
\left(\begin{array}{cccccccccccc}
 . &0& . & . & . & . & . & . & . & . & . & \\
 . & . &2& . & . & . & . & . & . & . & . & \\
 . & . & . &\sqrt{10}&0& . & . & . & . & . & . & \\
 . & . & . & . & . &3\sqrt{2}&0& . & . & . & . & \\ 
 . & . & . & . & . &0&2\sqrt{2}& . & . & . & . & \\
 . & . & . & . & . & . & . &2\sqrt{7}&0&0&0& \\
 . & . & . & . & . & . & . &0&3\sqrt{2}&0&0& \\
 . & . & . & . & . & . & . & . & . & . & . & \\
 . & . & . & . & . & . & . & . & . & . & . & \\
 . & . & . & . & . & . & . & . & . & . & . & \\
 . & . & . & . & . & . & . & . & . & . & . & \\
 & & & & & & & & & & &\ddots
\end{array}\right) $}}} & 
\end{eqnarray}
\begin{eqnarray}
&L_1 = \smat{0}\oplus\smat{2}\oplus\smat{\sqrt{10}&0}\oplus\smat{3\sqrt{2}&0\cr 0&2\sqrt{2}}
\oplus\smat{2\sqrt{7}&0&0&0\cr 0&3\sqrt{2}&0&0}
\oplus\smat{2\sqrt{10}&0&0&0\cr 0&\sqrt{30}&0&0\cr 0&0&2\sqrt{3}&0\cr 0&0&0&2\sqrt{3}} &
\nonumber\\[4pt]
&\oplus\smat{3\sqrt{6}&0&0&0&0&0&0\cr 0&2\sqrt{11}&0&0&0&0&0\cr 
0&0&\sqrt{26}&0&0&0&0\cr 0&0&0&\sqrt{26}&0&0&0\cr }
\oplus\smat{
\sqrt{70}&0&0&0&0&0&0&0\cr 
0&2\sqrt{15}&0&0&0&0&0&0\cr
0&0&\sqrt{42}&0&0&0&0&0\cr
0&0&0&\sqrt{42}&0&0&0&0\cr
0&0&0&0&4&0&0&0\cr
0&0&0&0&0&4&0&0\cr
0&0&0&0&0&0&4&0}\oplus\ldots &
\end{eqnarray} 
\begin{eqnarray} \label{L0_2}
& \mbox{}\hspace{-.25in}\mbox{} 
L_2\;=\;\mbox{\scriptsize{\mbox{$
\left( \begin{array}{ccccccccccccc}
.& \frac{1}{2} \sqrt{\frac{7}{5}} & .& .& .& . & . & 
   . & . & . & . & . &   \cr
. & . & . & 3\, \sqrt{\frac{2}{5}} & \sqrt{\frac{51}{10}} & . & 
   . & . & . & . & . & . &   \cr 
. & . & . & . & . & \frac{7}{\sqrt{5}} & \frac{1}{2} \sqrt{\frac{51}{5}}
    & . & . & . & . & . &   \cr 
. & . & . & . & . & . & . & 8\, \sqrt{\frac{2}{7}} & 
   \frac{1}{2} \sqrt{\frac{17}{3}} & \sqrt{\frac{26}{105}} & 0 & . &  \cr 
. & . & . & . & . & . & . & 0 & 2 & 
   2\, \sqrt{\frac{238}{65}} & \frac{27}{2\, \sqrt{65}} & . &  \cr 
. & . & . & . & . & . & . & . & . & . & . & 9\,
   \sqrt{\frac{5}{14}} &   \cr 
. & . & . & . & . & . & . & . & . & . & . & 0 
   &  \cr 
. & . & . & . & . & . & . & . & . & . & . & .
   &   \cr 
. & . & . & . & . & . & . & . & . & . & . & . 
   &  \cr 
. & . & . & . & . & . & . & . & . & . & . & . 
   &  \cr 
. & . & . & . & . & . & . & . & . & . & . & . 
   &  \cr 
. & . & . & . & . & . & . & . & . & . & . & . 
   &   \cr
  &   &   &   &   &   &   &   &   &   & 
     &   & \ddots  
\end{array} \right)$}}} & 
\end{eqnarray}
\begin{eqnarray}
& L_2= \smat{\frac{1}{2} \sqrt{\frac{7}{5}} } \oplus 
\smat{3\, \sqrt{\frac{2}{5}} &  \sqrt{\frac{51}{10}}} \oplus 
\smat{\frac{7}{\sqrt{5}} & \frac{1}{2} \sqrt{\frac{51}{5}} } \oplus 
\smat{8\, \sqrt{\frac{2}{7}} &\frac{1}{2}\sqrt{\frac{17}{3}} & \sqrt{\frac{26}{105}} & 0 \cr 
  0 & 2 & 2\, \sqrt{\frac{238}{65}} & \frac{27}{2\, \sqrt{65}}} \oplus \nonumber & \\[4pt]  
& \smat{9\,\sqrt{\frac{5}{14}} & \frac{1}{2}\sqrt{\frac{17}{5}} & \sqrt{\frac{13}{35}} &0  \cr
  0 & \frac{13}{\sqrt{15}} & 4\,\sqrt{\frac{119}{195}} & 9\,\sqrt{\frac{3}{130}} } \oplus
\smat{5\,\sqrt{\frac{5}{3}} & \frac{1}{2}\sqrt{\frac{119}{55}} & \sqrt{\frac{2}{5}} & 0 & 
  \sqrt{\frac{13}{462}} & 0 & 0 \cr 0 & 14\,\sqrt{\frac{6}{55}} & \frac{4}{13} 
  \sqrt{\frac{357}{5}} & \frac{27}{13}\sqrt{\frac{3}{10}} & \frac{10}{13}\sqrt{\frac{17}{143}} &
  \frac{1}{13}\sqrt{\frac{1718}{39}} & 0 \cr 0 & 0 & 3\,\sqrt{\frac{6}{13}} & 0 & 
  \frac{82}{13}\sqrt{\frac{22}{35}} & \frac{1}{26}\sqrt{\frac{102221}{15}} & 0 \cr
  0 & 0 & 0 & 3\,\sqrt{\frac{6}{13}} & 0 & 54\, \sqrt{\frac{6}{4295}} & 13\, 
  \sqrt{\frac{481}{4295}} }  \nonumber  & \\[4pt] 
& \oplus \smat{11\,\sqrt{\frac{7}{15}} & \sqrt{\frac{119}{330}} & 
  2\,\sqrt{\frac{2}{21}} & 
  0 & \frac{1}{2}\sqrt{\frac{65}{231}} & 0 & 0 & 0 \cr 0 & 5\,\sqrt{\frac{15}{11}} & 
  \frac{4}{13}\sqrt{51} & \frac{27}{13}\sqrt{\frac{3}{14}} & \frac{5}{13}\sqrt{\frac{255}{286}}
  & \frac{1}{26}\sqrt{\frac{4295}{13}} & 0 & \frac{1}{2 \sqrt{105}} \cr 0 & 0 & 19\,
  \sqrt{\frac{3}{91}} & 0 & \frac{41}{13}\sqrt{\frac{66}{35}} & \frac{1}{52}
  \sqrt{\frac{102221}{5}} & 0 & \frac{-7}{4}\sqrt{\frac{17}{195}} \cr 0 & 0 & 0 & 
  19\,\sqrt{\frac{3}{91}} & 0 & 81\,\sqrt{\frac{2}{4295}} & 
  \frac{13}{2}\sqrt{\frac{1443}{4295}} & 9 \sqrt{\frac{6}{455}}} \oplus \cdots & 
\end{eqnarray} 
The energy-momentum tensor is shown in Figure~8.

\subsection{Tricritical Ising model: $\V_{1/10}$}

Consider now the Virasoro module $\mathcal{V}_{1/10}$, corresponding to the 
field of dimension $h=h_{1,2}=1/10$. The character is
\begin{equation}
q^{-{1\over 10}}\chi_{1\over 10}(q)=1 + q + q^2 +2\,q^3 + 3\,q^4 + 4\,q^5 + 6\,q^6 + 8\,q^7 + 11\,q^8 + 
14\,q^9 + 19\,q^{10} + 24\,q^{11} + 32\,q^{12} +  \mbox{O}(q^{13})
\end{equation}
The matrices in the $L_1$ basis, correct to order $q^8$, are
\begin{eqnarray} \label{L0_h12}
& L_{0} =  \frac{1}{10}  + 
\mbox{Diagonal} \left( 0,1,2,3,3,4,4,4, \right. & \\[4pt] 
&  \left. 5,5,5,5,6,6,6,6,6,6,7,7,7,7,7,7,7,7,8,8,8,8,8,8,8,8,8,8,8, \ldots \right) &
\nonumber \end{eqnarray}
\begin{eqnarray} \label{L1_h12}
&\qquad L_1\;=\;\mbox{\scriptsize{\mbox{$
\left(\begin{array}{ccccccccccccc}
. & \frac{1}{\sqrt{5}} & . & . & . & . & . & . & . & . & . & . &   \cr 
. & . & 2\, \sqrt{\frac{3}{5}} & . & . & . & . & . & . & . & . & . &  \cr 
. & . & . & \sqrt{\frac{33}{5}} & 0 & . & . & . & . & . & . & . &  \cr 
. & . & . & . & . & \frac{8}{\sqrt{5}} & 0 & 0 & . & . & . & . &  \cr 
. & . & . & . & . & 0 & \sqrt{\frac{31}{5}} & 0 & . & . & . & .&   \cr 
. & . & . & . & . & . & . & . & \sqrt{21} & 0 & 0 & 0 &   \cr 
. & . & . & . & . & . & . & . & 0 & 6\,\sqrt{\frac{2}{5}} & 0 & 0&   \cr 
. & . & . & . & . & . & . & . & 0 & 0 & \sqrt{\frac{41}{5}} & 0&   \cr 
. & . & . & . & . & . & . & . & . & . & . & . &  \cr 
. & . & . & . & . & . & . & . & . & . & . & . &  \cr 
. & . & . & . & . & . & . & . & . & . & . & . &   \cr 
. & . & . & . & . & . & . & . & . & . & . & . &   \cr
 & & & & & & & & & & & &\ddots
\end{array}\right) $}}} &
\end{eqnarray}
\begin{eqnarray}
& L_1= \smat{\frac{1}{\sqrt{5}}} \oplus \smat{2 \sqrt{\frac{3}{5}}} \oplus
\smat{{\sqrt{\frac{33}{5}}} & 0 } \oplus
\smat{ \frac{8}{{\sqrt{5}}} & 0 & 0 \cr 0 & {\sqrt{\frac{31}{5}}} & 0 } \oplus
\smat{ \sqrt{21} & 0 & 0 & 0 \cr 0 & 6\,{\sqrt{\frac{2}{5}}} & 0 & 0 \cr 
   0 & 0 & {\sqrt{\frac{41}{5}}} & 0 } \oplus & \nonumber \\[4pt] &
\smat{2\,{\sqrt{\frac{39}{5}}} & 0 & 0 & 0 & 0 & 0 \cr 
   0 & {\sqrt{\frac{123}{5}}} & 0 & 0 & 0 & 0 \cr 0 & 0 & 2\,
   {\sqrt{\frac{23}{5}}} & 0 & 0 & 0 \cr 0 & 0 & 0 & {\sqrt{\frac{51}{5}}} & 0 & 0 } \oplus
\smat{{\sqrt{\frac{217}{5}}} & 0 & 0 & 0 & 0 & 0 & 0 & 0 \cr 
   0 & 2\,{\sqrt{\frac{46}{5}}} & 0 & 0 & 0 & 0 & 0 & 0 \cr 0 & 0 & 3\,
   {\sqrt{\frac{17}{5}}} & 0 & 0 & 0 & 0 & 0 \cr 0 & 0 & 0 & 4\,{\sqrt{\frac{7}{5}}} 
   & 0 & 0 & 0 & 0 \cr 0 & 0 & 0 & 0 & {\sqrt{\frac{61}{5}}} & 0 & 0 & 0 \cr 
   0 & 0 & 0 & 0 & 0 & {\sqrt{\frac{61}{5}}} & 0 & 0  } & \nonumber \\[4pt] & 
\oplus \smat{ 12\,\sqrt{\frac{2}{5}} & 0 & 0 & 0 & 0 & 0 & 0 & 0 & 0 & 0 & 0 \cr 
   0 & \sqrt{51} & 0 & 0 & 0 & 0 & 0 & 0 & 0 & 0 & 0 \cr
   0 & 0 & 4\,\sqrt{\frac{14}{5}} & 0 & 0 & 0 & 0 & 0 & 0 & 0 & 0 \cr 
   0 & 0 & 0 & \sqrt{\frac{183}{5}} & 0 & 0 & 0 & 0 & 0 & 0 & 0 \cr 
   0 & 0 & 0 & 0 & 2\,\sqrt{\frac{33}{5}} & 0 & 0 & 0 & 0 & 0 & 0 \cr 
   0 & 0 & 0 & 0 & 0 & 2\,\sqrt{\frac{33}{5}} & 0 & 0 & 0 & 0 & 0 \cr 
   0 & 0 & 0 & 0 & 0 & 0 & \sqrt{\frac{71}{5}} & 0 & 0 & 0 & 0 \cr 
   0 & 0 & 0 & 0 & 0 & 0 & 0 & \sqrt{\frac{71}{5}} & 0 & 0 & 0 }
 \oplus \cdots &
\end{eqnarray}
\begin{eqnarray} \label{L2_h12}
& \mbox{}\hspace{-.25in}\mbox{} L_2\;=\;\mbox{\scriptsize{\mbox{$
\left( \begin{array}{ccccccccccccc}
. & . & \frac{{\sqrt{3}}}{2} & . & . & . & . & . & . & . & . & . &  \cr 
. & . & . & \frac{13}{2\,\sqrt{11}} & \sqrt{\frac{10}{11}} & .&.& .&.&. & . & . &  \cr 
. & . & . & . & . & \frac{23}{4}\sqrt{\frac{3}{11}} & 2\,{\sqrt{\frac{30}{341}}} & 
 \frac{1}{4}\sqrt{\frac{65}{31}} & . & . & . & . & \cr 
. & . & . & . & . & . & . & . & \frac{11}{4}\sqrt{\frac{15}{7}} & {\sqrt{\frac{5}{31}}} & 
  \frac{1}{4}\sqrt{\frac{2145}{1271}} & 5\,{\sqrt{\frac{15}{3157}}}&  \cr 
. & . & . & . & . & . & . & . & 0 & \frac{1}{2}\sqrt{\frac{31}{2}} & 0 & 
  \frac{1}{2}\sqrt{\frac{861}{22}}&  \cr 
. & . & . & . & . & . & . & . & . & . & . & . & \cr 
. & . & . & . & . & . & . & . & . & . & . & . & \cr 
. & . & . & . & . & . & . & . & . & . & . & . & \cr 
. & . & . & . & . & . & . & . & . & . & . & . & \cr 
. & . & . & . & . & . & . & . & . & . & . & . & \cr 
. & . & . & . & . & . & . & . & . & . & . & . & \cr 
. & . & . & . & . & . & . & . & . & . & . & . &  \cr
 & & & & & & & & & & & &\ddots 
\end{array} \right) $ }}} & 
\end{eqnarray}
\begin{eqnarray}
&L_2 = \smat{\frac{\sqrt{3}}{2}}\oplus\smat{\frac{13}{2\,\sqrt{11}} & \sqrt{\frac{10}{11}}} \oplus 
\smat{\frac{23}{4}\sqrt{\frac{3}{11}} & 2\,\sqrt{\frac{30}{341}} & 
  \frac{1}{4}\sqrt{\frac{65}{31}} } \oplus 
\smat{\frac{11}{4}\sqrt{\frac{15}{7}} & \sqrt{\frac{5}{31}} & 
  \frac{1}{4}\sqrt{\frac{2145}{1271}} & 5\,{\sqrt{\frac{15}{3157}}} \cr 0 & 
  \frac{1}{2}\sqrt{\frac{31}{2}} & 0 & \frac{1}{2}\sqrt{\frac{861}{22}} } \oplus \nonumber &
\\[4pt] &
\smat{\frac{43}{2}\sqrt{\frac{5}{91}} & 8\,{\sqrt{\frac{5}{3813}}} & \sqrt{\frac{2145}{29233}} 
  & 40\,{\sqrt{\frac{5}{53669}}} & \frac{5}{4}\sqrt{\frac{274355}{6877299}} & 0 \cr 0 & 
  \frac{103}{2\,{\sqrt{246}}} & 0 & \frac{1}{2}\sqrt{\frac{8897}{374}} &
  -3610\,\sqrt{\frac{598}{39124724001}} & 3\,{\sqrt{\frac{32890}{1701001}}} \cr 0 & 0 & 
  \frac{3}{2}\sqrt{\frac{41}{23}} & 0 & \frac{2067}{4}\sqrt{\frac{697}{39123023}} & 
  200\,{\sqrt{\frac{345}{1701001}}} } \oplus & \\[4pt] &
\smat{\frac{53}{2}\sqrt{\frac{21}{403}} & 10\,{\sqrt{\frac{14}{29233}}} & 
  5\,{\sqrt{\frac{1001}{496961}}} & 50\,{\sqrt{\frac{3}{53669}}} & 
  \frac{25}{4}\sqrt{\frac{384097}{139838413}} & 0 & \frac{125}{2}\sqrt{\frac{239}{56932337}} 
  & 0 \cr 0 & 113\,{\sqrt{\frac{3}{1886}}} & 0 & \frac{3}{4}\sqrt{\frac{1271}{187}} &
  -14440\,{\sqrt{\frac{897}{795536054687}}} & 36\,{\sqrt{\frac{16445}{103761061}}} & 
  \frac{245}{2}\,{\sqrt{\frac{51}{3688487}}} & \frac{13}{4}\sqrt{\frac{165}{451949}} \cr 0 & 
  0 & \frac{133}{2\,{\sqrt{391}}} & 0 & \frac{84747}{4}\sqrt{\frac{17}{2386504403}} & 
  200\,\sqrt{\frac{14145}{103761061}} & \frac{-95}{2}\sqrt{\frac{299}{10161563}} & 
  486\,\sqrt{\frac{65}{18529909}} \cr 0 & 0 & 0 & \frac{3}{4}\sqrt{\frac{51}{7}} & 0 & 0 & 
  \frac{3}{2}\sqrt{\frac{739381}{68593}} & \frac{13}{4}\sqrt{\frac{2135}{9799}} } & 
\nonumber
\end{eqnarray}
The energy momentum tensor is shown in Figure~9.

Taking the direct sum of the two previous Virasoro modules, 
we can extract the field 
$\phi^{(1/10)}(z)$ whose block structure is shown in (\ref{directsum}). 
We present here the lower-left $10\times 10$ block of the field:
\begin{eqnarray}
&  \tpf{1/10}{1/10}{0} (z) = & \\[8pt] & \mbox{}\hspace{-5 mm}\mbox{}
\mbox{\scriptsize{\mbox{$
\left( \begin{array}{ccccccccccc}
1 & \frac{1}{\sqrt{35} z^2} & \frac{1}{{\sqrt{35}}\,z^3} & \frac{3}{5\,{\sqrt{14}}\,z^4} & 
   \frac{1}{5\,z^4} \sqrt{\frac{3}{238}} & \frac{2}{5\,{\sqrt{7}}\,z^5} & 
   \frac{1}{5\,z^5} \sqrt{\frac{3}{119}} & \frac{1}{7\,z^6} & \frac{1}{{\sqrt{714}}\,z^6} & 
   \frac{2}{35\,{\sqrt{195}}\,z^6} \\[6pt] 
\frac{z}{\sqrt{5}} & \frac{-9}{5\,{\sqrt{7}}\,z} & \frac{-4}{5\,{\sqrt{7}}\,z^2} & 
   \frac{-1}{5\,z^3}\sqrt{\frac{7}{10}} & \frac{-19}{5\,z^3}\sqrt{\frac{3}{1190}}
   & \frac{-3}{5\,{\sqrt{35}}\,z^4} & \frac{-2}{5\,z^4}\sqrt{\frac{21}{85}} & 
   \frac{-1}{7\,{\sqrt{5}}\,z^5} & \frac{-11}{{\sqrt{3570}}\,z^5} & 
   \frac{-58}{175\,{\sqrt{39}}\,z^5} \\[6pt] 
\frac{{\sqrt{3}}\,z^2}{5} & \frac{6}{5}\sqrt{\frac{3}{35}} & \frac{-9}{5\,z}\sqrt{\frac{3}{35}}
   & \frac{-17}{25\,z^2}\sqrt{\frac{3}{14}} & \frac{19}{25\,z^2} \sqrt{\frac{7}{34}} & 
   \frac{-8}{25\,z^3}  \sqrt{\frac{3}{7}} & \frac{38}{25\,{\sqrt{119}}\,z^3} & 
   \frac{-3\,{\sqrt{3}}}{35\,z^4} & \frac{1}{5\,z^4}\sqrt{\frac{2}{119}} & 
   \frac{232}{175\,{\sqrt{65}}\,z^4} \\[6pt] 
\frac{z^3}{5}\sqrt{\frac{11}{5}} & \frac{6\,z}{25\,{\sqrt{77}}} & \frac{96}{25\,{\sqrt{77}}} &
   \frac{-27}{25\,z}  \sqrt{\frac{11}{70}} & \frac{-57}{25\,z}\sqrt{\frac{21}{1870}} & 
   \frac{-13}{25\,z^2}   \sqrt{\frac{11}{35}} & \frac{38}{25\,z^2}\sqrt{\frac{21}{935}} & 
   \frac{-1}{7\,z^3} \sqrt{\frac{11}{5}} & \frac{19}{z^3}\sqrt{\frac{2}{19635}} & 
   \frac{-4408}{875\,{\sqrt{429}}\,z^3}  \\[6pt] 
0 & 2\,{\sqrt{\frac{10}{77}}}\,z & -{\sqrt{\frac{10}{77}}} & 0 & 
   \frac{-12}{z}\sqrt{\frac{3}{1309}} & 0 & \frac{-3}{z^2}\sqrt{\frac{6}{1309}} & 0 & 
   \frac{-2}{z^3}\sqrt{\frac{3}{1309}} & \frac{-1}{z^3} \sqrt{\frac{6}{715}}  \\[6pt] 
\frac{2 \,{\sqrt{11}}\,z^4}{25} & \frac{9\,z^2}{50\,{\sqrt{385}}} & 
   \frac{39\,z}{50\,{\sqrt{385}}} & \frac{1497}{250\,{\sqrt{154}}} & 
   \frac{57}{250}\sqrt{\frac{21}{374}} & \frac{-36}{125\,z} \sqrt{\frac{11}{7}} & 
   \frac{-114 }{125\,z}\sqrt{\frac{21}{187}} & \frac{-2\,{\sqrt{11}}}{25\,z^2} & 
   \frac{19}{25\,z^2}\sqrt{\frac{14}{561}} & \frac{1102}{125\,{\sqrt{2145}}\,z^2} \\[6pt]
0 & 12\,{\sqrt{\frac{2}{2387}}}\, z^2 & 19\,{\sqrt{\frac{2}{2387}}}\,z & 
   -10\,{\sqrt{\frac{5}{2387}}} & 48\,{\sqrt{\frac{3}{202895}}} & 0 & 
   \frac{-3}{z}\sqrt{\frac{186}{6545}} & 0 & \frac{-2}{z^2}\sqrt{\frac{93}{6545}} & 
   \frac{14 }{5\,z^2}\sqrt{\frac{6}{4433}} \\[6pt] 
0 &  \frac{z^2}{2} \sqrt{\frac{39}{217}} & \frac{-z}{2}\sqrt{\frac{39}{217}} & 
   \frac{1 }{2}\sqrt{\frac{39}{2170}} & \frac{-19}{2}\sqrt{\frac{13}{36890}} & 0 & 0 & 0 & 0 
   & \frac{-6}{5\,{\sqrt{31}}\,z^2} \\[6pt] 
\frac{2\,{\sqrt{231}}\,z^5}{125} & \frac{3\,z^3}{1750}\sqrt{\frac{33}{5}} & 
   \frac{54\,z^2}{875}\sqrt{\frac{3}{55}} & \frac{1149\,z}{8750}\sqrt{\frac{3}{22}} & 
   \frac{57\,z}{1250\,{\sqrt{374}}} & \frac{8541}{8750}\sqrt{\frac{3}{11}} & 
   \frac{741}{625\,{\sqrt{187}}} & \frac{-18 }{125\,z}\sqrt{\frac{33}{7}} & 
   \frac{-399}{125\,z}\sqrt{\frac{2}{187}} & \frac{-3306}{625\,{\sqrt{5005}}\,z} \\[6pt] 
0 & 2\,{\sqrt{\frac{11}{1085}}}\,z^3 & \frac{39\,z^2}{{\sqrt{11935}}} & 
   15\,{\sqrt{\frac{2}{2387}}}\,z & \frac{4\,z}{5}\sqrt{\frac{6}{40579}} & 
   \frac{-25}{{\sqrt{2387}}} & \frac{142}{5} \sqrt{\frac{3}{40579}} & 0 & 
   \frac{-6}{5\,z}\sqrt{\frac{186}{1309}} & \frac{-21}{5\,z}  \sqrt{\frac{3}{22165}}
\end{array} \right) $}}} & \nonumber 
\end{eqnarray}

\subsection{Tricritical Ising model: $\V_0 \oplus \V_{3/5}$}
Lastly, we present the mode $ \tpf{3/5}{3/5}{0}_{-3/5}$ of the field $\tpf{3/5}{3/5}{0} (z) $
\begin{eqnarray}
&\tpf{3/5}{3/5}{0}_{-3/5} = \smat{1} \oplus \smat{\frac{-4}{5} \sqrt{\frac{3}{385}} \\[5pt] 
\sqrt{\frac{65}{77}} }
\oplus \smat{\frac{-3}{25} {\sqrt{\frac{7}{11}}} \\[5pt] 5\,{\sqrt{\frac{2}{77}}} }
\oplus \smat{
\frac{-263}{875} \sqrt{\frac{3}{22}} & \frac{-12}{125}\sqrt{\frac{2}{187}} \\[5pt] 
25\,{\sqrt{\frac{2}{2387}}} & \frac{-4}{5} \sqrt{\frac{6}{40579}} \\[5pt] 
\frac{3}{7}\sqrt{\frac{183}{310}} & -148\, \sqrt{\frac{2}{160735}} \\[5pt]
0 & 5\,{\sqrt{\frac{299}{7259}}} } \oplus \smat{
\frac{-5177}{4375} \sqrt{\frac{3}{286}} & \frac{-372}{625} {\sqrt{\frac{2}{2431}}}\\[5pt]
25\,{\sqrt{\frac{5}{7161}}} & \frac{-2}{5}  {\sqrt{\frac{17}{11935}}} \\[5pt] 
\frac{9}{7}  {\sqrt{\frac{183}{1426}}} & -296\,{\sqrt{\frac{2}{739381}}} \\[5pt] 
0 & 10\,{\sqrt{\frac{65}{7259}}} \\[5pt]  
\frac{-2}{{\sqrt{897}}} & {\sqrt{\frac{17}{299}}} }
&  \nonumber \\[5pt]
& \oplus \smat{
\frac{-903 }{125} {\sqrt{\frac{7}{22165}}} & \frac{-736}{125} \sqrt{\frac{6}{376805}} & \frac{-1786}{40625}  {\sqrt{\frac{3}{2387}}} & \frac{-1064}{40625} {\sqrt{\frac{6}{5797}}} \\[5pt] 
\frac{125}{{\sqrt{41943}}} & \frac{-63}{25}  {\sqrt{\frac{14}{237677}}} & \frac{-42}
   {25\,{\sqrt{908765}}} & \frac{48}{25}  {\sqrt{\frac{14}{15449005}}} \\[5pt] 
9\,{\sqrt{\frac{305}{84847}}} & \frac{-296}{17} {\sqrt{\frac{30}{43493}}} & 
\frac{-6}{25}  {\sqrt{\frac{183}{1103011}}} & \frac{3304}{1275} {\sqrt{\frac{2}{1696227}}} \\[5pt] 
0 & \frac{50}{17} {\sqrt{\frac{39}{427}}} & 0 & \frac{-52}{51\,{\sqrt{6405}}} \\[5pt] 
-{\sqrt{\frac{10}{897}}} & \frac{3}{2}  {\sqrt{\frac{85}{2093}}} & 
\frac{-6}{65} {\sqrt{\frac{2}{23}}} & \frac{36}{65}{\sqrt{\frac{7}{391}}} \\[5pt] 
\frac{1}{7} {\sqrt{\frac{34698}{108035}}} & \frac{33001}{34\,{\sqrt{257256755}}} & 
\frac{-592642}{35}  {\sqrt{\frac{2}{1624392653}}} & 
\frac{-80468}{85} {\sqrt{\frac{7}{95552509}}} \\[5pt] 
0 & \frac{8}{17} {\sqrt{\frac{1495}{121443}}} & 4\,{\sqrt{\frac{782}{17349}}} & 
\frac{407}{17}  {\sqrt{\frac{23}{121443}}} }  
& \nonumber \\[5pt]
& \oplus \smat{\frac{-61541}{4375} {\sqrt{\frac{2}{22165}}} & \frac{-29684}{625\,{\sqrt{2637635}}} & \frac{-36613}{284375\,{\sqrt{1705}}} & \frac{-3116}{40625} {\sqrt{\frac{14}{28985}}} \\[5pt] 
250\,{\sqrt{\frac{5}{964689}}} & \frac{-16109}{125} {\sqrt{\frac{6}{191329985}}} & 
\frac{-784}{125}  {\sqrt{\frac{6}{20901595}}} & \frac{1792}{125} {\sqrt{\frac{21}{355327115}}} \\[5pt] 
\frac{15}{7}  {\sqrt{\frac{915}{12121}}} & \frac{-740}{17} \sqrt{\frac{30}{304451}} & \frac{-81}{175}   {\sqrt{\frac{183}{1575730}}} & \frac{708}{425} {\sqrt{\frac{21}{2827045}}} \\[5pt] 
0 & \frac{125}{119}  {\sqrt{\frac{39}{61}}} & 0 & \frac{-39}{595} {\sqrt{\frac{6}{61}}} \\[5pt] 
-10\,{\sqrt{\frac{10}{54717}}} & \frac{15}{2} {\sqrt{\frac{255}{127673}}} & 
\frac{-116}{65}  {\sqrt{\frac{3}{7015}}} & \frac{348}{65} {\sqrt{\frac{42}{119255}}} \\[5pt] 
\frac{2}{7} {\sqrt{\frac{57830}{237677}}} & \frac{33001}{34} {\sqrt{\frac{5}{565964861}}} & \frac{-1185284}{7\,{\sqrt{89341595915}}} & \frac{-80468}{17}  {\sqrt{\frac{14}{5255387995}}} \\[5pt] 
0 & \frac{40 }{17}{\sqrt{\frac{1495}{1335873}}} & 8\,{\sqrt{\frac{1955}{190839}}} & 
\frac{37}{17} {\sqrt{\frac{2530}{121443}}} \\[5pt] 
\frac{-9}{7} {\sqrt{\frac{5}{15433}}} & \frac{64}{9} {\sqrt{\frac{34}{540155}}} & 
\frac{821}{63} {\sqrt{\frac{2}{1003145}}} & \frac{-8 }{3}{\sqrt{\frac{7}{17053465}}} \\[5pt] 
0 & \frac{1}{63} {\sqrt{\frac{170}{11}}} & \frac{-4 }{9}{\sqrt{\frac{70}{143}}} & 
\frac{-1}{21} {\sqrt{\frac{55}{221}}}} 
\oplus \ldots & 
\end{eqnarray}
\clearpage

\begin{figure}[p]
\mbox{}\hspace{1.0in}\mbox{}
\rotatebox{90}{
$  T(z)= \smat{ 
0 & \frac{1}{2\,z^4} {\sqrt{\frac{7}{5}}} & \frac{1}{z^5} {\sqrt{\frac{7}{5}}} & 
  \frac{1}{z^6} {\sqrt{\frac{7}{2}}} & 0 & \frac{{\sqrt{7}}}{z^7} & 0 & \frac{7}{2\,z^8} & 
  0 & 0 & 0 & \frac{7}{z^9} {\sqrt{\frac{2}{5}}} & 0 & 0 & 0 &  \\[5pt] 
\frac{1}{2}  {\sqrt{\frac{7}{5}}} & \frac{2}{z^2} & \frac{2}{z^3} & 
    \frac{3}{z^4} {\sqrt{\frac{2}{5}}} & \frac{1}{z^4} {\sqrt{\frac{51}{10}}} & 
    \frac{4}{{\sqrt{5}}\,z^5} & \frac{1}{z^5}  {\sqrt{\frac{51}{5}}} & 
    \frac{2}{z^6}     {\sqrt{\frac{5}{7}}} & \frac{1}{z^6} {\sqrt{\frac{85}{6}}} & 
    \frac{2}{z^6} {\sqrt{\frac{13}{21}}} & 0 & \frac{3}{z^7} {\sqrt{\frac{2}{7}}} & 
    \frac{{\sqrt{17}}}{z^7} & \frac{2}{z^7} {\sqrt{\frac{13}{7}}} & 0 & \\[5pt]
{\sqrt{\frac{7}{5}}}\,z & \frac{2}{z} & \frac{3}{z^2} & \frac{{\sqrt{10}}}
   {z^3} & 0 & \frac{7}{{\sqrt{5}}\,z^4} & \frac{1}{2\,z^4} {\sqrt{\frac{51}{5}}} & 
   \frac{18}{{\sqrt{35}}\,z^5} & \frac{2}{z^5} {\sqrt{\frac{34}{15}}} & 
   \frac{-2}{z^5} {\sqrt{\frac{13}{21}}} & 0 & \frac{11}{{\sqrt{14}}\,z^6} & 
   \frac{{\sqrt{17}}}{z^6} & - \frac{1}{z^6}  {\sqrt{\frac{13}{7}}} & 0 & \\[5pt] 
{\sqrt{\frac{7}{2}}}\,z^2 & 3\,{\sqrt{\frac{2}{5}}} & \frac{{\sqrt{10}}}{z} & 
   \frac{4}{z^2} & 0 & \frac{3\,{\sqrt{2}}}{z^3} & 0 & \frac{8}{z^4} {\sqrt{\frac{2}{7}}} & 
   \frac{1}{2\,z^4} {\sqrt{\frac{17}{3}}} & \frac{1}{z^4} {\sqrt{\frac{26}{105}}} & 0 & 
   \frac{5}{z^5}  {\sqrt{\frac{5}{7}}} & \frac{1}{z^5} {\sqrt{\frac{34}{5}}} & - 
   \frac{1}{z^5}  {\sqrt{\frac{26}{35}}}   & 0 & \\[5pt] 
0 & {\sqrt{\frac{51}{10}}} & 0 & 0 & \frac{4}{z^2} & 0 & \frac{2\,{\sqrt{2}}}{z^3} & 0 & 
   \frac{2}{z^4} & \frac{2}{z^4} {\sqrt{\frac{238}{65}}} & \frac{27}{2\,{\sqrt{65}}\,z^4} &
   0 & \frac{4}{z^5} {\sqrt{\frac{2}{15}}} & \frac{4}{z^5} {\sqrt{\frac{238}{195}}} & 
   \frac{9 }{z^5}{\sqrt{\frac{3}{65}}} & \\[5pt] 
{\sqrt{7}}\,z^3 & \frac{4\,z}{{\sqrt{5}}} & \frac{7}{{\sqrt{5}}} & 
   \frac{3}{z} {\sqrt{2}} & 0 & \frac{5}{z^2} & 0 & \frac{2}{z^3} {\sqrt{7}} & 0 & 0 & 0 &
   \frac{9}{z^4}   {\sqrt{\frac{5}{14}}} & \frac{1}{2\,z^4} {\sqrt{\frac{17}{5}}} & 
   \frac{1}{z^4} {\sqrt{\frac{13}{35}}} & 0 & \\[5pt] 
0 & {\sqrt{\frac{51}{5}}}\,z & \frac{1}{2} {\sqrt{\frac{51}{5}}} & 0 & 
   \frac{2\,{\sqrt{2}}}{z} & 0 & \frac{5}{z^2} & 0 & \frac{3\,{\sqrt{2}}}{z^3} & 0 & 0 & 0 &   \frac{13} {{\sqrt{15}}\,z^4} & \frac{4}{z^4} {\sqrt{\frac{119}{195}}} & 
   \frac{9}{z^4} {\sqrt{\frac{3}{130}}} & \\[5pt] 
\frac{7\,z^4}{2} & 2\,{\sqrt{\frac{5}{7}}}\,
   z^2 & \frac{18\,z}{{\sqrt{35}}} & 8\,{\sqrt{\frac{2}{7}}} & 0 & \frac{2\,{\sqrt{7}}}{z} & 0 & \frac{6}{z^2} & 0 & 0 & 0 & \frac{2\,{\sqrt{10}}}
   {z^3} & 0 & 0 & 0 & \\[5pt] 0 & {\sqrt{\frac{85}{6}}}\,z^2 & 2\,{\sqrt{\frac{34}{15}}}\,z & \frac{1}{2} {\sqrt{\frac{17}{3}}} & 2 & 0 & \frac{3\,{\sqrt{2}}}
   {z} & 0 & \frac{6}{z^2} & 0 & 0 & 0 & \frac{{\sqrt{30}}}{z^3} & 0 & 0 & \\[5pt] 0 & 2\,{\sqrt{\frac{13}{21}}}\,z^2 & -2\,{\sqrt{\frac{13}{21}}}\,z & {\sqrt{
       \frac{26}{105}}} & 2\,{\sqrt{\frac{238}{65}}} & 0 & 0 & 0 & 0 & \frac{6}{z^2} & 0 & 0 & 0 & \frac{2\,{\sqrt{3}}}{z^3} & 0 & \\[5pt] 0 & 0 & 0 & 0 & \frac{27}
   {2\,{\sqrt{65}}} & 0 & 0 & 0 & 0 & 0 & \frac{6}{z^2} & 0 & 0 & 0 & \frac{2\,{\sqrt{3}}}{z^3} & \\[5pt] 7\,{\sqrt{\frac{2}{5}}}\,z^5 & 3\,{\sqrt{\frac{2}{7}}}\,
   z^3 & \frac{11\,z^2}{{\sqrt{14}}} & 5\,{\sqrt{\frac{5}{7}}}\,z & 0 & 9\,{\sqrt{\frac{5}{14}}} & 0 & \frac{2\,{\sqrt{10}}}{z} & 0 & 0 & 0 & \frac{7}
   {z^2} & 0 & 0 & 0 & \\[5pt] 0 & {\sqrt{17}}\,z^3 & {\sqrt{17}}\,z^2 & {\sqrt{\frac{34}{5}}}\,z & 4\,{\sqrt{\frac{2}{15}}}\,z & \frac{1}{2} {\sqrt{\frac{17}{5}}}
    & \frac{13}{{\sqrt{15}}} & 0 & \frac{{\sqrt{30}}}{z} & 0 & 0 & 0 & \frac{7}{z^2} & 0 & 0 & \\[5pt] 0 & 2\,{\sqrt{\frac{13}{7}}}\,z^3 & - 
     {\sqrt{\frac{13}{7}}}\,z^2   & - {\sqrt{\frac{26}{35}}}\,z   & 4\,{\sqrt{\frac{238}{195}}}\,z & {\sqrt{\frac{13}{35}}} & 4\,
   {\sqrt{\frac{119}{195}}} & 0 & 0 & \frac{2\,{\sqrt{3}}}{z} & 0 & 0 & 0 & \frac{7}{z^2} & 0 & \\[5pt] 
0 & 0 & 0 & 0 & 9\,{\sqrt{\frac{3}{65}}}\,z & 0 & 9\,
   {\sqrt{\frac{3}{130}}} & 0 & 0 & 0 & \frac{2\,{\sqrt{3}}}{z} & 0 & 0 & 0 & \frac{7}{z^2} 
 & \\[5pt]
  & &  & & & & & & & & & & & & & \ddots }
$   
}
\mbox{}\hspace{.4in}\mbox{}
\rotatebox{90}{\parbox{9.3in}{Figure~8: Energy-momentum tensor (in the $L_1$ basis) of the tricritical Ising model in the vacuum $h=0$ sector. Notice that the squares of the 
coefficients in the first column give the expansion of $(1-u)^{-4}$ in agreement with the two-point function (\ref{Ttaylor}).}}
\thispagestyle{empty}
\end{figure}

\clearpage

\begin{figure}[p]
\mbox{}\hspace{1.2in}\mbox{}
\rotatebox{90}{
$T(z)= \smat{
\frac{1}{10\,z^2} & \frac{1}{{\sqrt{5}}\,z^3} & \frac{1}{2\,z^4} {\sqrt{3}} & 
   \frac{2}{z^5} {\sqrt{\frac{5}{11}}} & - \frac{1}{z^5} {\sqrt{\frac{2}{11}}}  & 
   \frac{25}{4\,{\sqrt{11}}\,z^6} & \frac{-5}{z^6} {\sqrt{\frac{10}{341}}} & 
   \frac{1}{4\,z^6} {\sqrt{\frac{39}{155}}} & \frac{25}{2\,z^7} {\sqrt{\frac{3}{77}}} & 
   \frac{-25}{{\sqrt{341}}\,z^7} & \frac{3 }{2\,z^7}{\sqrt{\frac{39}{1271}}} & 
   - \frac{1}{z^7} {\sqrt{\frac{3}{287}}}  & \\[6pt] 
\frac{1}{{\sqrt{5}}\,z} & \frac{11}{10\,z^2} & \frac{2}{z^3} {\sqrt{\frac{3}{5}}} & 
   \frac{13}{2\,{\sqrt{11}}\,z^4} & \frac{1}{z^4} {\sqrt{\frac{10}{11}}} & 
   \frac{7}{2\,z^5}  {\sqrt{\frac{5}{11}}} & \frac{19}{z^5} {\sqrt{\frac{2}{341}}} & 
   \frac{-1}{2\,z^5} {\sqrt{\frac{39}{31}}} & \frac{25}{4\,z^6} {\sqrt{\frac{15}{77}}} & 
   \frac{15}{z^6} {\sqrt{\frac{5}{341}}} & \frac{-49}{4\,z^6} {\sqrt{\frac{39}{6355}}} & 
   \frac{3}{z^6} {\sqrt{\frac{15}{287}}} & \\[6pt] 
\frac{{\sqrt{3}}}{2} & \frac{2}{z} {\sqrt{\frac{3}{5}}} & \frac{21}{10\,z^2} & 
   \frac{1 }{z^3}{\sqrt{\frac{33}{5}}} & 0 & \frac{23}{4\,z^4} {\sqrt{\frac{3}{11}}} & 
   \frac{2}{z^4}{\sqrt{\frac{30}{341}}} & \frac{1}{4\,z^4} {\sqrt{\frac{65}{31}}} & 
   \frac{30}{{\sqrt{77}}\,z^5} & \frac{13}{z^5} {\sqrt{\frac{3}{341}}} & 
   \frac{2}{z^5} {\sqrt{\frac{13}{1271}}} & \frac{-15}{{\sqrt{287}}\,z^5} & \\[6pt] 
2\,{\sqrt{\frac{5}{11}}}\,z & \frac{13}{2\,{\sqrt{11}}} & \frac{1}{z} {\sqrt{\frac{33}{5}}}
   & \frac{31}{10\,z^2} & 0 & \frac{8}{{\sqrt{5}}\,z^3} & 0 & 0 & 
   \frac{11}{4\,z^4} {\sqrt{\frac{15}{7}}} & \frac{1}{z^4} {\sqrt{\frac{5}{31}}} & 
   \frac{1}{4\,z^4} {\sqrt{\frac{2145}{1271}}} & \frac{5}{z^4} {\sqrt{\frac{15}{3157}}} & \\[6pt] 
-{\sqrt{\frac{2}{11}}}\,z  & {\sqrt{\frac{10}{11}}} & 0 & 0 & \frac{31}{10\,z^2} & 0 & 
   \frac{1}{z^3} {\sqrt{\frac{31}{5}}} & 0 & 0 & \frac{1}{2\,z^4} {\sqrt{\frac{31}{2}}} & 
   0 & \frac{1}{2\,z^4} {\sqrt{\frac{861}{22}}} & \\[6pt] 
\frac{25\,z^2}{4\,{\sqrt{11}}} & \frac{7\,z}{2} {\sqrt{\frac{5}{11}}} & 
   \frac{23}{4} {\sqrt{\frac{3}{11}}} & \frac{8}{{\sqrt{5}}\,z} & 0 & \frac{41}{10\,z^2} & 
   0 & 0 & \frac{{\sqrt{21}}}{z^3} & 0 & 0 & 0 & \\[6pt] 
-5\,{\sqrt{\frac{10}{341}}}\,z^2 & 19\,{\sqrt{\frac{2}{341}}}\,z & 
   2\,{\sqrt{\frac{30}{341}}} & 0 & \frac{1}{z} {\sqrt{\frac{31}{5}}} & 0 & 
   \frac{41}{10\,z^2} & 0 & 0 & \frac{6}{z^3} {\sqrt{\frac{2}{5}}} & 0 & 0 & \\[6pt] 
\frac{1 \,z^2}{4} {\sqrt{\frac{39}{155}}} & \frac{1\,z }{2}  {\sqrt{\frac{39}{31}}} & 
   \frac{1}{4} {\sqrt{\frac{65}{31}}} & 0 & 0 & 0 & 0 & \frac{41}{10\,z^2} & 0 & 0 & 
   \frac{1}{z^3} {\sqrt{\frac{41}{5}}} & 0 & \\[6pt] 
\frac{25\,z^3}{2} {\sqrt{\frac{3}{77}}} & \frac{25 \,z^2}{4}{\sqrt{\frac{15}{77}}} & 
   \frac{30\,z}{{\sqrt{77}}} & \frac{11 }{4}{\sqrt{\frac{15}{7}}} & 0 & 
   \frac{{\sqrt{21}}}{z} & 0 & 0 & \frac{51}{10\,z^2} & 0 & 0 & 0 & \\[6pt] 
\frac{-25\,z^3}{{\sqrt{341}}} & 15\,{\sqrt{\frac{5}{341}}}\,z^2 & 13\,{\sqrt{\frac{3}{341}}}\,z & {\sqrt{\frac{5}{31}}} & \frac{1}{2} {\sqrt{\frac{31}{2}}} & 0 & 
   \frac{6}{z}  {\sqrt{\frac{2}{5}}} & 0 & 0 & \frac{51}{10\,z^2} & 0 & 0 & \\[6pt] 
\frac{3\,z^3}{2} {\sqrt{\frac{39}{1271}}} & \frac{-49 \,z^2}{4}{\sqrt{\frac{39}{6355}}} & 
   2\,{\sqrt{\frac{13}{1271}}}\,z & \frac{1}{4} {\sqrt{\frac{2145}{1271}}} & 0 & 0 & 0 & 
   \frac{1 }{z}{\sqrt{\frac{41}{5}}} & 0 & 0 & \frac{51}{10\,z^2} & 0 & \\[6pt] 
- {\sqrt{\frac{3}{287}}}\,z^3   & 3\,{\sqrt{\frac{15}{287}}}\,z^2 & \frac{-15\,z}{{\sqrt{287}}} & 5\,
   {\sqrt{\frac{15}{3157}}} & \frac{1}{2} {\sqrt{\frac{861}{22}}} & 0 & 0 & 0 & 0 & 0 & 0 & \frac{51}{10\,z^2} & \\[6pt] 
  & & & & & & & & & & & & \ddots }
$
}
\mbox{}\hspace{.4in}\mbox{}
\rotatebox{90}{Figure~9: Energy-momentum tensor (in the $L_1$ basis) of the tricritical Ising model in the $h=1/10$ sector.}
\thispagestyle{empty}
\end{figure}
\clearpage

\setlength{\arraycolsep}{1mm}
\begin{figure}[p]
\mbox{}\hspace{.7in}\mbox{}
\rotatebox{90}{
$\mbox{}\hspace{-2mm} T(z)= \smat{\frac{3}{5\,z^2} & \frac{{\sqrt{\frac{6}{5}}}}{z^3} & \frac{3\,{\sqrt{\frac{3}{11}}}}{z^4} & \frac{{\sqrt{\frac{13}{11}}}}{2\,z^4} & \frac{3\,
     {\sqrt{\frac{5}{11}}}}{z^5} & \frac{{\sqrt{\frac{10}{11}}}}{z^5} & \frac{25\,{\sqrt{\frac{3}{77}}}}{2\,z^6} & \frac{25}{{\sqrt{341}}\,z^6} & \frac{3\,
     {\sqrt{\frac{183}{1085}}}}{2\,z^6} & 0 & \frac{75\,{\sqrt{\frac{3}{2002}}}}{z^7} & \frac{25\,{\sqrt{\frac{5}{1023}}}}{z^7} & \frac{9\,
     {\sqrt{\frac{183}{9982}}}}{z^7} & 0 & \frac{-2\,{\sqrt{\frac{7}{897}}}}{z^7} & \\[6pt] \frac{{\sqrt{\frac{6}{5}}}}{z} & \frac{8}{5\,z^2} & \frac{{\sqrt{\frac{22}
        {5}}}}{z^3} & 0 & \frac{7\,{\sqrt{\frac{3}{22}}}}{z^4} & \frac{{\sqrt{\frac{3}{11}}}}{2\,z^4} & \frac{17\,{\sqrt{\frac{5}{154}}}}{z^5} & \frac{2\,
     {\sqrt{\frac{30}{341}}}}{z^5} & \frac{-3\,{\sqrt{\frac{61}{434}}}}{z^5} & 0 & \frac{50\,{\sqrt{\frac{5}{1001}}}}{z^6} & \frac{25}
   {{\sqrt{682}}\,z^6} & \frac{-33\,{\sqrt{\frac{61}{24955}}}}{z^6} & 0 & \frac{{\sqrt{\frac{70}{299}}}}{z^6} & \\[6pt] 3\,{\sqrt{\frac{3}{11}}} & \frac{{\sqrt{
         \frac{22}{5}}}}{z} & \frac{13}{5\,z^2} & 0 & \frac{4\,{\sqrt{\frac{3}{5}}}}{z^3} & 0 & \frac{19}{2\,{\sqrt{7}}\,z^4} & \frac{{\sqrt{\frac{3}{31}}}}
   {2\,z^4} & \frac{3\,{\sqrt{\frac{305}{2387}}}}{2\,z^4} & 0 & \frac{55}{{\sqrt{182}}\,z^5} & \frac{{\sqrt{\frac{5}{31}}}}{z^5} & \frac{-3\,
     {\sqrt{\frac{61}{109802}}}}{z^5} & 0 & \frac{-10\,{\sqrt{\frac{7}{3289}}}}{z^5} & \\[6pt] \frac{{\sqrt{\frac{13}{11}}}}{2} & 0 & 0 & \frac{13}
   {5\,z^2} & 0 & \frac{{\sqrt{\frac{26}{5}}}}{z^3} & 0 & \frac{3\,{\sqrt{\frac{13}{31}}}}{z^4} & \frac{4\,{\sqrt{\frac{1365}{20801}}}}{z^4} & \frac{{\sqrt
        {\frac{759}{122}}}}{z^4} & 0 & \frac{2\,{\sqrt{\frac{65}{93}}}}{z^5} & \frac{40\,{\sqrt{\frac{546}{478423}}}}{z^5} & \frac{2\,{\sqrt{\frac{165}{61}}}}
   {z^5} & \frac{13\,{\sqrt{\frac{7}{759}}}}{z^5} & \\[6pt] 3\,{\sqrt{\frac{5}{11}}}\,z & 7\,{\sqrt{\frac{3}{22}}} & \frac{4\,{\sqrt{\frac{3}{5}}}}{z} & 0 & \frac{18}
   {5\,z^2} & 0 & \frac{2\,{\sqrt{\frac{21}{5}}}}{z^3} & 0 & 0 & 0 & \frac{8\,{\sqrt{\frac{30}{91}}}}{z^4} & \frac{1}{{\sqrt{93}}\,z^4} & \frac{3\,
     {\sqrt{\frac{1830}{54901}}}}{z^4} & 0 & \frac{5\,{\sqrt{\frac{35}{9867}}}}{2\,z^4} & \\[6pt] {\sqrt{\frac{10}{11}}}\,z & \frac{{\sqrt{\frac{3}{11}}}}
   {2} & 0 & \frac{{\sqrt{\frac{26}{5}}}}{z} & 0 & \frac{18}{5\,z^2} & 0 & \frac{{\sqrt{\frac{62}{5}}}}{z^3} & 0 & 0 & 0 & \frac{22\,{\sqrt{\frac{2}{93}}}}
   {z^4} & \frac{52\,{\sqrt{\frac{105}{478423}}}}{z^4} & \frac{{\sqrt{\frac{429}{122}}}}{z^4} & - \frac{{\sqrt{\frac{455}{1518}}}}{z^4}   & \\[6pt] 
    \frac{25\,{\sqrt{\frac{3}{77}}}\,z^2}{2} & 17\,{\sqrt{\frac{5}{154}}}\,z & \frac{19}{2\,{\sqrt{7}}} & 0 & \frac{2\,{\sqrt{\frac{21}{5}}}}{z} & 0 & \frac{23}
   {5\,z^2} & 0 & 0 & 0 & \frac{{\sqrt{26}}}{z^3} & 0 & 0 & 0 & 0 & \\[6pt] \frac{25\,z^2}{{\sqrt{341}}} & 2\,{\sqrt{\frac{30}{341}}}\,z & \frac{{\sqrt{\frac{3}
        {31}}}}{2} & 3\,{\sqrt{\frac{13}{31}}} & 0 & \frac{{\sqrt{\frac{62}{5}}}}{z} & 0 & \frac{23}{5\,z^2} & 0 & 0 & 0 & \frac{6\,{\sqrt{\frac{3}{5}}}}
   {z^3} & 0 & 0 & 0 & \\[6pt] \frac{3\,{\sqrt{\frac{183}{1085}}}\,z^2}{2} & -3\,{\sqrt{\frac{61}{434}}}\,z & \frac{3\,{\sqrt{\frac{305}{2387}}}}{2} & 4\,
   {\sqrt{\frac{1365}{20801}}} & 0 & 0 & 0 & 0 & \frac{23}{5\,z^2} & 0 & 0 & 0 & \frac{{\sqrt{\frac{46}{5}}}}{z^3} & 0 & 0 & \\[6pt] 0 & 0 & 0 & {\sqrt{\frac{759}
      {122}}} & 0 & 0 & 0 & 0 & 0 & \frac{23}{5\,z^2} & 0 & 0 & 0 & \frac{{\sqrt{\frac{46}{5}}}}{z^3} & 0 & \\[6pt] 75\,{\sqrt{\frac{3}{2002}}}\,z^3 & 50\,
   {\sqrt{\frac{5}{1001}}}\,z^2 & \frac{55\,z}{{\sqrt{182}}} & 0 & 8\,{\sqrt{\frac{30}{91}}} & 0 & \frac{{\sqrt{26}}}{z} & 0 & 0 & 0 & \frac{28}
   {5\,z^2} & 0 & 0 & 0 & 0 & \\[6pt] 25\,{\sqrt{\frac{5}{1023}}}\,z^3 & \frac{25\,z^2}{{\sqrt{682}}} & {\sqrt{\frac{5}{31}}}\,z & 2\,{\sqrt{\frac{65}{93}}}\,
   z & \frac{1}{{\sqrt{93}}} & 22\,{\sqrt{\frac{2}{93}}} & 0 & \frac{6\,{\sqrt{\frac{3}{5}}}}{z} & 0 & 0 & 0 & \frac{28}{5\,z^2} & 0 & 0 & 0 & \\[6pt] 9\,
   {\sqrt{\frac{183}{9982}}}\,z^3 & -33\,{\sqrt{\frac{61}{24955}}}\,z^2 & -3\,{\sqrt{\frac{61}{109802}}}\,z & 40\,{\sqrt{\frac{546}{478423}}}\,z & 3\,
   {\sqrt{\frac{1830}{54901}}} & 52\,{\sqrt{\frac{105}{478423}}} & 0 & 0 & \frac{{\sqrt{\frac{46}{5}}}}{z} & 0 & 0 & 0 & \frac{28}
   {5\,z^2} & 0 & 0 & \\[6pt] 0 & 0 & 0 & 2\,{\sqrt{\frac{165}{61}}}\,z & 0 & {\sqrt{\frac{429}{122}}} & 0 & 0 & 0 & \frac{{\sqrt{\frac{46}{5}}}}
   {z} & 0 & 0 & 0 & \frac{28}{5\,z^2} & 0 & \\[6pt] -2\,{\sqrt{\frac{7}{897}}}\,z^3 & {\sqrt{\frac{70}{299}}}\,z^2 & -10\,{\sqrt{\frac{7}{3289}}}\,z & 13\,
   {\sqrt{\frac{7}{759}}}\,z & \frac{5\,{\sqrt{\frac{35}{9867}}}}{2} & -{\sqrt{\frac{455}{1518}}} & 0 & 0 & 0 & 0 & 0 & 0 & 0 & 0 & \frac{28}{5\,z^2} \\[6pt]
& & & & & & & & & & & & & & & \ddots}
$   
}
\mbox{}\hspace{.4in}\mbox{}
\rotatebox{90}{Figure~10: Energy-momentum tensor (in the $L_1$ basis) of the tricritical Ising model in the $h=3/5$ sector.}
\thispagestyle{empty}
\end{figure}
\clearpage

\setlength{\topmargin}{-25mm} 
\setlength{\arraycolsep}{1mm}
\begin{figure}[p]
\mbox{}\hspace{-11mm}\mbox{}
\rotatebox{90}{
$ \hspace{-1mm} \tpf{3/5}{3/5}{0} (z)=\smat{
  1 & \frac{6}{{\sqrt{35}}\,z^2} & \frac{6}{{\sqrt{35}}\,z^3} & \frac{9}{5\,z^4}\sqrt{\frac{2}{7}} & 
    \frac{8}{5\,z^4} {\sqrt{\frac{6}{119}}} & 
    \frac{12}{5\,{\sqrt{7}}\,z^5} & \frac{16}{5\,z^5} {\sqrt{\frac{3}{119}}} & \frac{6}{7\,z^6} & 
    \frac{8}{z^6}{\sqrt{\frac{2}{357}}} & \frac{47}{35\,{\sqrt{195}}\,z^6} & \frac{4}{5\,z^6}\sqrt{\frac{14}{3315}} & \\[5pt] 
{\sqrt{\frac{6}{5}}}\,z & \frac{-4\,{\sqrt{\frac{6}{7}}}}{5\,z} & \frac{{\sqrt{
         \frac{6}{7}}}}{5\,z^2} & \frac{8\,{\sqrt{\frac{3}{35}}}}{5\,z^3} & \frac{-16\,{\sqrt{\frac{7}{85}}}}{5\,z^3} & \frac{{\sqrt{\frac{42}{5}}}}
   {5\,z^4} & \frac{-72\,{\sqrt{\frac{2}{595}}}}{5\,z^4} & \frac{4\,{\sqrt{\frac{6}{5}}}}{7\,z^5} & \frac{-16}{{\sqrt{595}}\,z^5} & \frac{-188\,
     {\sqrt{\frac{2}{13}}}}{175\,z^5} & \frac{-32\,{\sqrt{\frac{7}{221}}}}{25\,z^5} & \\[5pt] 
\frac{{\sqrt{33}}\,z^2}{5} & \frac{-4\,{\sqrt{\frac{3}{385}}}}{5} & 
    \frac{-4\,{\sqrt{\frac{33}{35}}}}{5\,z} & \frac{-{\sqrt{\frac{66}{7}}}}{25\,z^2} & \frac{72\,{\sqrt{\frac{14}{187}}}}{25\,z^2} & \frac{2\,
     {\sqrt{\frac{33}{7}}}}{25\,z^3} & \frac{-16\,{\sqrt{\frac{7}{187}}}}{25\,z^3} & \frac{2\,{\sqrt{33}}}{35\,z^4} & \frac{-64\,{\sqrt{\frac{2}{1309}}}}
   {5\,z^4} & \frac{3572}{175\,{\sqrt{715}}\,z^4} & \frac{304\,{\sqrt{\frac{14}{12155}}}}{25\,z^4} & \\[5pt] 0 & {\sqrt{\frac{65}{77}}} & 0 & 0 & \frac{2\,
     {\sqrt{\frac{78}{1309}}}}{z^2} & 0 & \frac{2\,{\sqrt{\frac{39}{1309}}}}{z^3} & 0 & \frac{{\sqrt{\frac{78}{1309}}}}{z^4} & \frac{2\,{\sqrt{\frac{3}{55}}}}
   {z^4} & \frac{-16\,{\sqrt{\frac{6}{6545}}}}{z^4} & \\[5pt] 
\frac{4\,{\sqrt{\frac{11}{5}}}\,z^3}{5} & \frac{-6\,z}{25\,{\sqrt{77}}} & \frac{-3\,
     {\sqrt{\frac{7}{11}}}}{25} & \frac{-24\,{\sqrt{\frac{22}{35}}}}{25\,z} & \frac{-24\,{\sqrt{\frac{42}{935}}}}{25\,z} & \frac{-12\,{\sqrt{\frac{11}{35}}}}
   {25\,z^2} & \frac{12\,{\sqrt{\frac{231}{85}}}}{25\,z^2} & 0 & \frac{4\,{\sqrt{\frac{14}{2805}}}}{z^3} & \frac{-1786}{125\,{\sqrt{429}}\,z^3} & \frac{-1064\,
     {\sqrt{\frac{14}{7293}}}}{125\,z^3} & \\[5pt] 
0 & 3\,{\sqrt{\frac{2}{77}}}\,z & 5\,{\sqrt{\frac{2}{77}}} & 0 & \frac{-8\,{\sqrt{\frac{3}{6545}}}}{z} & 0 & 
    \frac{{\sqrt{\frac{66}{595}}}}{z^2} & 0 & \frac{16\,{\sqrt{\frac{3}{6545}}}}{z^3} & \frac{-14\,{\sqrt{\frac{6}{143}}}}{5\,z^3} & \frac{32\,
     {\sqrt{\frac{21}{2431}}}}{5\,z^3} & \\[5pt] 
\frac{2\,{\sqrt{231}}\,z^4}{25} & \frac{-{\sqrt{\frac{33}{5}}}\,z^2 }{175} & \frac{-31\,
     {\sqrt{\frac{3}{55}}}\,z}{175} & \frac{-263\,{\sqrt{\frac{3}{22}}}}{875} & \frac{-12\,{\sqrt{\frac{2}{187}}}}{125} & \frac{-16\,{\sqrt{33}}}{125\,z} & 
    \frac{-504}{125\,{\sqrt{187}}\,z} & \frac{-4\,{\sqrt{\frac{33}{7}}}}{25\,z^2} & \frac{168\,{\sqrt{\frac{2}{187}}}}{25\,z^2} & \frac{2679}
   {125\,{\sqrt{5005}}\,z^2} & \frac{1596\,{\sqrt{\frac{2}{12155}}}}{125\,z^2} & \\[5pt] 
0 & 3\,{\sqrt{\frac{11}{1085}}}\,z^2 & 12\,{\sqrt{\frac{5}{2387}}}\,z & 25\,
   {\sqrt{\frac{2}{2387}}} & \frac{-4\,{\sqrt{\frac{6}{40579}}}}{5} & 0 & \frac{-4\,{\sqrt{\frac{93}{1309}}}}{5\,z} & 0 & \frac{3\,{\sqrt{\frac{186}{1309}}}}
   {5\,z^2} & \frac{126\,{\sqrt{\frac{3}{22165}}}}{5\,z^2} & \frac{-144\,{\sqrt{\frac{42}{376805}}}}{5\,z^2} & \\[5pt] 
0 & \frac{3\,{\sqrt{\frac{183}{31}}}\,z^2}
   {7} & \frac{-3\,{\sqrt{\frac{183}{31}}}\,z}{7} & \frac{3\,{\sqrt{\frac{183}{310}}}}{7} & -148\,{\sqrt{\frac{2}{160735}}} & 0 & 0 & 0 & 0 & \frac{9\,
     {\sqrt{\frac{61}{2821}}}}{5\,z^2} & \frac{-1652\,{\sqrt{\frac{2}{417911}}}}{15\,z^2} & \\[5pt] 
0 & 0 & 0 & 0 & 5\,
   {\sqrt{\frac{299}{7259}}} & 0 & 0 & 0 & 0 & 0 & \frac{26\,{\sqrt{\frac{115}{7259}}}}{3\,z^2} & \\[5pt] 
\frac{2\,{\sqrt{6006}}\,z^5}{125} & \frac{-8\,
     {\sqrt{\frac{66}{65}}}\,z^3}{875} & \frac{-41\,{\sqrt{\frac{33}{130}}}\,z^2}{875} & \frac{-1564\,{\sqrt{\frac{3}{143}}}\,z}{4375} & \frac{-72\,z}
   {625\,{\sqrt{2431}}} & \frac{-5177\,{\sqrt{\frac{3}{286}}}}{4375} & \frac{-372\,{\sqrt{\frac{2}{2431}}}}{625} & \frac{-8\,{\sqrt{\frac{858}{7}}}}
   {125\,z} & \frac{-168\,{\sqrt{\frac{13}{187}}}}{125\,z} & \frac{-5358\,{\sqrt{\frac{2}{385}}}}{8125\,z} & \frac{-6384}{8125\,{\sqrt{935}}\,z} & \\[5pt] 
0 & 
    \frac{8\,{\sqrt{\frac{11}{651}}}\,z^3}{5} & {\sqrt{\frac{33}{217}}}\,z^2 & 5\,{\sqrt{\frac{30}{2387}}}\,z & \frac{-4\,{\sqrt{\frac{2}{202895}}}\,z}
   {5} & 25\,{\sqrt{\frac{5}{7161}}} & \frac{-2\,{\sqrt{\frac{17}{11935}}}}{5} & 0 & \frac{-12\,{\sqrt{\frac{62}{6545}}}}{5\,z} & \frac{-84}
   {25\,{\sqrt{4433}}\,z} & \frac{96\,{\sqrt{\frac{14}{75361}}}}{25\,z} & \\[5pt] 
0 & \frac{24\,{\sqrt{\frac{366}{3565}}}\,z^3}{7} & \frac{-3\,
     {\sqrt{\frac{183}{7130}}}\,z^2}{7} & \frac{-66\,{\sqrt{\frac{183}{713}}}\,z}{35} & \frac{-888\,z}{5\,{\sqrt{739381}}} & \frac{9\,{\sqrt{\frac{183}{1426}}}}
   {7} & -296\,{\sqrt{\frac{2}{739381}}} & 0 & 0 & \frac{-18\,{\sqrt{\frac{122}{324415}}}}{5\,z} & \frac{6608}
   {15\,{\sqrt{48059765}}\,z} & \\[5pt] 
0 & 0 & 0 & 0 & 3\,{\sqrt{\frac{130}{7259}}}\,z & 0 & 10\,{\sqrt{\frac{65}{7259}}} & 0 & 0 & 0 & \frac{-52\,
     {\sqrt{\frac{2}{7259}}}}{3\,z} & \\[5pt] 
0 & 4\,{\sqrt{\frac{5}{897}}}\,z^3 & -2\,{\sqrt{\frac{15}{299}}}\,z^2 & 2\,{\sqrt{\frac{6}{299}}}\,z & -2\,
   {\sqrt{\frac{34}{299}}}\,z & \frac{-2}{{\sqrt{897}}} & {\sqrt{\frac{17}{299}}} & 0 & 0 & \frac{-24\,{\sqrt{\frac{7}{115}}}}{13\,z} & \frac{504\,
     {\sqrt{\frac{2}{1955}}}}{13\,z} & \\[5pt] 
\frac{2\,{\sqrt{\frac{31031}{5}}}\,z^6}{125} & \frac{-24\,{\sqrt{\frac{11}{403}}}\,z^4}{625} & \frac{-408\,
     {\sqrt{\frac{11}{403}}}\,z^3}{4375} & \frac{-2589\,{\sqrt{\frac{11}{4030}}}\,z^2}{4375} & \frac{-12\,{\sqrt{\frac{66}{34255}}}\,z^2}{625} & \frac{-38991\,z}
   {4375\,{\sqrt{22165}}} & \frac{-984\,{\sqrt{\frac{3}{376805}}}\,z}{625} & \frac{-903\,{\sqrt{\frac{7}{22165}}}}{125} & \frac{-736\,{\sqrt{\frac{6}{376805}}}}
   {125} & \frac{-1786\,{\sqrt{\frac{3}{2387}}}}{40625} & \frac{-1064\,{\sqrt{\frac{6}{5797}}}}{40625} & \\[5pt] 
0 & \frac{4\,{\sqrt{\frac{231}{6355}}}\,z^4}
   {5} & 32\,{\sqrt{\frac{11}{133455}}}\,z^3 & 5\,{\sqrt{\frac{66}{8897}}}\,z^2 & \frac{-2\,{\sqrt{\frac{22}{151249}}}\,z^2}{25} & 100\,
   {\sqrt{\frac{3}{97867}}}\,z & \frac{-142\,z}{25\,{\sqrt{1663739}}} & \frac{125}{{\sqrt{41943}}} & \frac{-63\,{\sqrt{\frac{14}{237677}}}}{25} & \frac{-42}
   {25\,{\sqrt{908765}}} & \frac{48\,{\sqrt{\frac{14}{15449005}}}}{25} & \\[5pt] 
0 & \frac{72\,{\sqrt{\frac{61}{12121}}}\,z^4}{5} & \frac{216\,
     {\sqrt{\frac{61}{12121}}}\,z^3}{35} & \frac{-801\,{\sqrt{\frac{61}{121210}}}\,z^2}{35} & \frac{-1628\,{\sqrt{\frac{6}{217465}}}\,z^2}{85} & \frac{-171\,
     {\sqrt{\frac{61}{60605}}}\,z}{7} & \frac{-1184\,{\sqrt{\frac{3}{217465}}}\,z}{17} & 9\,{\sqrt{\frac{305}{84847}}} & \frac{-296\,{\sqrt{\frac{30}{43493}}}}
   {17} & \frac{-6\,{\sqrt{\frac{183}{1103011}}}}{25} & \frac{3304\,{\sqrt{\frac{2}{1696227}}}}{1275} & \\[5pt] 
0 & 0 & 0 & 0 & \frac{11\,{\sqrt{\frac{39}{427}}}\,z^2}
   {17} & 0 & \frac{20\,{\sqrt{\frac{78}{427}}}\,z}{17} & 0 & \frac{50\,{\sqrt{\frac{39}{427}}}}{17} & 0 & \frac{-52}{51\,{\sqrt{6405}}} & \\[5pt] 
0 & 
   {\sqrt{\frac{42}{299}}}\,z^4 & -2\,{\sqrt{\frac{14}{897}}}\,z^3 & -2\,{\sqrt{\frac{21}{1495}}}\,z^2 & -11\,{\sqrt{\frac{17}{10465}}}\,z^2 & 2\,
   {\sqrt{\frac{42}{1495}}}\,z & -17\,{\sqrt{\frac{17}{20930}}}\,z & -{\sqrt{\frac{10}{897}}} & \frac{3\,{\sqrt{\frac{85}{2093}}}}{2} & \frac{-6\,
     {\sqrt{\frac{2}{23}}}}{65} & \frac{36\,{\sqrt{\frac{7}{391}}}}{65} & \\[5pt] 
0 & {\sqrt{\frac{34698}{151249}}}\,z^4 & -2\,{\sqrt{\frac{34698}{151249}}}\,
   z^3 & 6\,{\sqrt{\frac{17349}{756245}}}\,z^2 & \frac{99003\,z^2}{17\,{\sqrt{257256755}}} & -2\,{\sqrt{\frac{34698}{756245}}}\,z & \frac{-99003\,z}
   {17\,{\sqrt{514513510}}} & \frac{{\sqrt{\frac{34698}{108035}}}}{7} & \frac{33001}{34\,{\sqrt{257256755}}} & \frac{-592642\,{\sqrt{\frac{2}{1624392653}}}}
   {35} & \frac{-80468\,{\sqrt{\frac{7}{95552509}}}}{85} & \\[5pt] 
0 & 0 & 0 & 0 & \frac{16}{17}{\sqrt{\frac{4485}{40481}}}\,z^2 & 0 & \frac{-8}{17} {\sqrt{\frac{8970}{40481}}}\,z & 0 & \frac{8}{17} {\sqrt{\frac{1495}{121443}}} & 4\,{\sqrt{\frac{782}{17349}}} & 
   \frac{407}{17}\sqrt{\frac{23}{121443}} \\[5pt]
& & & & & & & & & & & \ddots} 
$   
}
\mbox{}\hspace{.02in}\mbox{}
\rotatebox{90}{{Figure~11: Primary field $\tpf{3/5}{3/5}{0} (z)$ (in the $L_1$ basis) of the tricritical Ising model. The squares of the first column entries
agree with $(1-u)^{-6/5}$.
}}
\thispagestyle{empty}
\end{figure}
\clearpage

\newpage
\setlength{\topmargin}{-14mm} 
\section{3-state Potts model}
\setcounter{figure}{18}

In this section we consider the 3-state Potts (hard hexagon) model which is an example of a parafermion model. This theory is dual to the tricritical Ising  theory $\M(4,5)$ and is obtained in the $A_4$ ABF model~\cite{ABF} by taking the sign of the spectral parameter $u$  to be negative.

The groundstate sector can be understood as the sum  $\mathcal{V}_0 \oplus \mathcal{V}_3$ of two Virasoro modules or as a single module of the $W_3$ algebra associated with the identity field
$\phi_{(\ell,m)}$ with $\ell=m=0$. The first viewpoint is more convenient in the following matrix representations. 
The relevant character expression is 
\begin{eqnarray}
 \chi_{0}(q)+\chi_{3}(3) & = & 1 + q^2 + q^3 + 2\,q^4 + 2\,q^5 + 4\,q^6 + 4\,q^7 + 7\,q^8 + 8\,q^9 + 
  O(q^{10})  \\
&+ &  q^3 ( 1 + q + 2\,q^2 + 3\,q^3 + 4\,q^4 + 5\,q^5 + 8\,q^6 + 10\,q^7 + 14\,q^8  + 
 O(q^{9}) ) \nonumber  \\
&= & 1 + q^2 + 2\,q^3 + 3\,q^4 + 4\,q^5 + 7\,q^6 + 8\,q^7 + 12\,q^8 + 16\,q^9 +O(q^{10})  \nonumber
\end{eqnarray}
For the Virasoro module $\mathcal{V}_0$ we have
\begin{equation}
L_0=\mbox{Diagonal}(0,2, 3, 4, 4, 5, 5, 6, 6, 6, 6, 7, 7, 7, 7, 8, 8, 8, 8, 8, 8, 8,\ldots )
\end{equation}
and in the $L_1$ basis we find
\setlength{\arraycolsep}{1.2mm}
\begin{equation}
L_1=\smat{
   . & . & . & . & . & . & . & . & . & . & . & . & . & . & . & \\[5pt] 
   . & . & 2 & . & . & . & . & . & . & . & . & . & . & . & . & \\[5pt] 
   . & . & . & {\sqrt{10}} & 0 & . & . & . & . & . & . & . & . & . & . & \\[5pt] 
   . & . & . & . & . & 3\,{\sqrt{2}} & 0 & . & . & . & . & . & . & . & . & \\[5pt] 
   . & . & . & . & . & 0 & 2\,{\sqrt{2}} & . & . & . & . & . & . & . & . & \\[5pt] 
   . & . & . & . & . & . & . & 2\,{\sqrt{7}} & 0 & 0 & 0 & . & . & . & . & \\[5pt] 
   . & . & . & . & . & . & . & 0 & 3\,{\sqrt{2}} & 0 & 0 & . & . & . & . & \\[5pt] 
   . & . & . & . & . & . & . & . & . & . & . & 2\,{\sqrt{10}} & 0 & 0 & 0 & \\[5pt] 
   . & . & . & . & . & . & . & . & . & . & . & 0 & {\sqrt{30}} & 0 & 0 & \\[5pt] 
   . & . & . & . & . & . & . & . & . & . & . & 0 & 0 & 2\,{\sqrt{3}} & 0 & \\[5pt] 
   . & . & . & . & . & . & . & . & . & . & . & 0 & 0 & 0 & 2\,{\sqrt{3}} & \\[5pt]  
    & & & & & & & & & & & & & & & \ddots }
\end{equation}
\setlength{\arraycolsep}{1.2mm}
\begin{equation}
L_2=\smat{
 . & {\sqrt{\frac{2}{5}}} & . & . & . & . & . & . & . & . & . & . & . & . & . & \\[6pt]
 . & . & . & 3\,{\sqrt{\frac{2}{5}}} & {\sqrt{\frac{26}{5}}} & . & . & . & . & . & . & . 
   & . & . & . & \\[6pt] 
 . & . & . & . & . & \frac{7}{{\sqrt{5}}} & {\sqrt{\frac{13}{5}}} & . & . & . & . & . & . 
   & . & . & \\[6pt] 
 . & . & . & . & . & . & . & 8\,{\sqrt{\frac{2}{7}}} & \frac{{\sqrt{13}}}{3} & 
   \frac{1}{3} {\sqrt{\frac{17}{7}}} & 0 & . & . & . & . & \\[6pt] 
 . & . & . & . & . & . & . & 0 & 2 & \frac{8}{5}{\sqrt{\frac{91}{17}}} & 
   \frac{6}{5} {\sqrt{\frac{46}{17}}} & . & . & . & . & \\[6pt] 
 . & . & . & . & . & . & . & . & . & . & . & 9\,{\sqrt{\frac{5}{14}}} & 
   {\sqrt{\frac{13}{15}}} & {\sqrt{\frac{17}{42}}} & 0 & \\[6pt] 
 . & . & . & . & . & . & . & . & . & . & . & 0 & \frac{13}{{\sqrt{15}}} & 
   \frac{8}{5}{\sqrt{\frac{182}{51}}} & \frac{4}{5} {\sqrt{\frac{69}{17}}} & \\[6pt]
   & & & & & & & & & & & & & & & \ddots  }
\end{equation}
\setlength{\arraycolsep}{1.8mm}

For the Virasoro module $\mathcal{V}_3$ we have
\begin{equation}
L_0=3+\mbox{Diagonal}(0, 1, 2, 2, 3, 3, 3, 4, 4, 4, 4, 5, 5, 5, 5, 5, 
    6, 6, 6, 6, 6, 6, 6, 6,\ldots) 
\end{equation}
and in the $L_1$ basis we find
\begin{equation}
L_1=\smat{
   . & {\sqrt{6}} & . & . & . & . & . & . & . & . & . & . & . & . & . & . & \\[5pt]
   . & . & {\sqrt{14}} & 0 & . & . & . & . & . & . & . & . & . & . & . & . & \\[5pt] 
   . & . & . & . & 2\,{\sqrt{6}} & 0 & 0 & . & . & . & . & . & . & . & . & . & \\[5pt] 
   . & . & . & . & 0 & {\sqrt{10}} & 0 & . & . & . & . & . & . & . & . & . & \\[5pt] 
   . & . & . & . & . & . & . & 6 & 0 & 0 & 0 & . & . & . & . & . & \\[5pt] 
   . & . & . & . & . & . & . & 0 & {\sqrt{22}} & 0 & 0 & . & . & . & . & . & \\[5pt] 
   . & . & . & . & . & . & . & 0 & 0 & 2\,{\sqrt{3}} & 0 & . & . & . & . & . & \\[5pt] 
   . & . & . & . & . & . & . & . & . & . & . & 5\,{\sqrt{2}} & 0 & 0 & 0 & 0 & \\[5pt] 
   . & . & . & . & . & . & . & . & . & . & . & 0 & 6 & 0 & 0 & 0 & \\[5pt] 
   . & . & . & . & . & . & . & . & . & . & . & 0 & 0 & {\sqrt{26}} & 0 & 0 & \\[5pt] 
   . & . & . & . & . & . & . & . & . & . & . & 0 & 0 & 0 & {\sqrt{14}} & 0 & \\[5pt]
   & & & & & & & & & & & & & & & & \ddots  } 
\end{equation}
\setlength{\arraycolsep}{1.1mm}
\begin{equation}
L_2=\smat{
   . & . & 3\,{\sqrt{\frac{3}{7}}} & {\sqrt{\frac{299}{35}}} & . & . & . & . & . & . & . & 
     . & . & . & . & . & \\[6pt] 
   . & . & . & . & 5\,{\sqrt{\frac{3}{7}}} & \frac{1}{5} {\sqrt{\frac{897}{7}}}
     & \frac{{\sqrt{14}}}{5} & . & . & . & . & . & . & . & . & . & \\[6pt] 
   . & . & . & . & . & . & . & \frac{11}{{\sqrt{6}}} & \frac{1}{5} {\sqrt{\frac{897}{11}}} & 
     \frac{7}{5\,{\sqrt{3}}} & 3\,{\sqrt{\frac{3}{154}}} & . & . & . & . & . & \\[6pt] 
   . & . & . & . & . & . & . & 0 & 3\,{\sqrt{\frac{5}{11}}} & 0 & 4\,{\sqrt{\frac{598}{385}}} 
     & . & . & . & . & . & \\[6pt] 
   . & . & . & . & . & . & . & . & . & . & . & 4\,{\sqrt{2}} & 
     \frac{1}{5}\sqrt{\frac{598}{11}} & \frac{14}{5\,{\sqrt{13}}} & 
     \frac{9} {7} \sqrt{\frac{2}{11}} & \frac{1}{7} {\sqrt{\frac{23}{13}}} & \\[6pt] 
   . & . & . & . & . & . & . & . & . & . & . & 0 & 8\,{\sqrt{\frac{2}{11}}} & 0 & 
     \frac{4}{7} {\sqrt{\frac{598}{11}}} & \frac{13}{35} & \\[6pt] 
   . & . & . & . & . & . & . & . & . & . & . & 0 & 0 & 3\,{\sqrt{\frac{6}{13}}} & 0 & 
     \frac{7}{5}\sqrt{\frac{138}{13}} & \\[6pt] 
   & & & & & & & & & & & & & & & & \ddots  }
\end{equation}
\setlength{\arraycolsep}{1.8mm}

\section{Yang-Lee theory}
In this section we consider the Yang-Lee theory $\M(2,5)$ with $c=-22/5$ which is the simplest example of a non-unitary minimal theory.

\subsection{Yang-Lee theory: $\V_0$}

In the groundstate sector $\V_0$, the character is
\begin{equation}
\chi_{0}(q)= 1 + q^2 + q^3 + q^4 + q^5 + 2\,q^6 + 2\,q^7 + 3\,q^8 + 3\,q^9 + 
4\,q^{10} + 4\,q^{11} + 6\,q^{12} + O(q^{13})
\end{equation}
and the energy is
\begin{equation}
L_0=\mbox{Diagonal} (0, 2, 3, 4, 5, 6, 6, 7, 7, 8, 8, 8, 9, 9, 9,  \ldots )
\end{equation}
In the $L_1$ basis we find
\begin{equation}
L_1=\smat{
   . & . & . & . & . & . & . & . & . & . & . & . & . & . & . & \\[4pt] 
   . & . & 2 & . & . & . & . & . & . & . & . & . & . & . & . & \\[4pt] 
   . & . & . & { \sqrt{10}} & . & . & . & . & . & . & . & . & . & . & . & \\[4pt] 
   . & . & . & . & 3\, {\sqrt{2}} & . & . & . & . & . & . & . & . & . & . & \\[4pt] 
   . & . & . & . & . & 2\, {\sqrt{7}} & 0 & . & . & . & . & . & . & . & . & \\[4pt] 
   . & . & . & . & . & . & . & 2\,{\sqrt{10}} & 0 & . & . & . & . & . & . & \\[4pt] 
   . & . & . & . & . & . & . & 0 & 2\,{\sqrt{3}} & . & . & . & . & . & . & \\[4pt] 
   . & . & . & . & . & . & . & . & . & 3\,{\sqrt{6}} & 0 & 0 & . & . & . & \\[4pt] 
   . & . & . & . & . & . & . & . & . & 0 & {\sqrt{26}} & 0 & . & . & . & \\[4pt] 
   . & . & . & . & . & . & . & . & . & . & . & . & {\sqrt{70}} & 0 & 0 & \\[4pt] 
   . & . & . & . & . & . & . & . & . & . & . & . & 0 & {\sqrt{42}} & 0 & \\[4pt] 
   . & . & . & . & . & . & . & . & . & . & . & . & 0 & 0 & 4 & \\
   & & & & & & & & & & & & & & & \ddots }
\end{equation}
\setlength{\arraycolsep}{1.1mm}
\begin{equation}
L_2=\smat{
   . & i \, {\sqrt{\frac{11}{5}}} & . & . & . & . & . & . & . & . & . & . & . & . & . & \cr 
   . & . & . & 3\, {\sqrt{\frac{2}{5}}} & . & . & . & . & . & . & . & . & . & . & . & \cr 
   . & . & . & . & \frac{7}{{\sqrt{5}}} & . & . & . & . & . & . & . & . & . & . & \cr 
   . & . & . & . & . & 8\,{\sqrt{\frac{2}{7}}} & i \,
   {\sqrt{\frac{31}{35}}} & . & . & . & . & . & . & . & . & \cr 
   . & . & . & . & . & . & . & 9\,{\sqrt{\frac{5}{14}}} & i \,
   {\sqrt{\frac{93}{70}}} & . & . & . & . & . & . & \cr 
   . & . & . & . & . & . & . & . & . & 5\,{\sqrt{\frac{5}{3}}} & i \,
   {\sqrt{\frac{93}{65}}} & i \, {\sqrt{\frac{41}{273}}} & . & . & . & \cr 
   . & . & . & . & . & . & . & . & . & 0 & 3\,{\sqrt{\frac{6}{13}}} & {\sqrt{\frac{7626}
      {455}}} & . & . & . & \cr 
   . & . & . & . & . & . & . & . & . & . & . & . & 11\,{\sqrt{\frac{7}{15}}} & 2\,i \,
   {\sqrt{\frac{31}{91}}} & i \,  {\sqrt{\frac{205}{546}}} & \cr 
   . & . & . & . & . & . & . & . & . & . & . & . & 0 & 19\,{\sqrt{\frac{3}{91}}} & 3\,
   {\sqrt{\frac{1271}{910}}} & \cr
   & & & & & & & & & & & & & & & \ddots }
\end{equation}
Notice that the first null vector appears at level 4, as is clear from Figure~2.
Notice also that the choice of signs of the square roots here is fixed by our convention and does not necessarily agree with the principal value from Figure~2.

\subsection{Yang-Lee theory: $\V_{-1/5}$}
The character in this sector is
\begin{equation}
q^{{1\over 5}}\chi_{-\frac{1}{5}}(q)= 1 + q + q^2 + q^3 + 2\,q^4 + 2\,q^5 + 3\,q^6 + 3\,q^7 + 
4\,q^8 + 5\,q^9 + 6\,q^{10} + 7\,q^{11} + 9\,q^{12} + O(q^{13})
\end{equation}
with energy
\begin{equation}
L_0=\mbox{Diagonal} (0,1,2, 3, 4,4, 5,5, 6, 6,6, 7, 7,7,8, 8, 8, 8,  \ldots )
-\frac{1}{5}
\end{equation}
In the $L_1$ basis we find
\begin{equation}
L_1=\frac{1}{\sqrt{5}} \smat{
 . & i \, \sqrt{2} & . & . & . & . & . & . & . & . & . & . & . & . & \\[4pt] 
 . & . & {\sqrt{6}} & . & . & . & . & . & . & . & . & . & . & . & \\[4pt] 
 . & . & . & 2\, \sqrt{6} & . & . & . & . & . & . & . & . & . & . & \\[4pt] 
 . & . & . & . & 2\,{\sqrt{13}} & 0 & . & . & . & . & . & . & . & . & \\[4pt] 
 . & . & . & . & . & . & 3\, {\sqrt{10}} & 0 & . & . & . & . & . & . & \\[4pt] 
 . & . & . & . & . & . & 0 & {\sqrt {38}} & . & . & . & . & . & . & \\[4pt] 
 . & . & . & . & . & . & . & . & {\sqrt{138}} & 0 & 0 & . & . & . & \\[4pt] 
 . & . & . & . & . & . & . & . & 0 & {\sqrt{86}} & 0 & . & . & . & \\[4pt] 
 . & . & . & . & . & . & . & . & . & . & . & 14 & 0 & 0 & \\[4pt] 
 . & . & . & . & . & . & . & . & . & . & . & 0 & 12 & 0 & \\[4pt] 
 . & . & . & . & . & . & . & . & . & . & . & 0 & 0 & {\sqrt{58}} & \\[4pt] 
   & & & & & & & & & & & & & & \ddots }
\end{equation}
\begin{equation}
L_2=\smat{
 . & . & i \, {\sqrt{3}} & . & . & . & . & . & . & . & . & . & . & . & \\[6pt] 
 . & . & . & 1 & . & . & . & . & . & . & . & . & . & . & \\[6pt] 
 . & . & . & . & 7\,{\sqrt{\frac{3}{26}}} & i \,
    {\sqrt{\frac{95}{26}}} & . & . & . & . & . & . & . & . & \\[6pt] 
 . & . & . & . & . & . & 4\,{\sqrt{\frac{10}{13}}} & i \,
    {\sqrt{\frac{30}{13}}} & . & . & . & . & . & . & \\[6pt] 
 . & . & . & . & . & . & . & . & 17\,{\sqrt{\frac{5}{69}}} & 
    2\,i \,{\sqrt{\frac{15}{43}}} & 5\,i \,
    {\sqrt{\frac{2755}{77142}}} & . & . & . & \\[6pt] 
 . & . & . & . & . & . & . & . & 0 & 3\,{\sqrt{\frac{19}{43}}} & 3\,
    {\sqrt{\frac{667}{1118}}} & . & . & . & \\[6pt] 
 . & . & . & . & . & . & . & . & . & . & . & 11\,{\sqrt{\frac{6}{23}}} & 
    5\,i \,{\sqrt{\frac{3}{86}}} & 25\,i \, {\sqrt{\frac{57}{25714}}} & \\[6pt] 
    . & . & . & . & . & . & . & . & . & . & . & 0 &
    \frac{31}{{\sqrt{86}}} & 3\,{\sqrt{\frac{437}{1118}}} & \\[6pt]
 & & & & & & & & & & & & & & \ddots  }
\end{equation}
The field $\phi^{(-1/5)}(z)$ is shown in Figure~13.


\begin{figure}[p]
\mbox{}\hspace{.4in}\mbox{}
\rotatebox{90}{
$ \tpf{-1/5}{-1/5}{0} (z) = \smat{
 1 & \frac{i }{{\sqrt{55}}\,z^2} & \frac{i }
   {{\sqrt{55}}\,z^3} & \frac{\frac{3\,i }{5}}{{\sqrt{22}}\,z^4} & \frac{
      \frac{2\,i }{5}}{{\sqrt{11}}\,z^5} & \frac{i }
   {{\sqrt{77}}\,z^6} & \frac{-1}{5\,{\sqrt{23870}}\,z^6} & \frac{3\,i }
   {{\sqrt{770}}\,z^7} & \frac{-{\sqrt{\frac{3}{23870}}}}{5\,z^7} & \\[6pt] i \,
   {\sqrt{\frac{2}{5}}}\,z & \frac{-6\,{\sqrt{\frac{2}{11}}}}{5\,z} & \frac{-7}
   {5\,{\sqrt{22}}\,z^2} & \frac{-8}{5\,{\sqrt{55}}\,z^3} & \frac{-9}
   {5\,{\sqrt{110}}\,z^4} & \frac{-2\,{\sqrt{\frac{2}{385}}}}{z^5} & \frac{\frac
       {-16\,i }{25}}{{\sqrt{2387}}\,z^5} & \frac{-{\sqrt{\frac{11}{7}}}}
   {10\,z^6} & \frac{\frac{-27\,i }{50}\,{\sqrt{\frac{3}{2387}}}}{z^6} & \\[6pt] 
   \frac{i }{5}\,{\sqrt{3}}\,z^2 & \frac{14\,{\sqrt{\frac{3}{55}}}}{5} & 
    \frac{-6\,{\sqrt{\frac{3}{55}}}}{5\,z} & \frac{-13\,{\sqrt{\frac{3}{22}}}}
   {25\,z^2} & \frac{-7\,{\sqrt{\frac{3}{11}}}}{25\,z^3} & \frac{-3\,
     {\sqrt{\frac{3}{77}}}}{5\,z^4} & \frac{\frac{72\,i }{25}\,
     {\sqrt{\frac{6}{11935}}}}{z^4} & \frac{-4\,{\sqrt{\frac{6}{385}}}}
   {5\,z^5} & \frac{\frac{136\,i }{25}\,{\sqrt{\frac{2}{11935}}}}{z^5} & \\[6pt] 
   \frac{2\,i }{5}\,{\sqrt{\frac{2}{5}}}\,z^3 & \frac{-7\,{\sqrt{\frac{2}{11}}}\,
     z}{25} & \frac{91}{25\,{\sqrt{22}}} & \frac{-36}{25\,{\sqrt{55}}\,z} & 
    \frac{-19\,{\sqrt{\frac{2}{55}}}}{25\,z^2} & \frac{-8\,{\sqrt{\frac{2}{385}}}}
   {5\,z^3} & \frac{\frac{-72\,i }{125}\,{\sqrt{\frac{11}{217}}}}{z^3} & 
    \frac{-3\,{\sqrt{\frac{7}{11}}}}{25\,z^4} & \frac{\frac{-36\,i }{125}\,
     {\sqrt{\frac{21}{341}}}}{z^4} & \\[6pt] \frac{i }{25}\,{\sqrt{26}}\,z^4 & 
    \frac{-21\,z^2}{25\,{\sqrt{1430}}} & \frac{-161\,z}{25\,{\sqrt{1430}}} & 
    \frac{2527}{250\,{\sqrt{143}}} & \frac{-12\,{\sqrt{\frac{26}{11}}}}
   {125\,z} & \frac{-{\sqrt{\frac{26}{77}}}}{5\,z^2} & \frac{\frac{612\,i }
      {125}\,{\sqrt{\frac{11}{14105}}}}{z^2} & \frac{-13\,{\sqrt{\frac{13}{385}}}}
   {25\,z^3} & \frac{\frac{-108\,i }{125}\,{\sqrt{\frac{33}{14105}}}}
   {z^3} & \\[6pt] 0 & i \,{\sqrt{\frac{57}{286}}}\,z^2 & -i \,
   {\sqrt{\frac{57}{286}}}\,z & \frac{i }{2}\,
   {\sqrt{\frac{57}{715}}} & 0 & 0 & \frac{-7\,{\sqrt{\frac{399}{4433}}}}
   {5\,z^2} & 0 & \frac{-7\,{\sqrt{\frac{133}{4433}}}}{5\,z^3} & \\[6pt] \frac{6\,i }
    {125}\,{\sqrt{13}}\,z^5 & \frac{-28\,z^3}{125\,{\sqrt{715}}} & \frac{-21\,
     {\sqrt{\frac{11}{65}}}\,z^2}{250} & \frac{-84\,{\sqrt{\frac{26}{11}}}\,z}
   {625} & \frac{13363}{1250\,{\sqrt{143}}} & \frac{-36\,{\sqrt{\frac{13}{77}}}}
   {125\,z} & \frac{\frac{-1224\,i }{625}\,{\sqrt{\frac{22}{14105}}}}{z} & 
    \frac{-93\,{\sqrt{\frac{13}{770}}}}{125\,z^2} & \frac{\frac{102\,i }{625}\,
     {\sqrt{\frac{858}{1085}}}}{z^2} & \\[6pt] 0 & 4\,i \,{\sqrt{\frac{3}{715}}}\,
   z^3 & \frac{7\,i }{2}\,{\sqrt{\frac{3}{715}}}\,z^2 & \frac{-i }{5}\,
   {\sqrt{\frac{78}{11}}}\,z & \frac{3\,i }{2}\,
   {\sqrt{\frac{3}{143}}} & 0 & \frac{42\,{\sqrt{\frac{42}{22165}}}}
   {5\,z} & 0 & \frac{-7\,{\sqrt{\frac{91}{3410}}}}{5\,z^2} & \\[6pt] \frac{i }{125}\,
   {\sqrt{\frac{1794}{5}}}\,z^6 & \frac{-14\,{\sqrt{\frac{26}{759}}}\,z^4}
   {625} & \frac{-602\,{\sqrt{\frac{2}{9867}}}\,z^3}{625} & \frac{-4109\,
     {\sqrt{\frac{3}{16445}}}\,z^2}{1250} & \frac{-4193\,{\sqrt{\frac{11}{8970}}}\,z}
   {625} & \frac{13831\,{\sqrt{\frac{7}{98670}}}}{125} & \frac{408\,i }{3125}\,
   {\sqrt{\frac{231}{9269}}} & \frac{-18\,{\sqrt{\frac{897}{77}}}}{625\,z} & 
    \frac{\frac{-1224\,i }{3125}\,{\sqrt{\frac{253}{2821}}}}{z} & \\[6pt] 0 & 
   \frac{2\,i }{5}\,{\sqrt{\frac{78}{473}}}\,z^4 & \frac{34\,i }{5}\,
   {\sqrt{\frac{6}{6149}}}\,z^3 & \frac{97\,i }{10}\,{\sqrt{\frac{3}{30745}}}\,
   z^2 & -81\,i \,{\sqrt{\frac{3}{61490}}}\,z & 3\,i \,
   {\sqrt{\frac{105}{12298}}} & \frac{-294\,{\sqrt{\frac{21}{190619}}}}
   {25} & 0 & \frac{42\,{\sqrt{\frac{301}{4433}}}}{25\,z} & \\[6pt] 0 & i \,
   {\sqrt{\frac{551}{32637}}}\,z^4 & -2\,i \,{\sqrt{\frac{551}{32637}}}\,
   z^3 & i \,{\sqrt{\frac{3306}{54395}}}\,z^2 & -2\,i \,
   {\sqrt{\frac{551}{163185}}}\,z & i \,{\sqrt{\frac{551}{1142295}}} & \frac
     {139\,{\sqrt{\frac{1653}{4721486}}}}{5} & 0 & 0 & \\[6pt] \frac{2\,i }{625}\,
   {\sqrt{1794}}\,z^7 & \frac{-6\,{\sqrt{\frac{78}{1265}}}\,z^5}{625} & \frac{-53\,
     {\sqrt{\frac{26}{3795}}}\,z^4}{625} & \frac{-5672\,z^3}
   {3125\,{\sqrt{9867}}} & \frac{-7697\,{\sqrt{\frac{3}{6578}}}\,z^2}{3125} & 
    \frac{-29977\,{\sqrt{\frac{2}{69069}}}\,z}{625} & \frac{-408\,i }{3125}\,
   {\sqrt{\frac{33}{324415}}}\,z & \frac{32168\,{\sqrt{\frac{11}{31395}}}}
   {625} & \frac{20196\,i }{3125}\,{\sqrt{\frac{11}{324415}}} & \\[6pt] 0 & 
   \frac{3\,i }{5}\,{\sqrt{\frac{78}{2365}}}\,z^5 & \frac{9\,i }{5}\,
   {\sqrt{\frac{39}{4730}}}\,z^4 & \frac{174\,i }{25}\,{\sqrt{\frac{3}{6149}}}\,
   z^3 & \frac{2\,i }{5}\,{\sqrt{\frac{6}{6149}}}\,z^2 & -7\,i \,
   {\sqrt{\frac{42}{6149}}}\,z & \frac{49\,{\sqrt{\frac{21}{953095}}}\,z}{25} & 
   \frac{5\,i }{2}\,{\sqrt{\frac{105}{6149}}} & \frac{-3577\,
     {\sqrt{\frac{7}{953095}}}}{50} & \\[6pt] 0 & 3\,i \,{\sqrt{\frac{114}{54395}}}\,
   z^5 & -8\,i \,{\sqrt{\frac{38}{163185}}}\,z^4 & \frac{-26\,i }{5}\,
   {\sqrt{\frac{19}{32637}}}\,z^3 & \frac{14\,i }{5}\,
   {\sqrt{\frac{114}{10879}}}\,z^2 & \frac{-71\,i }{5}\,
   {\sqrt{\frac{38}{228459}}}\,z & \frac{-139\,{\sqrt{\frac{57}{11803715}}}\,z}
   {5} & 2\,i \,{\sqrt{\frac{95}{228459}}} & 417\,
   {\sqrt{\frac{19}{11803715}}} & \\[6pt] 
   & & & & & & & & & \ddots }
$
}
\mbox{}\hspace{.1in}\mbox{}
\rotatebox{90}{\parbox{9.3in}{Figure~13: Primary field $\tpf{-1/5}{-1/5}{0} (z)$ (in the $L_1$ basis) of the Yang-Lee theory. Notice that the squares of the 
coefficients in the first column give the expansion of $(1-u)^{2/5}$ in agreement with the two-point function (\ref{laur-tay2}).}}
\thispagestyle{empty}
\end{figure}


\section{Discussion}

In this paper we have presented a general  level-by-level algorithm to build matrix representations of the Virasoro algebra and fields for the $s\ell(2)$ minimal and parafermion models. We expect this algorithm to work generally for rational CFTs. 
Our results, however, are far from complete. 
Our algorithm appears to be consistent level-by-level, but we have no proof that the infinite matrices actually give a representation of the Virasoro algebra. Although it is probably not difficult to establish the convergence of our truncated matrices to order $q^N$, it would be highly desirable to have closed form expressions for the infinite matrices. Likewise, it is highly desirable to obtain explicit expressions for the Virasoro generators in terms of the fermion operators introduced in FPI~\cite{FPI}.  Unfortunately, both of these problems seem to be very difficult. 

In treating the fields we have for simplicity worked with one chiral half of the bulk fields. It is possible~\cite{FP} to extend this to a full treatment of the bulk fields for the periodic system but it would also be nice to give a proper treatment of the boundary fields on the cylinder. 
Lastly, in the case of the tricritical Ising and 3-state Potts models, there are extended chiral algebras, namely, the superconformal algebra and $W_3$ algebra. It would be nice to work out, at least level-by-level, the matrix representatives of the generators of these higher symmetry algebras.


\goodbreak

\section*{Acknowledgements} 
\label{sec:Acknowledgements}
This research is supported by the Australian Research Council. We thank Omar Foda, Vladimir Rittenberg and Ole Warnaar for discussions and Jean-Bernard Zuber for comments on the manuscript.


\begin{thebibliography}{99}
\bibitem{ABF} G. E. Andrews, R. J. Baxter and P. J. Forrester, J. Stat. Phys. {\bf 35}, 193--266 (1984); P. J. Forrester and R. J. Baxter, J. Stat. Phys. {\bf 38}, 435--472 (1985).
\bibitem{BPZ} A. A. Belavin, A. M. Polyakov and A. B. Zamolodchikov, Nucl. Phys. {\bf B241}, 333--380 (1984).
\bibitem{FMS} P. Di Francesco, P. Mathieu and D. S\'en\'echal, {\it Conformal Field Theory}, Springer (1996).
\bibitem{FPI} G. Feverati and P. A. Pearce, {\it Critical RSOS and Minimal Models I: Paths, Fermionic Algebras and Virasoro Modules} (2002).
\bibitem{BaxBook} R. J. Baxter, {\it Exactly Solved Models in Statistical Mechanics}, Academic Press, London (1982).
\bibitem{FNO} B. L. Feigin, T. Nakanishi and H. Ooguri, Int. J. Mod. Phys. {\bf A7} (Suppl. {\bf 1A}), 217--238 (1992).
\bibitem{RosgenV} M. R\"osgen and R. Varnhagen, Phys. Lett. {\bf B350} 203--11 (1995).
\bibitem{ItoyamaThacker} H. Itoyama and H. B. Thacker, Phys. Rev. Lett. {\bf 58}, 1395--1398 (1987); H. B. Thacker and H. Itoyama, Nucl. Phys. {\bf B5A} (Proc. Suppl.), 9--14 (1988).
\bibitem{KooS} W. M. Koo and H. Saleur, Nucl. Phys. {\bf B426}, 459--504 (1994).
\bibitem{ZamFat} A. B. Zamolodchikov and V. A. Fateev, Sov. Phys. JEPT {\bf 62}, 215--225 (1985).
\bibitem{BaxHH} R. J. Baxter, J. Phys. {\bf A13}, L61--L70 (1980).
\bibitem{spinons} F. D. M. Haldane, Z. N. C. Ha, J. C. Talstra, D. Bernard and V. Pasquier, Phys. Rev. Lett. {\bf 69}, 2021 (1992);
D. Bernard, V. Pasquier and D. Serban, Nucl. Phys. {\bf B428}, 612--628 (1994); 
P. Bouwknegt, A. W. W. Ludwig and K. Schoutens, Phys. Lett. {\bf B359}, 304--312 (1995);
P. Bouwknegt, L. Chim and D. Ridout, Nucl. Phys. {\bf B572}, 547--573 (2000).
\bibitem{Wolfram} Mathematica is a Copyright of Wolfram Research, Inc. 1988-2002.
\bibitem{FP} G. Feverati and P. A. Pearce, in preparation (2002).


\end{thebibliography}
\end{document}